\documentclass[aps, prd, twocolumn, showpacs, superscriptaddress, groupedaddress]{revtex4-1} 
\usepackage{graphicx}	
\usepackage{amssymb}
\usepackage{mathrsfs}
\usepackage{dcolumn}
\usepackage{color}
\usepackage{subfigure, rotating, bm, array}
\usepackage[pagebackref=false, colorlinks=true]{hyperref}
\hypersetup{
linkcolor=blue,     
citecolor=blue,     
urlcolor=blue} 

\newcommand{\cg}{\textnormal{\textsl{g}}}

\newcommand{\be}{\begin{equation}}
\newcommand{\ee}{\end{equation}}
\newcommand{\bea}{\begin{eqnarray}}
\newcommand{\eea}{\end{eqnarray}}
\newcommand{\bee}{\begin{enumerate}}
\newcommand{\eee}{\end{enumerate}}
\newcommand{\bei}{\begin{itemize}}
\newcommand{\eei}{\end{itemize}}
\def\mn{{\mu\nu}}

\def\boxx{\Box}
\def\mn{{\mu\nu}}

\def\p#1{{\partial}_{#1}}

\def\rhom{{\rho_{\rm m}}}
\def\rhor{{\rho_{\rm r}}}
\def\drhom{{\dot{\rho}_{\rm m}}}
\def\drhor{{\dot{\rho}_{\rm r}}}

\def\Om0L{\Omega_{\rm m0}^{\Lambda}}
\def\Or0L{\Omega_{\rm r,0}^{\Lambda}}
\def\Ol0L{\Omega_{\Lambda,0}^{\Lambda}}
\def\H0L{H_{0}^{\Lambda}}

\def\Omf0{\Omega_{\rm m0}^{f(R)}}
\def\Orf0{\Omega_{\rm r,0}^{f(R)}}
\def\Olf0{\Omega_{\Lambda,0}^{f(R)}}
\def\Hf0{H_{0}^{f(R)}}

\def\OmL0i{\Omega_{i,0}^{\Lambda}}
\def\Omf0i{\Omega_{i,0}^{f(R)}}
\usepackage{natbib}
\usepackage{comment}
\usepackage{amsmath}
\usepackage{graphicx}
\usepackage{soul}
\usepackage[normalem]{ulem}
\usepackage{cleveref}

\newcommand{\multiref}[2]{\ref{#1}-\ref{#2}}

\begin{document}
\title{Investigating the accelerated expansion of the Universe through updated constraints on viable $f(R)$ models within the metric formalism}

\author{Kumar Ravi$^{\,1}$}
\email[Corresponding Author: ]{cimplyravi@gmail.com}
\author{Anirban Chatterjee$^{\,2}$}
\email{anirbanc@iitk.ac.in}
\author{Biswajit Jana$^{\,1}$}
\email{vijnanachaitanya2020@gmail.com}
\author{Abhijit Bandyopadhyay$^{\,1}$}
\email{abhijit@rkmvu.ac.in}
\affiliation{$^{1}$Department of Physics, Ramakrishna Mission Vivekananda Educational and Research Institute, Belur Math 711202, West Bengal, India\\$^{2}$Department of Physics, Indian Institute of Technology Kanpur, Kanpur 208016, UP,  India}

\date{\today}

\begin{abstract}
Modified theories of gravity encompass a class of $f(R)$-models that seek to 
elucidate the observed late time accelerated expansion of the universe. 
In this study, we examine a set of viable $f(R)$ models (Hu-Sawicki: two cases, Satrobinsky, Tsujikawa, exponential and arcTanh models) in metric formalism, using recent cosmological data sets: type Ia supernovae data, cosmic chronometer observations, baryonic acoustic oscillations data,  data from H\textsc{ii} starburst galaxies, and local measurements of the Hubble parameter ($H_0$). We re-parameterize the $f(R)$ models using a distortion/deviation parameter ($b$) which is a measure of their deviation from the standard $\Lambda$CDM model. The model parameters are constrained using a Bayesian analysis with the Monte Carlo Markov Chain method. We employ statistical tools such as the Akaike Information Criterion, Bayesian Information Criterion, and reduced chi-square statistics to conduct a comparative investigation of these models. We determine the transition redshift, the evolution of total equation-of-state (EoS) parameter, and the EoS for the component responsible for current accelerated expansion to characterize the expansion's evolution. Taking into account the ``Hubble tension," we perform the study with and without a Gaussian prior for $H_0$ from local measurements. Our findings are as follows: (i) in many cases the $f(R)$ models are strongly favored over the standard $\Lambda$CDM model, (ii) the deviation parameter ($b$) significantly deviates from zero in several cases, (iii) the inclusion of local $H_0$ not only increases the fitted value of $H_0$ (as expected) but also affects the gap between predictions of $f(R)$ models and the $\Lambda$CDM model, and (iv) the relevant quantities characterizing the (accelerated) expansion of the universe obtained in our models are consistent with those obtained in a model-independent way by others. Our investigation and results present a compelling case for pursuing further research on $f(R)$ models with future observations to come.
\end{abstract}




\maketitle
\section{Introduction}
\label{intro}
In the last two decades, cosmological investigations have revealed  an accelerated expansion of the current universe and have also identified a transition from a decelerating phase to the current accelerated phase occurring during the late time phase of cosmic evolution.
The first empirical aspect of discovering this
phenomenon came from the interpretation of
luminosity distance and redshift measurements
of type Ia supernovae (SNe Ia) events
 \cite{SupernovaCosmologyProject:1997zqe, SupernovaCosmologyProject:1998vns, SupernovaSearchTeam:1998bnz, SupernovaSearchTeam:1998fmf}.
Furthermore, observation of baryon acoustic oscillations   \cite{SDSS:2005xqv, Percival:2006gs}, analysis of cosmic microwave background radiation   \cite{WMAP:2006bqn,WMAP:2008ydk,Planck:2013pxb,Planck:2015fie}, and examination of the power spectrum of matter distributions in the universe   \cite{2dFGRS:2005yhx,Percival:2006gt} 
have  substantiated  evidence of this late-time cosmic acceleration. 
A general label describing the origin of the observed
late time cosmic acceleration  is ``Dark energy" which refers to a theoretical  unclustered form of energy exerting a negative pressure to counteract the gravitational attraction and thereby causing the cosmic acceleration.\\

Despite extensive research over many years, the true nature and origin of dark energy remains an enigma. Various theoretical perspectives have emerged, each attempting to construct models that can explain the observed cosmic acceleration. 
The introduction of the $\Lambda$ term into Einstein's equation, known as the ``$\Lambda$ Cold Dark Matter ($\Lambda$CDM) model", is the simplest model that can explain the present accelerated expansion of the universe. However, this model encounters several challenges, such as the cosmic coincidence issue \cite{Zlatev:1998tr} and the fine-tuning problem \cite{Martin:2012bt} when considered in the context of particle physics.
This motivates development and exploration of alternative dark energy models from a range of perspectives.
An important class comprises field theoretic models that involve the incorporation of a scalar field within the energy-momentum tensor of the Einstein equation. The  scalar field plays the role in generating the necessary negative pressure to drive cosmic acceleration, either through the  slowly varying potentials (quintessence models \cite{Tsujikawa:2013fta}) or by means of their kinetic energy (k-essence models \cite{Armendariz-Picon:2000ulo}).

Within the context of this study, another crucial category of models under consideration is modified gravity models which attempt to explain the acceleration through the geometry itself
without modifying the energy-momentum tensor of the Einstein equation.
These models primarily involve modifications to the geometric component of Einstein's equation, which may result from higher-order corrections to the Einstein-Hilbert action.
By introducing suitable modifications, it becomes feasible to induce cosmic acceleration.   
The most straightforward types of modifications involve the extension of the Ricci scalar $R$ to  an arbitrary function $f(R)$. The meticulous choice with appropriate justification of this particular arbitrary function play a crucial role in all modified gravity models. 
Theoretical considerations, like the need for a ghost-free theory with stable perturbations and the presence of Noether symmetries, impose initial constraints on the forms of the arbitrary function $f(R)$.
However, ability to reproduce the observed features of cosmic evolution, the behavior of local (solar) systems etc. also serve as the primary tool to further constrain the 
$f(R)$-models.\\

In this study, our objective is to provide updated constraints on viable $f(R)$ gravity models using the latest cosmological data, including supernovae type Ia(SNIa) from Pantheonplus compilation, cosmic chronometer(CC) observations, baryonic acoustic oscillations(BAO) data, 
H\textsc{ii} starburst galaxies(H\textsc{ii}G) data, and 
data from local measurements of the Hubble parameter ($H_0$).  
Specifically, we focus on six viable $f(R)$ models: the Hu-Sawicki model(two cases), the Starobinsky model, the exponential model, the Tsujikawa model, and the arcTanh model. By considering these models, we obtained an update of best-fit values and uncertainties  at different confidence levels
of the associated parameters of the models.
We have chosen these particular $f(R)$ gravity models because most of the modified gravity models either fail to explain the matter-dominated era or have already been ruled out by observational data sets. The selected models, mentioned above, represent a few remaining options to study the impact of modified gravity theory within the framework of metric or Palatini formalism. For our investigation, we adopt the metric formalism techniques  in 
the context of a homogeneous and isotropic universe.
Rather than working with original forms of these $f(R)$ models(which for some models give a false impression that they are non-reducible to the $\Lambda$CDM model), we chose to re-parameterize these models in terms of what is called ``deviation/distortion parameter($b$)''. In this form, with $b\rightarrow0$, a $f(R)$ model tends to the $\Lambda$CDM model.

The current state of research in this specific context, which involves constraining various $f(R)$ models within the metric formalism using cosmologically relevant data, is experiencing a high level of activity. Here we refer to several such relevant previous works. In \cite{Basilakos:2013nfa}, the authors obtained constraints on the Hu-Sawicki and the Starobinsky models by utilizing approximate analytical solutions derived from Taylor series expansions for the Hubble parameter. The constraints were derived using SNIa data from Union 2.1 compilation, BAO data, cosmic microwave background (CMB) shift parameter data and growth rate data.
In \cite{Nunes:2016drj}, the constraints on Hu-Sawicki, Starobinsky, exponential, and Tsujikawa models were obtained by utilizing CC data, local measurements of $H_{0}$, SNIa data from the Joint-Lightcurve-Analysis (JLA) compilation, and BAO data.
Furthermore, \cite{Odintsov:2017qif} constrained the exponential model using SNIa (Union 2.1), CC, BAO, and CMB data, while also discussing the viability of this model in describing the entire cosmological history.
Exploration of various $f(R)$ models, including the Hu-Sawicki and the exponential models, with data sets such as SNIa (JLA), CC, BAO, CMB, local $H_{0}$, and growth rate, was conducted in \cite{Perez-Romero:2017njc}.
Other significant works that investigated Hu-Sawicki and/or exponential models, utilizing various data sets including gravitational lensing data, are \cite{DAgostino:2019hvh,DAgostino:2020dhv,Odintsov:2020qzd}.
In \cite{Farrugia:2021zwx}, constraints on Hu-Sawicki, Starobinsky, exponential, and Tsujikawa models were examined in both flat and non-flat spacetimes, utilizing various cosmological data sets.
Recently, \cite{Leizerovich:2021ksf} have advocated for the use of quasar X-ray and UV fluxes data to investigate $f(R)$ models and have constrained Hu-Sawicki and exponential models accordingly.
Moreover, the Taylor series approach from \cite{Basilakos:2013nfa} was extended to include the arcTanh model in \cite{Sultana:2022qzn}. Using Gaussian process reconstruction of the Hubble diagram with CC and H\textsc{ii}G data, the parameters of these three $f(R)$ models were determined.\\

This paper is organised as follows.
In Section \ref{frcosmo} we derive the modified Friedmann equations and other relevant equations from the action for $f(R)$ gravity.
Section \ref{frmodels} covers discussions of the conditions that any viable $f(R)$ models must satisfy, along with brief introductions of the specific $f(R)$ models investigated in this work.
In Section \ref{ocd}, we introduce the cosmological data sets and the corresponding equations that establish connection between theory and data. Additionally, we discuss the statistical procedures employed to obtain constraints on the model parameters.
The obtained constraints on the model parameters for all the models are presented in Section \ref{results}.
The performance of the different models is assessed using statistical tools in Section \ref{comparison}.
Section \ref{accel} is dedicated to deriving the relevant quantities that characterize expansion history of the universe based on the model constraints.
In Section \ref{concl} we provide the concluding remarks for this work.
Due to need for careful considerations in solving the modified Friedmann equations, such as avoiding numerical instabilities and minimizing computation time, Appendix \ref{odesol} is included to discuss the method used for numerically solving the Friedmann equations.
The data for baryonic acoustic oscillations and cosmic chronometer, collected from different sources, are tabulated in Appendix in the Tables \ref{BAOdata} and \ref{CCdata}, respectively. \\

Unless otherwise mentioned we have set $c=1$ (where, $c$ denotes speed of light in vacuum), and value of the Hubble parameter is expressed in the unit km\,s$^{-1}$Mpc$^{-1}$.

\section{$\lowercase{f}(R)$ cosmology in Metric formalism}
\label{frcosmo}
The $f(R)$ theories of gravity involve generalisation 
of Lagrangian by making it an  arbitrary function of the Ricci scalar $R$ where $f(R)=R$ corresponds to the standard Einstein theory of gravity. Algebraic expressions for $f(R)$,
different from $f(R)=R$, define different $f(R)$\,-\,models. 
The generalized  Einstein-Hilbert action for $f(R)$ theories is  
\begin{equation}
S = \frac{1}{2\kappa}\int\,d^{4}x\,\sqrt{-g}\,f(R) + S_{\rm m} + S_{\rm r}\,,
\label{actionP}
\end{equation}
where $\kappa=8\pi{G}$, $g$ is the determinant of the metric tensor, $S_{\rm m}$  and $S_{\rm r}$ are the actions for matter fields and radiation fields, respectively. 
The variation of this action   with respect to the metric 
gives the corresponding field equation in metric formalism  as 
\begin{equation}
FR_{\mn} - \frac{1}{2}f\cg_{\mn} - \left(\nabla_{\mu}\nabla_{\nu} - \cg_{\mn}\boxx \right)F  = {\kappa}T_{\mn}\,,
\label{FE00}
\end{equation}
where $F = \p{}f/\p{}R$, $\boxx\equiv\cg^{\mn}\nabla_{\mu}\nabla_{\nu}$ is the covariant 
D'Alambertian and  $T_{\mn}$ is the energy-momentum tensor of matter and radiation. 
This field equation can also be expressed as 
\begin{eqnarray}
&&FG_{\mn} - F^{\prime}\nabla_{\mu}\nabla_{\nu}R - F^{\prime\prime}(\nabla_{\mu}R)(\nabla_{\nu}R)\nonumber\\
&&+\cg_{\mn}\left[\frac{1}{2}(RF-f) + F^{\prime}\boxx{R} + F^{\prime\prime}(\nabla{R})^{2} \right]  = {\kappa}T_{\mn}\,,
\label{FE0}
\end{eqnarray}
where $G_{\mn}\equiv R_{\mn}-\frac{1}{2}R\cg_{\mn}$ is Einstein tensor and prime($^{\prime}$) denotes derivative with respect to $R$.  Taking  trace of both sides of Eq. \ref{FE00}  
we obtain
\begin{equation}
\boxx{R} = \frac{1}{3F^{\prime}}\left[\kappa{T} - 3F^{\prime\prime}(\nabla{R})^{2} + 2f - RF \right]\,,
\label{TrEq0}
\end{equation}
using which in  Eq. \ref{FE0} we rewrite the 
field equation as 
\begin{eqnarray}
G_{\mn} &=& \frac{1}{F}\bigg[F^{\prime\,}\nabla_{\mu}\nabla_{\nu}R + F^{\prime\prime}(\nabla_{\mu}R)(\nabla_{\nu}R)\nonumber\\
&&-\frac{\cg_{\mn}}{6}\left(RF+f + 2\kappa{T}\right) + {\kappa}T_{\mn}\bigg]\,,
\label{FE}
\end{eqnarray}
where $T\equiv T^{\mu}_{\mu}$ is the trace of energy-momentum tensor.

\par
In this work we consider the universe to be isotropic and homogeneous at
large scale and described by spatially flat Friedmann-Lema\^itre-Robertson-Walker (FLRW) metric
\begin{equation}
ds^{2} = -dt^{2} + a(t)^{2}\left[dr^{2} + r^{2}(d\theta^{2}+\sin^{2}\theta\,d\phi^{2}) \right]\,,
\label{metricEq}
\end{equation}
where $a$ is the FLRW scale factor. The Hubble parameter($H$) is defined 
as $H \equiv \dot{a}/a$
where we use a generic notation $\dot{X}$ to denote
derivative of any quantity $X$  with respect to cosmological time ($t$).
 In this formalism, for flat FRLW geometry, the Ricci scalar relates to the Hubble parameter by the   relation
\begin{equation}
R = 6\left(2H^{2} + \dot{H}\right)\,.
\label{Ricci}
\end{equation}
We consider the  content of universe to be a perfect fluid comprising of two 
components: radiation and matter in the form of pressureless dust (non-relativistic).
For the epochs($0\leq{z}<10^{4}$) when any interaction between matter and radiation could be ignored, the energy-momentum tensor for the fluid can be written as $T^{\mu}_{\nu}=\text{diag}(-\rho,p,p,p)$ with $\rho=\rho_{\rm m}+\rho_{\rm r}$ and $p=p_{\rm r}$ (with radiation
pressure $p_{\rm r}=\rho_{\rm r}/3$), where the subscripts  `m'  and  `r'  stand  for matter and radiation, respectively. Each  of the non-interacting
components separately follows the continuity equations 
\be
\drhom + 3H\rhom = 0,\quad\drhor + 4H\rhor=0\,.
\label{consvEq}
\ee
The solution to these conservation equations are $\rho_{\rm m} = 
\rho_{\rm m0}/a^{3}$, $\rho_{\rm r} = \rho_{\rm r0}/a^{4}$ where subscript 
`0' denotes values at present epoch. 
\par

In the context of the FLRW universe filled with an ideal perfect fluid characterised 
by $\rho$ and $p$, Eq. \ref{TrEq0} reduces to
\be
F^{\prime}\ddot{R} + F^{\prime\prime}\dot{R}^{2} = - \left[\frac{\kappa(3p-\rho)}{3} + \frac{2f-RF}{3} \right] - 3HF^{\prime}\dot{R}\,.
\label{TrEq}
\ee
using which  
the `00' and `$ii$' components of the field  Eq. \ref{FE} take following respective 
forms:
\begin{eqnarray}
-3H^{2} &=& -\frac{1}{F}\bigg[{\kappa}\rho +\frac{RF-f}{2} - 3H{F^{\prime}}\dot{R}\bigg]\,,
\label{FE111}\\
-2\dot{H}  &=& \frac{1}{F}\bigg[\kappa(\rho+p) + F^{\prime\prime}\dot{R}^2 + (\ddot{R} - H\dot{R})F^{\prime}\bigg]\,.
\label{FE222}
\end{eqnarray}
The Eqs. \ref{FE111} and \ref{FE222} are modified form of the Friedmann equations
for $f(R)$-models. The 
temporal profile of the Hubble parameter $H$  is
commonly expressed in the form  $H(z)$,
$z$ being the redshift related to the FLRW scale factor by $1 + z = 1/a$ (where, $a$
is normalised to unity at the present epoch). We require this profile of $H(z)$
from the cosmological data sets for obtaining observational constraints 
on  $f(R)$\,-\,models. For this purpose one may solve either the system of: (i) Eqs. \ref{TrEq}, \ref{FE111} and \ref{FE222} or, (ii) Eqs. \ref{Ricci}, \ref{FE111} and \ref{FE222}.
We take the path (ii) to solve the system.
Eq. \ref{FE111} serves as a constraint equation which fixes the initial conditions and must be satisfied at every integration step during the process of finding solutions. Finding analytical solutions is almost impossible, and therefore numerical methods are employed. However, solving this system of ordinary differential equations (ODEs) using a naive approach often leads to numerical instability. Additional details on how to solve this system
of ODEs are discussed in Appendix \ref{odesol}.\\

If we compare the modified Friedmann Eqs. \ref{FE111} and \ref{FE222} to the 
usual Friedmann equations with a dark energy component
characterised by energy density $\rho_{\rm DE}$  and pressure $p_{\rm DE}$, \textit{i.e.}
with the equations $3H^{2} = \kappa\left(\rho_{\rm m}+\rho_{\rm r}+\rho_{\rm DE}\right)$ and $-2\dot{H} = 
\kappa\left(\rho_{\rm m}+\rho_{\rm r}+\rho_{\rm DE}+p_{\rm m}+p_{\rm r}+p_{\rm DE}\right)$, 
we can deduce the ``effective (geometric) dark energy'' with density  and pressure
corresponding to $f(R)$-theory as 
\be
\rho_{\rm DE} = \frac{1}{\kappa}\left[\frac{RF-f}{2} - 3HF^{\prime}\dot{R} + 3(1-F)H^{2}\right]\,,
\label{rho_DE}
\ee
and,
\be
\begin{split}
p_{\rm DE} &= \frac{1}{\kappa}\bigg[ \frac{f-RF}{2} +F^{\prime}\ddot{R} + 2HF^{\prime}\dot{R} +F^{\prime\prime} \dot{R}^{2}\\
&\hspace{1cm}-(1-F)(2\dot{H}+3H^{2})\bigg]\,,
\end{split}
\label{p_DE}
\ee 
with the equation-of-state parameter for this effective dark energy  defined as
\be
w_{\rm DE} = \frac{p_{\rm DE}}{\rho_{DE}}\,.
\label{w_DE0}
\ee
Using Eqs. \ref{FE111} and \ref{FE222}, we may recast Eq. \ref{w_DE0}  into a more computationally advantageous form  (which we use later) as 
\be
w_{\rm DE} = \frac{w_{\rm tot}-{\kappa}p_{\rm r}/(3H^{2})}{1- \kappa\left(\rho_{\rm m}+\rho_{\rm r}\right)/(3H^{2})}\,,
\label{w_DE1}
\ee
where
\be
w_{\rm tot} = -1 + \frac{2(1+z)}{3H}\frac{dH}{dz}\,,
\label{w_tot}
\ee
and we have taken $p_{\rm m}=0$ for the pressureless matter.
\par 
The relevant quantities obtained from observations,  depending on cosmological models or   even through cosmography (a model-independent kinematical approach), indicate that the Universe has recently undergone a transition  from a phase of decelerated expansion to accelerated expansion. These quantities include $w_{\rm tot}|{z=0}\sim-0.7$, $w_{\rm DE}|_{z=0}\sim-1$, and a transition redshift $(z{\rm t})\sim 0.5-1$. The transition redshift signifies the redshift at which the transition from decelerated to accelerated expansion occurred and is determined by the zero-crossing of the deceleration parameter ($q(z)$) given by
\be
q(z) \equiv -\frac{\ddot{a}}{aH^{2}} = -1 + \frac{(1+z)}{H}\frac{dH}{dz}\,.
\ee
In the $\Lambda$CDM model, the value of $w_{\rm DE}$ is fixed at $-1$ for all redshifts due to the presence of the cosmological constant ($\Lambda$) term. However, in $f(R)$ models, the source of $w_{\rm DE}|_{z=0}\sim-1$ arises from the underlying geometry itself. Unlike the $\Lambda$CDM model, there is no need to invoke the existence of ``dark energy" in $f(R)$ 
theories of gravity.

\section{The Specific $\lowercase{f}(R)$ Models And Their Viability Conditions}
\label{frmodels}
In this section we provide a brief introduction to the models examined in this study, both in their originally proposed forms and their subsequent transformations into more generalized representations.
These transformations aim to highlight how these models can be more readily reduced to the 
$\Lambda$CDM model under appropriate conditions. However, before going 
into the details of each model, it is essential to discuss the viability conditions.
\par 
In the metric formalism (unlike Palatini formalism), any viable $f(R)$ cosmological model must satisfy the following set of stringent  theoretical conditions (see \cite{DeFelice:2010aj,Amendola:2015ksp} for detailed discussions): 
\begin{eqnarray}
&& F>0\,,\quad\mathrm{for}\,\,\,R\geq{R}_{0}>0\,,
\label{vcond1}\\
&& F^{\prime}>0\,,\quad\mathrm{for}\,\,\,R\geq{R}_{0}>0\,,
\label{vcond2}\\
&& 
f(R) \approx R - 2\Lambda\,,\quad\mathrm{for}\,\,\,R\gg{R}_{0}\,,
\label{vcond3}\\
\mbox{and, }&& 
0<\frac{RF^{\prime}}{F}(r)<1\quad\mathrm{at}\,\,\,r = -\frac{RF}{f}=-2\,,
 \label{vcond4}
\end{eqnarray}
%
where $R_{0}$ denotes the Ricci scalar at the present epoch. \\

These conditions arise from various considerations. Firstly, the effective gravitational
constant ($G_{\rm eff}=G/F$) must be positive, ensuring the avoidance of anti-gravity (as 
expressed in Eq. \ref{vcond1}). Besides, any acceptable $f(R)$-model should exhibit stability 
under perturbations and avoid the instability of Dolgov-Kawasaki type (Eq. \ref{vcond2}). 
Similar to the $\Lambda$CDM model, a viable $f(R)$ model must also be consistent with local 
gravity tests (Eqs. \ref{vcond2} and \ref{vcond3}). Furthermore, the existence of a 
matter-dominated epoch in cosmological dynamics necessitates that an $f(R)$ model satisfies the 
conditions given by Eqs \ref{vcond2} and \ref{vcond3}. Lastly, Eq. \ref{vcond3} 
ensures the stability of the late-time de-Sitter solution, 
from which the late-time accelerated expansion of the 
Universe is usually inferred.\\

The viability conditions mentioned above can be more 
easily assessed for an $f(R)$-model if we can somehow 
transform that model into the following form:
\be
f(R) = R - 2\Lambda{y(R,b,\Lambda)}\,,
\label{devEq}
\ee
where the function ${y(R,b,\Lambda)}$ serves to measure  the deviation from the $\Lambda$CDM model, with the parameter $b$ (referred to as the ``distortion/deviation parameter") specifically quantifying the extent of the deviation \cite{Basilakos:2013nfa}. For this reason, all the models examined in this study are recast into the form of Eq. \ref{devEq}, allowing for a clearer evaluation of their compliance with the viability conditions.\\

More specifically, the viability conditions  in Eqs. \ref{vcond2} and \ref{vcond3} can be alternatively written as
$\displaystyle \lim_{R\to\infty} f(R) = R + C_{0}$, where $C_{0}$ represents a constant value. In order for a candidate $f(R)$-model to exhibit asymptotic behavior similar to the standard $\Lambda$CDM model (which is supported by observations of the cosmic microwave background), we can identify that $C_{0}=-2\Lambda$, 
where $\Lambda$ corresponds to the cosmological constant introduced by Einstein and Hilbert in their action.
This amounts to writing symbolically $\Lambda^{\Lambda} = \Lambda^{f(R)}$ where the superscripts $\Lambda$ 
and $f(R)$ denote  quantities in reference to the $\Lambda$CDM model and any viable $f(R)$\,-\,model, respectively.
This relation may also be written as
\be
\Ol0L(\H0L)^{2} = \Olf0(\Hf0)^{2}\,,
\label{LfR1}
\ee
where subscript `0' denotes values at present time. 
Furthermore, considering the definition of the energy-momentum
tensor and the resulting conservation equation(Eq. \ref{consvEq}), we can
infer that both the $\Lambda$CDM model and any
viable $f(R)$-model  yield identical matter density
and radiation density at the current epoch \textit{i.e.}
\be
\OmL0i(\H0L)^{2} = \Omf0i(\Hf0)^{2} = \frac{8\pi{G}}{3}\rho_{i,0}\,,
\label{LfR2}
\ee
where the subscript $i=(\mbox{m, r})$ stands for 
(matter, radiation). 
Also  any viable $f(R)$-model is expected to
exhibit deviations from the standard $\Lambda$CDM model predominantly during  the late times, 
in order that the explanation for accelerated expansion at present epoch comes from vanishing $\Lambda$. In general terms, this implies 
\be
\OmL0i \neq \Omf0i\,,\quad \H0L \neq \Hf0\,.
\label{LfR3}
\ee
Based on Eqs. \ref{LfR1}-\ref{LfR3} and the initial condition requirements for solving the ODE system described in Eqs. \ref{Ricci}, \ref{FE111}, and \ref{FE222} (see Appendix \ref{odesol}), we utilize parameters ($\Om0L,\,b,\,\H0L$) for model-fitting to the data. However, while reporting our findings, we express the results in terms of the parameters ($\Omega_{\rm m0}^{f(R)},\,b,\,\Hf0$), which are determined using Eq. \ref{LfR2}.

\subsection{The Hu-Sawicki Model}
The Hu-Sawicki model, initially proposed in \cite{Hu:2007nk}, is described by the following equation:
\be
f(R)_{\rm HS} = R - \mu^{2}\frac{c{1}(R/\mu^{2})^{n_{{\rm HS}}}}{1+ c{2}(R/\mu^{2})^{n_{{\rm HS}}}}\,,
\label{HSmodel0}
\ee
where $c_{1}$ and $c_{2}$ are dimensionless parameters, ${n_{{\rm HS}}}$ is a positive constant typically assumed to be an integer, and $\mu^{2}\approx\Omega_{\rm m0}H_{0}^{2}$. By defining $\mu^{2}c_{1}/2c_{2} \equiv \Lambda$ and $2\left(c_{2}^{1-1/{n_{{\rm HS}}}}\right)/c_{1} \equiv b$, we can express Eq. \ref{HSmodel0} as   \cite{Basilakos:2013nfa}:
\be
f(R)_{\rm HS} = R - 2\Lambda
\left[1 - \left\{1+\left(\frac{R}{b\Lambda}\right)^{{n_{\rm HS}}}\right\}^{-1}\right]\,.
\label{HSmodel1}
\ee
We identify the parameter $\Lambda$ as the usual cosmological constant, and $b$ as the deviation parameter, which indicates the
 model's deviation from the $\Lambda$CDM model.
In order to satisfy the viability conditions $F>0$ and 
$F^{\prime}>0$ for $R\geq{R}_{0}$, it is necessary to consider $b>0$ (especially when ${n_{{\rm HS}}}$ is an odd integer). 
Although many researchers have constrained the scenarios where 
${n_{{\rm HS}}}=1$ and/or $2$ 
\cite{Basilakos:2013nfa,Nunes:2016drj,DAgostino:2019hvh,DAgostino:2020dhv,Farrugia:2021zwx,Leizerovich:2021ksf,Sultana:2022qzn}, 
they have acknowledged computational challenges as a hindrance to explore the case of ${n_{{\rm HS}}}=3$ 
(further elaboration on this point is provided in Sec. \ref{results}). 
In this investigation, we have also imposed constraints on the case ${n{_{\rm HS}}}=3$.

\subsection{The Starobinsky Model}
The model proposed by Starobinsky \cite{Starobinsky:2007hu}  is
\be
f(R)_{\rm ST} = R - {\lambda}R_{\rm S}\left[1 - \left(1 + \frac{R^{2}}{R_{\rm S}^{2}}\right)^{-n_{_{\rm S}}}\right]\,,
\label{STmodel0}
\ee
where ${n_{_{\rm S}}}$ is a positive constant, $\lambda(>0)$ and $R_{\rm S}\approx R_{0}$ are free 
parameters (where $R_{0}$ denotes the Ricci scalar at present epoch).
 This model too can be reformulated in a more general form as \cite{Basilakos:2013nfa}
\be
f(R)_{\rm ST} = R - 2\Lambda\left[1 - \left\{1+\left(\frac{R}{b\Lambda}\right)^{2}\right\}^{-{{n_{_{\rm S}}}}}\right]
\label{STmodel1}
\ee
with $\Lambda = {\lambda}R_{\rm S}/2$ and $b=2/\lambda$. 
In this study, we have imposed constraints on the cases where ${{n_{_{\rm S}}}}=1$
and the reason for not exploring higher values of ${{n_{_{\rm S}}}}$ 
is explained later in Sec.\ \ref{results}. 
Note that the Hu-Sawicki model (Eq. \ref{HSmodel1}) with 
${{n_{_{\rm HS}}}}=2$ and the Starobinsky model (Eq. \ref{STmodel1}) 
with ${{n_{_{\rm S}}}}=1$ are equivalent.
Unlike the Hu-Sawicki model (with $n_{_{\rm HS}}=1$), the viability condition for the Starobinsky model does not require $b>0$. 
Based on the algebraic form of the Starobinsky model (Eq. \ref{STmodel1}), we can infer that regardless of the data used, the parameter $b$ must exhibit a symmetric distribution (centered around $b=0$) from the MCMC fitting procedure. 
Since our interest lies in investigating deviations from the $\Lambda$CDM model supported by the data, we considered $b>0$ without loss of generality.

\subsection{The Exponential Model}
The exponential model, initially proposed as a viable $f(R)$ model in \cite{Cognola:2007zu}, has been further investigated in \cite{Linder:2009jz,Chen:2014tdy,Odintsov:2017qif,Leizerovich:2021ksf} (also see references therein), is given by  
\be
f(R)_{\rm E} = R + \alpha\left[\exp(-\beta{R}) - 1\right]\,,
\label{expmodel0}
\ee
where $\alpha$ and $\beta$ are the parameters of this model.
For large  $R$, an acceptable $f(R)$-model must approximately resemble the $\Lambda$CDM, which is achievable only when $\alpha>0$ and $\beta>0$. By substituting  
$\Lambda = \alpha/2$ and $b=2/(\alpha\beta)$
the exponential model can be expressed as  
\be
f(R)_{\rm E} = R - 2\Lambda\left[1 - \exp\left(-\frac{R}{b\Lambda}\right)\right]\,,
\label{expmodel1}
\ee
from where it becomes evident that when $R$ becomes significantly larger than $b\Lambda$
($R\gg{b}\Lambda$), the function $f(R)_{\rm E}$ approaches $R - 2\Lambda$.

\subsection{The Tsujikawa Model}
Tsujikawa  proposed an alternative model \cite{Tsujikawa:2007xu} as 
\be
f(R)_{\rm T} = R - {\xi}R_{\rm T}\tanh\left(\frac{R}{R_{\rm T}}\right)\,,
\ee
where $\xi(>0)$ and $R_{\rm T}(>0)$ are the model parameters.
With $\Lambda = {\xi}R_{\rm T}/2$ and $b=2/\xi$, the model can be rewritten as 
\be
f(R)_{\rm T} = R - 2\Lambda\tanh\left(\frac{R}{b\Lambda}\right)\,.
\label{tsujimodel1}
\ee
We see clearly that when the parameter $b \to 0$ (which corresponds to $\xi \to \infty$, $R_{\rm T} \to 0$, while ${\xi}R_{\rm T}$ remains finite), the model 
reduces to $f(R)_{\rm T} = R - 2\Lambda$.

\subsection{The ArcTanh Model}
In this study we also examined a model proposed in \cite{Perez-Romero:2017njc}
as
\be
f(R)_{\rm aTanh} = R - \frac{2\Lambda}{1+b\,\text{arctanh}\left(\frac{\Lambda}{R}\right)}\,,
\label{aTanhmodel}
\ee
where the parameter $b$ is required to be positive, in order to 
prevent any occurrence of future singularities.


\section{Observed Cosmological Data}
\label{ocd}
In this section we present a concise overview of the cosmological data sets 
utilized in this study. For any given $f(R)$ model, the system of   Eqs. \ref{Ricci}, \ref{FE111} and \ref{FE222}   is solved, primarily yielding the function $H(z)$. 
So, it becomes necessary to outline the theoretical equations 
connecting $H(z)$ with various observed quantities. Additionally, at the end of this section, a brief introduction is provided on the statistical techniques employed to extract model parameters from the data.

\subsection{Type Ia Supernova Data}

Type Ia Supernovae (SNIa), known as standard candles \cite{Phillips:1993ng}, have significantly contributed to our comprehension of cosmology. The SNIa observations \cite{SupernovaCosmologyProject:1998vns,SupernovaSearchTeam:1998fmf} played a pivotal role in the discovery of the accelerated expansion of the present-day universe. 
In this study we used the apparent magnitude data for SNIa obtained from the recently 
released PantheonPlus compilation \cite{Scolnic:2021amr}. This compilation comprises of
1701 distinct light curves of 1550 unique spectroscopically confirmed SNIa, 
sourced from 18 surveys. This compilation provides SNIa data within the range 
$0.00122 < z_{\rm HD} <2.26137$, where $z_{\rm HD}$ denotes the Hubble diagram redshift. 
It offers a considerably larger number of low-redshift data compared to the 
previous Pantheon compilation \cite{Pan-STARRS1:2017jku}. 
In our current work, we employ the apparent magnitude at maximum brightness ($m_{\rm b}$), heliocentric redshift ($z_{\rm hel}$), cosmic microwave background corrected redshift ($z_{\rm cmb}$), and the total (statistical + systematic) covariance matrix  from this compilation \cite{Scolnic:2021amr}.\\

The theoretical definition of the apparent magnitude involves  the luminosity distance, which is given by the integral expression: 
 \begin{eqnarray}
d_{\rm L}(z) &=& c(1 + z_{\rm hel})\int_{0}^{z_{\rm cmb}}\frac{dz'}{H(z')}\,,
\label{d_L}
\end{eqnarray}
where 
the function $H(z)$ is obtained by solving the 
system of  
 Eqs. \ref{Ricci}, \ref{FE111} and \ref{FE222}.
 The apparent magnitude is defined as
\begin{equation}
m_{\rm th} = M +5\log_{10}\left(\frac{c/H_{0}}{\text{Mpc}}\right) + 5\log_{10}\left(D_{\rm L}(z)\right) + 25\,,
\label{mth}
\end{equation} 
where $M$ represents the absolute magnitude, and $D_{\rm L}(z) \equiv H_{0}d_{\rm L}(z)/c$ is the dimensionless Hubble-free luminosity distance.
The computation of apparent magnitude involves a degeneracy between the absolute magnitude ($M$) and the Hubble parameter at the current epoch ($H_{0}$), as evident from the Eq. \ref{mth}.
Therefore, after marginalizing over these nuisance parameters ($M$ and $H_{0}$), the appropriate residual that needs to be minimized for model fitting is given by \cite{SNLS:2011lii}   
\begin{equation}
\tilde{\chi}^{2}_{\rm sn} = A - \frac{B^{2}}{D} + \log\frac{D}{2\pi}\,,
\label{chiSN}
\end{equation} 
where $A = (m_{\rm b}-m_{\rm th})^{T}C^{-1}(m_{\rm b}-m_{\rm th})$, $B = (m_{\rm b}-m_{\rm th})^{T}C^{-1}\boldsymbol{1}$, and $D = \boldsymbol{1}^{T}C^{-1}\boldsymbol{1}$ and $C$ represents the total covariance matrix  of the data (provided in the Pantheon+ compilation). $\boldsymbol{1}$ is an array of ones of length equal to the number of data points.


\subsection{Cosmic Chronometers Data}

By examining the differential age evolution \cite{Jimenez:2001gg, Wei:2018cov, Simon:2004tf} of old elliptical galaxies, where star formation and interactions with other galaxies have ceased, previous studies have provided 32 data points for $H(z)$ in the redshift range of 0.07-1.965 \cite{Simon:2004tf,Moresco:2016mzx, Moresco:2015cya, Zhang:2012mp, Stern:2009ep, Moresco:2012jh, Ratsimbazafy:2017vga,Borghi:2021rft} (compiled in the Table \ref{CCdata} of the Appendix). This so called differential age method, employed in these studies, uses the   relation 
\begin{equation}
H(z) = -\frac{1}{1+z}\frac{dz}{dt} \simeq -\frac{1}{1+z}\frac{\Delta{z}}{\Delta{t}}\,,
\end{equation}
to obtain $H(z)$. 
The parameters of any model are estimated by minimizing the following residual:
\begin{equation}
\chi^{2}_{_{\rm CC}} = \sum_{i=1}^{32}\frac{\left(H_{\rm obs}(z_{i}) - H_{\rm th}(z_{i})\right)^{2}}{\sigma_{H,i}^{2}}\,,
\label{chiCC}
\end{equation}
where $H_{\rm obs}(z_{i})$'s are the observed values of the Hubble parameter function at redshift $z_{i}$, while $\sigma_{H,i}^{2}$ denotes the corresponding uncertainties associated with the measurements of $H_{\rm obs}(z_{i})$. The theoretical Hubble function at redshift $z_{i}$, which is model-dependent and  obtained from the solutions of Eqs. \ref{Ricci}, \ref{FE111}, and \ref{FE222}, is denoted by $H_{\rm th}(z_{i})$ in the above Eq. \ref{chiCC}.

\subsection{BAO Data}

In the Big Bang Model of the universe, prior to decoupling of matter and radiation components, the contents of the Universe were evenly distributed, albeit with small  fluctuations. 
Photons and baryons were strongly  coupled through Thomson scattering. 
 As the universe expanded and cooled, resulting in a decrease in temperature and density, the fluctuations were amplified by gravity. 
  The gravitational pull caused the tightly bound photon-baryon mixture  to condense in regions with higher densities, resulting in compressions and rarefactions in the form of acoustic waves known as Baryonic Acoustic Oscillations (BAO).  
 Matter and radiation were then decoupled and this epoch which is marked
by the release of baryons from the  Compton drag of photons is known as
drag epoch ($z_{\rm d}$), after which the photons travelled freely, whereas 
acoustic waves remained frozen in the baryons.
The length scale 
characterizing the maximum distance traveled by the acoustic wave before 
decoupling is known as sound horizon at the epoch of drag ($r_{\rm d}$).  BAO, therefore,
 holds the status of standard ruler for length scale in 
Cosmology \cite{Hu:1995en,  Eisenstein:1997ik}.\\

We have compiled a collection of 30 data points representing various BAO observables from a range of surveys, as documented in the literature 
\cite{Beutler:2011hx,SDSS:2009ocz,Padmanabhan:2012hf,Tojeiro:2014eea,Anderson:2012sa,Bautista:2017wwp,Seo:2012xy,Sridhar:2020czy,deSainteAgathe:2019voe,Carter:2018vce,Ross:2014qpa,Kazin:2014qga,Ata:2017dya,DES:2017rfo,Bautista:2017zgn,duMasdesBourboux:2017mrl,BOSS:2013igd,BOSS:2016wmc,Bautista:2020ahg,Neveux:2020voa}. These data points, which are used 
in our current study, are listed in the Table \ref{BAOdata}(in the Appendix). In our calculations for the 
drag epoch $z_{d}$ and the sound horizon at the epoch of drag $r_{d}$, we employ improved 
fits from \cite{Aizpuru:2021vhd}. The BAO observables include the Hubble distance,
which is defined as 
$D_{\rm H} = c/H(z)$, the transverse comoving distance ($D_{\rm M}(z)$), the angular 
diameter distance ($D_{\rm A}(z)$), and the volume-averaged distance ($D_{\rm V}(z)$) 
which are  defined as 
\begin{equation}
D_{A}(z) =\frac{c}{1+z}\int_{0}^{z}\frac{dz'}{H(z')}\,,
\label{D_A}
\end{equation}
\be
D_{M}(z) = (1+z)D_{A}(z)\,,
\label{D_M}
\ee
and,
\begin{equation}
D_{V}(z) = \left[(1+z)^{2}D_{A}^{2}(z)\frac{z}{H(z)}\right]^{1/3}\,.
\label{D_V}
\end{equation}
Once again $H(z)$ in the above relations is obtained by solving Eqs. \ref{Ricci}, \ref{FE111}, and \ref{FE222}.\\

For fitting any model to the uncorrelated data points of the BAO data, the following residual has been used
\begin{equation}
\chi^{2}_{_{\rm BAO-UC}} = \sum_{i=1}^{20}\left[\frac{A_{\rm th}(z_{i}) - A_{\rm obs}(z_{i})}{\sigma_{i}}\right]^{2}\,,
\label{chiBAO-UC}
\end{equation}
where $A_{\rm obs}(z_{i})$ and $\sigma_{i}$ respectively
 denote the observed values and their uncertainties
at redshift $z_{i}$ and $A_{\rm th}$ denotes the theoretical prediction from model under consideration. These quantities are given in the columns 2-4 of the Table \ref{BAOdata}. 
For correlated data points the appropriate residual to be minimized is
\begin{equation}
\chi^{2}_{_{\rm BAO-C}} = \sum_{j=1}^{4}\left[\left({\bf A}_{\rm th} - {\bf A}_{\rm obs}\right)_{j}^{T}{\bf C}_{j}^{-1}
\left({\bf A}_{\rm th} - {\bf A}_{\rm obs}\right)_{j}\right]\,,
\label{chiBAO-C}
\end{equation}
where ${\bf C}_{j}$'s are  the covariance matrices of the 4 different data-sets and $\left({\bf A}_{\rm th} - {\bf A}_{\rm obs}\right)_{j}$ denotes an array of difference between observed and theoretical values for each of the 4 data-sets. These data and the
covariance matrix can be found from the corresponding source papers referred in the last column of Table \ref{BAOdata}.
For the whole data-set of BAO, the $\chi^{2}$ which is to be minimized is 
\be
\chi^{2}_{_{\rm BAO}} = \chi^{2}_{_{\rm BAO-UC}} + \chi^{2}_{_{\rm BAO-C}}\,.
\label{chiBAO}
\ee

\subsection{H\,$\textsc{\textmd{ii}}$ starburst Galaxies Data}

H\textsc{ii} galaxies (H\textsc{ii}G) are   massive and compact starburst structures
surrounded by ionized hydrogen gas.  Their optical emission spectra exhibit strong and narrow Balmer ${\rm H}\alpha$ and ${\rm H}\beta$ lines, along with a weak continuum. 
The cosmological significance of   H\textsc{ii}G observation comes from the empirically established correlation between the luminosity ($L$) of the ${\rm H}\beta$ lines and the velocity dispersion ($\sigma$, which is a measure of the width of the spectral lines). This correlation is attributed to the fact that an increase in the mass of the starburst component leads to a simultaneous increase in the number of ionized photons (and thus the luminosity $L$) and the turbulent velocity (hence, the velocity dispersion $\sigma$) (see \cite{Gonzalez-Moran:2021drc} and  references therein).\\

In this study, we have used a total of 181 data points from \cite{Erb:2006tg, Maseda:2014gea, Masters:2014gna, Gonzalez-Moran:2019uij, Gonzalez-Moran:2021drc} corresponding to the emission of the Balmer $\rm{H}\beta$ line. These data points span a redshift range of $0.0088<z<2.5449$. It is important to note that the redshift coverage provided by the latest SNIa data is as follows: we have 8 data points for $z_{\rm SN}>1.4$, while for $z_{\rm H\textsc{ii}G}>1.4$ we have 69 data points, and for $z_{\rm H\textsc{ii}G}>z_{\rm SN,\,max}$ we have 11 observations of H\textsc{ii}G. Consequently, the observations of H\textsc{ii}G not only explore  hitherto unexplored   higher redshift regions compared to SNIa but also provide a denser coverage of these higher redshift regions in comparison to SNIa.\\

The empirical relation between the luminosity $L$ and the dispersion of velocity $\sigma$ is given by
\be
\log\left[\frac{L}{\mathrm{erg/s}}\right] = \beta\log\left[\frac{\sigma}{\mathrm{km/s}}\right]+\alpha\,,
\ee
where we take $\beta = 5.022\pm0.058$ and $\alpha=33.268\pm0.083$ from \cite{Gonzalez-Moran:2019uij,Gonzalez-Moran:2021drc}. From the luminosity distance of H\textsc{ii}G, $d_{\rm L} = \left[L/(4{\pi}F)\right]^{1/2}$, we can derive the distance modulus as
\be
\mu_{o} = 2.5\left(\alpha + \beta\log\left[\frac{\sigma}{\mathrm{km/s}}\right] - \log\left[\frac{\sigma}{\mathrm{erg/s/cm^{2}}}\right] - 40.08\right).\label{mu_HIIG}
\ee

Using Eq. \ref{mu_HIIG}, we compute the distance moduli of H\textsc{ii}G from the observables 
$L$, $\sigma$, and $F$. For any given cosmological model with theoretical distance moduli $\mu_{\theta}$ (represented as $m_{\rm th}-M$ in Eq. \ref{mth}), the parameters associated with that model can be constrained by minimising the following $\chi^{2}$ function
\be
\chi^{2}_{_{\mathrm{H}\textsc{ii}\mathrm{G}}} = \sum_{i=1}^{181}\frac{\left[\mu_{o} - \mu_{\theta}\right]^{2}}{\epsilon^{2}}\,,\label{chiHIIG}
\ee
where
\be
\epsilon^{2} = \epsilon_{\mu_{o,\mathrm{stat}}}^{2} + \epsilon_{\mu_{\theta,\mathrm{stat}}}^{2} + \epsilon_{\mathrm{sys}}^{2}\,,
\ee
with
\be
\epsilon_{\mu_{o,\mathrm{stat}}}^{2} = 6.25\left(\epsilon_{\log{F}}^{2} +  \beta^{2}\epsilon_{\log{\sigma}}^{2} + \epsilon_{\beta}^{2}(\log{\sigma})^{2} + \epsilon_{\alpha}^{2}\right)\,,
\ee
\be
\epsilon_{\mu_{\theta,\mathrm{stat}}}^{2} = \left[\frac{5}{\ln10}\left(\frac{c(1+z)}{d_{L}(z)H(z)} + \frac{1}{1+z} \right)\sigma_{z}\right]^{2},
\ee
and $\epsilon_{\mathrm{sys}}=0.0.257$ as suggested in \cite{Chavez:2016epc,Gonzalez-Moran:2019uij,Gonzalez-Moran:2021drc}. 
The uncertainty contribution of $\epsilon_{\mu_{\theta,\mathrm{stat}}}$ in the distance modulus is due to the uncertainties in redshifts($\sigma_{z}\sim10^{-4}$) of H\textsc{ii}G and is derived from simple error propagation theory.

\subsection{Local Measurement of $H_{0}$}

A contentious issue remains regarding the current value of the Hubble parameter $H_{0}$. The Planck constraints provide a value of $H_{0}=67.4\pm 0.5$ \cite{Planck:2018vyg}, while the locally measured value by the SH0ES collaboration yields $H_{0}=73.04\pm 1.04$ \cite{Riess:2021jrx}. The discrepancy between these two measurements, known as the Hubble tension, is significant at the $4-5\sigma$ level. To address this tension, we have included
in our study both scenarios - with  vis-a-vis  without the SH0ES prior for $H_{0}$.
The contribution of this data point to the total $\chi^{2}$  
is given by
\be
\chi^{2}_{\rm SH0ES} = \frac{(H_{0}-73.04)^{2}}{1.04^{2}}\,.
\label{chiSH0ES}
\ee

\subsection{Methodology}
We employed the well-known Markov Chain Monte Carlo (MCMC) analysis for parameter estimations of the models. This involves maximizing the likelihood function, given by
\begin{equation}
\mathcal{L} = \exp\left(-\sum_{i}\chi^{2}_{i}/2\right)\,,
\end{equation}
where the subscript `$i$', depending on the combination of different data-sets used, stands for one or more from the set: $\{$ `SN', `CC', `BAO', `H\textsc{ii}G', `SH0ES'$\}$. These $\chi^{2}_{i}$'s are defined in Eqs. \ref{chiSN}, \ref{chiCC}, \ref{chiBAO}, \ref{chiHIIG} and \ref{chiSH0ES}.  To get posterior probability distribution of the model parameters we also need to set priors on them. For the reasons mentioned in the last section and will further be discussed in the Appendix \ref{odesol}, for all the $f(R)$ models in this work we do data-fitting in terms of the parameters: $(\Om0L,b,\H0L)$. We used the uniform priors $\Om0L\in[0,1]$ and 
$\H0L\in[50,90]$ for all the models, whereas priors for $b$ require further considerations. 
From the algebraic expressions of these six $f(R)$\,-\,models we  worked out and established
a hierarchy of similarity of these models with the standard $\Lambda$CDM model   using some ``measures-of-similarity''(\textit{e.g.} mean-square-error, correlation, dynamic time warping \textit{etc.}). A ballpark hierarchy of similarity we find is: the most similar $f(R)$ models are the Tsujikawa model(TSUJI) and the Hu-Sawicki model($n_{_{\rm HS}}=3$)(HS3), followed by the Starobinsky($n_{_{\rm S}}=1$)(ST1) and the exponential(EXP) models, and the least similar ones are the Hu-Sawicki model($n_{_{\rm HS}}=1$)(HS1) and the arcTanh(aTanh) models. To allow for the exploration that whether the data supports a model different from the 
$\Lambda$CDM model, an $f(R)$ model which is more similar to the $\Lambda$CDM requires a bigger change in $b$ to make them noticeably different from the $\Lambda$CDM model. Consequently we use uniform priors $b\in[0,b_{\rm max}]$ where $b_{\rm max}$ is 7, 7, 5, 5, 3 and 3 for the models HS3, TSUJI, ST1, EXP, HS1 and aTanh, respectively.\\ 

We developed our own PYTHON codes using the publicly available PYTHON modules: (i) for solving stiff ODEs  \cite{numbalsoda}, (ii) to perform MCMC analysis --- EMCEE \cite{Foreman-Mackey:2012any}, and (iii) for plotting of posterior probability distributions of the parameters --- GetDist \cite{lewis2019getdist}.

\section{RESULTS: Observational constraints on $\lowercase{f}(R)$ models}

\label{results}
\begin{figure}
\centering
\includegraphics[scale=0.4]{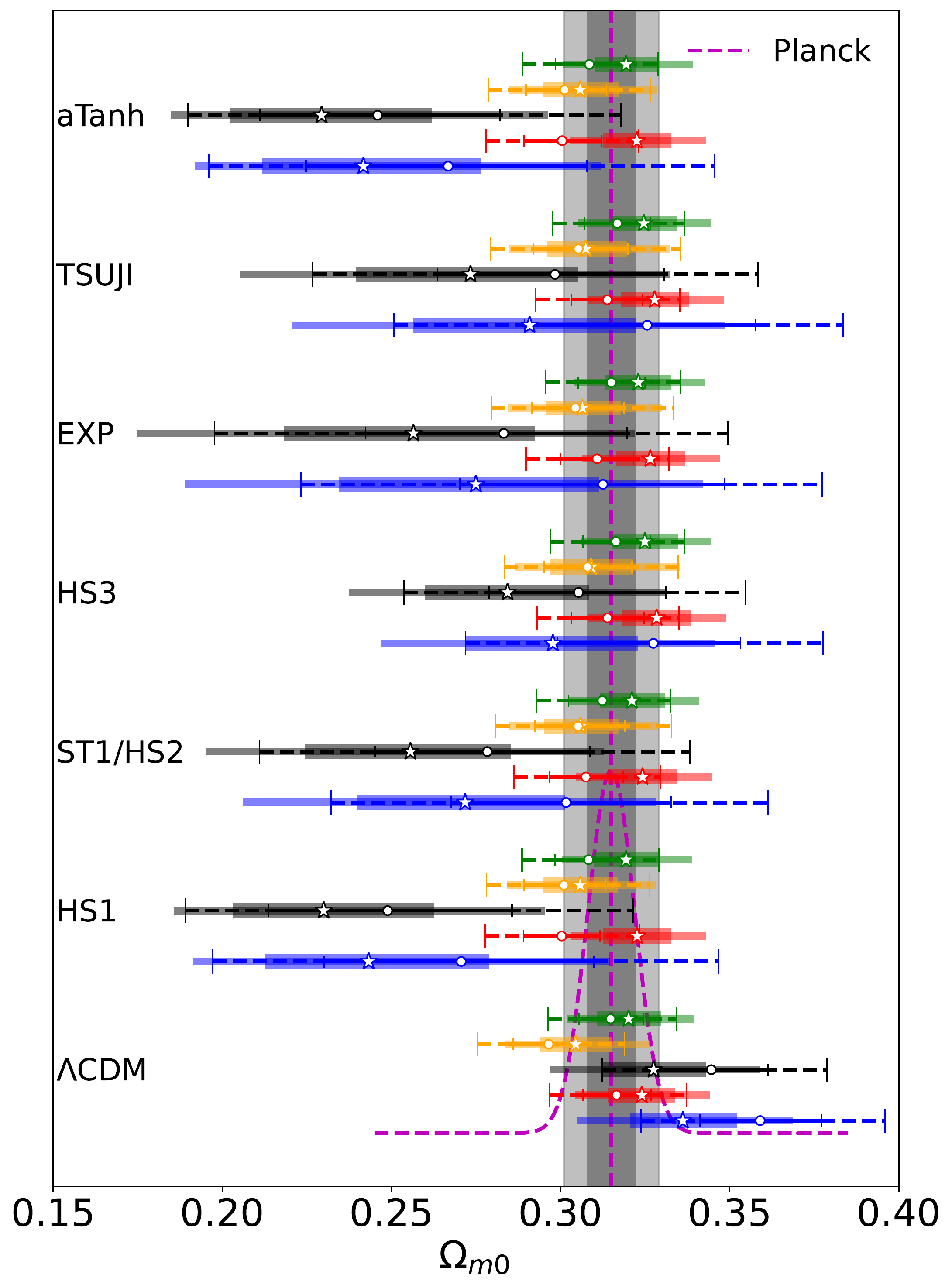}
\caption{This figure is helpful in getting a sense of variations of $\Omega_{m0}$ from model to model and with data for a given model. Different colors represent different data-set combinations and these are same as in any of the parameter distribution plots(e.g. Fig. \ref{HS1dist}) or see 2nd paragraph of Sec.\ref{acc}.
The blank star and blank circle markers represent median values for the cases with and without SH0ES prior for $H_{0}$, respectively. The thick and thin horizontal colored bars represent 1-sigma(68.26\%) and 2-sigma(95.44\%) confidence intervals for the cases with SH0ES prior. The colored continuous/dashed lines represent 1-sigma(68.26\%, with smaller cap size) and 2-sigma(95.44\%, with bigger cap size) confidence intervals for the cases without SH0ES prior.}
\label{Om0tension}
\end{figure}

\begin{figure}
\centering
\includegraphics[scale=0.4]{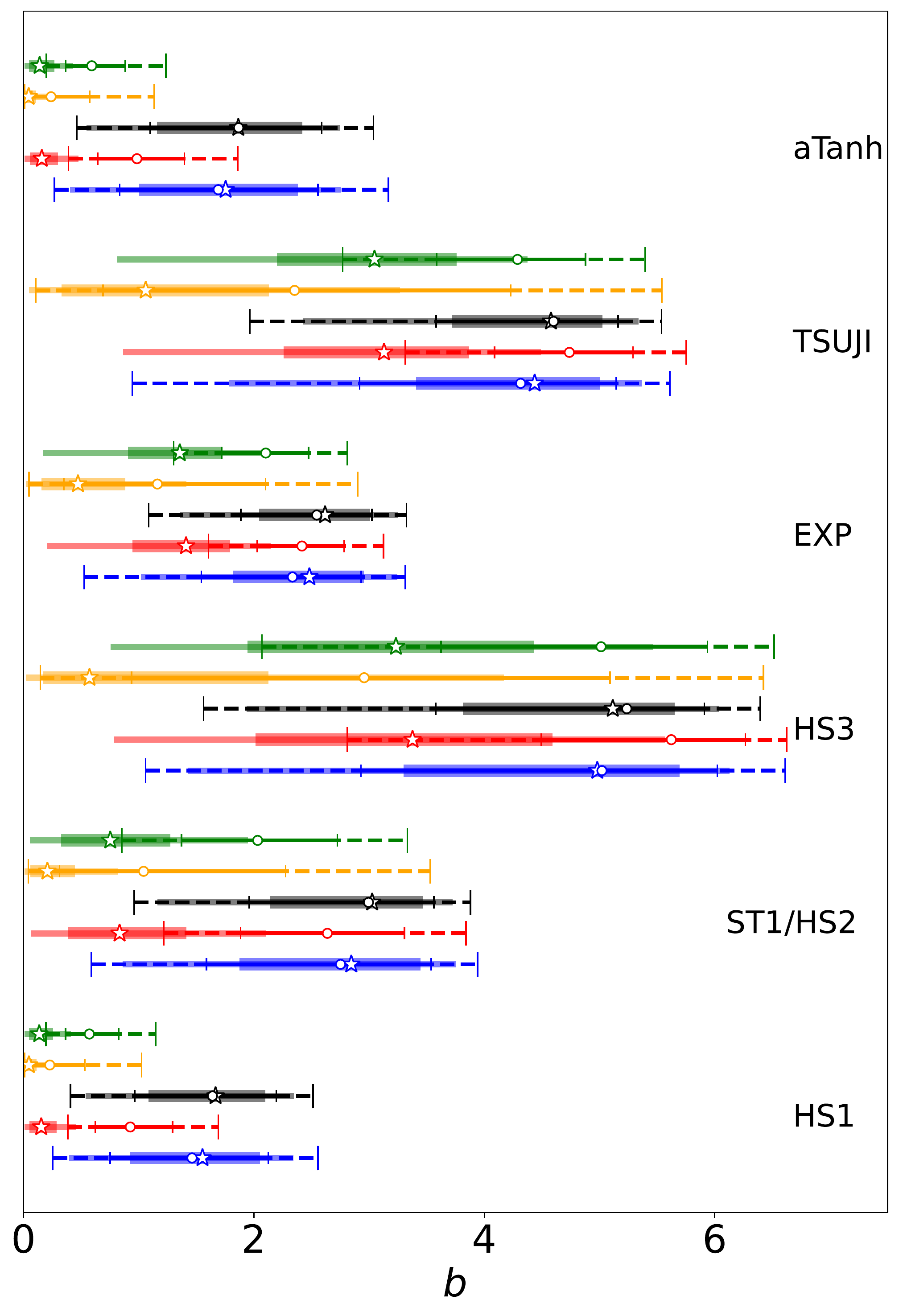}
\caption{This figure is helpful in getting a sense of variations of the deviation parameter $b$ from model to model and with data for a given model. Different colors represent different data-set combinations and these are same as in any of the parameter distribution plots(e.g. Fig. \ref{HS1dist}) or see 2nd paragraph of Sec.\ref{acc}. The blank star and blank circle markers represent median values for the cases with and without SH0ES prior for $H_{0}$, respectively. The thick and thin horizontal colored bars represent 1-sigma(68.26\%) and 2-sigma(95.44\%) confidence intervals for the cases with SH0ES prior. The colored continuous/dashed lines represent 1-sigma(68.26\%, with smaller cap size) and 2-sigma(95.44\%, with bigger cap size) confidence intervals for the cases without SH0ES prior.}
\label{btension}
\end{figure}

\begin{figure}
\centering
\includegraphics[scale=0.4]{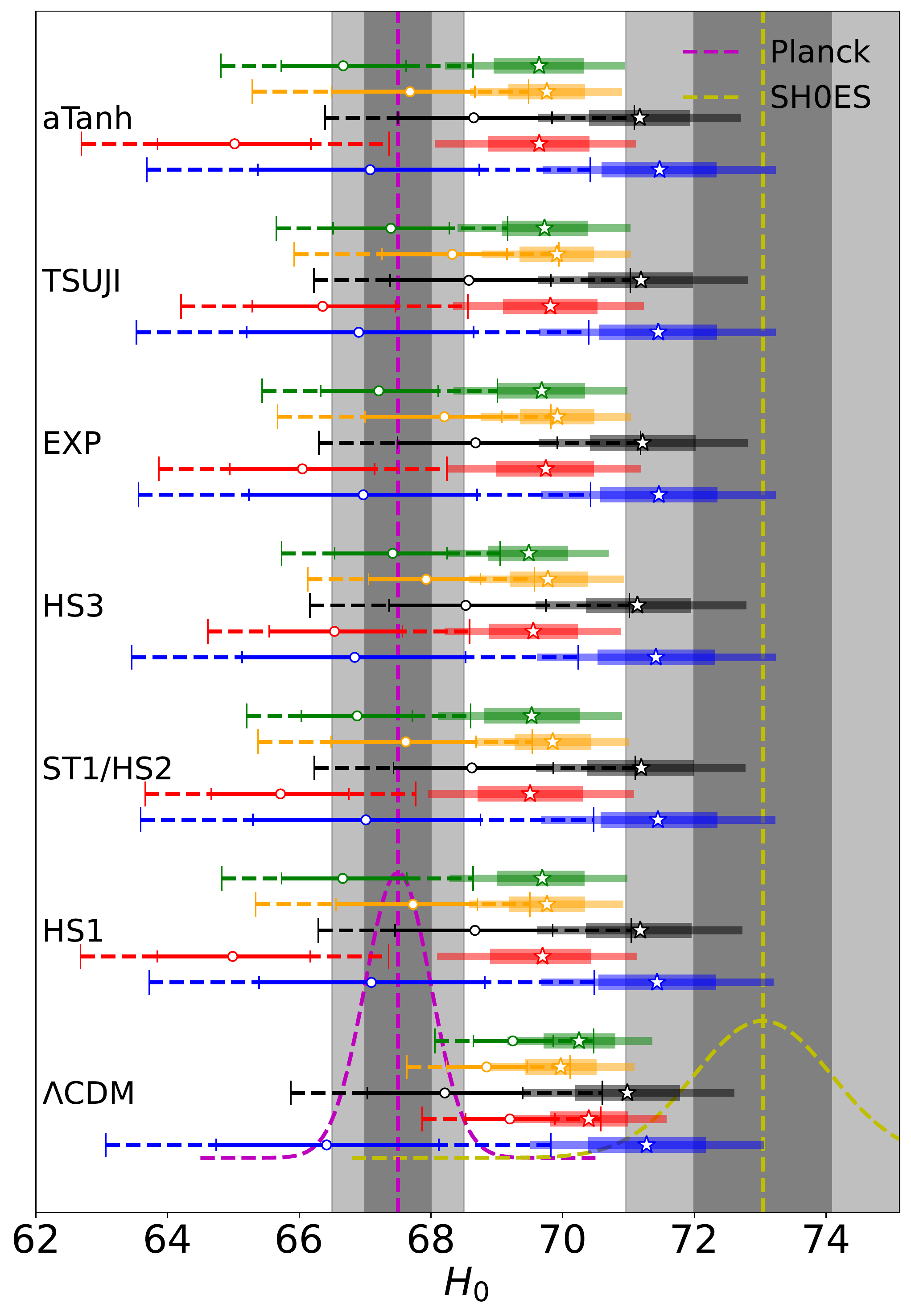}
\caption{Hubble Tension: Different colors represent different data-set combinations and these are same as in any of the parameter distribution plots(e.g. Fig. \ref{HS1dist}) or see 2nd paragraph of Sec.\ref{acc}. The blank star and blank circle markers represent median values for the cases with and without SH0ES prior for $H_{0}$, respectively. The thick and thin horizontal colored bars represent 1-sigma(68.26\%) and 2-sigma(95.44\%) confidence intervals for the cases with SH0ES prior. The colored continuous/dashed lines represent 1-sigma(68.26\%, with smaller cap size) and 2-sigma(95.44\%, with bigger cap size) confidence intervals for the cases without SH0ES prior.}
\label{H0tension}
\end{figure}
In this section we will discuss our findings regarding the parameters of the six $f(R)$ models and the $\Lambda$CDM model. We have considered five combinations of data sets: SNIa+CC (SC), SNIa+BAO+CC (SBC), SNIa+CC+H$\textsc{ii}$G (SCH$\textsc{ii}$), BAO+CC+HIIG (BCH$\textsc{ii}$), and SNIa+BAO+CC+HIIG (SBCH$\textsc{ii}$). We also 
separately included a SH0ES prior for the Hubble parameter $H_{0}$ for each of these combinations. While each data set individually can constrain any model, they may not necessarily provide tight constraints on the parameters. The choice of data combinations is also guided by considerations of goodness-of-fit.

\setcounter{secnumdepth}{6}
\paragraph*{Acronym and Color convention:}\label{acc}
In addition to the above mentioned acronyms for different data sets, when including the SH0ES prior for $H_{0}$, we append the notations as follows: SC$H_{0}$ to denote SNIa+CC+$H_{0}$ and SC($H_{0})$ to denote SNIa+CC or SNIa+CC+$H_{0}$, depending on the case. Similar notations are used for the other four data sets. Unless otherwise specified, the following color and data set correspondences are used in the upcoming figures: (i) SC($H_{0})$—Blue, (ii) SBC($H_{0})$—Red, (iii) SCH$\textsc{ii}(H_{0})$—Black, (iv) BCH$\textsc{ii}(H_{0})$—Orange, and (v) SBCH$\textsc{ii}(H_{0})$—Green.\\
\par
We generated MCMC samples of size 875,000 (with 25 walkers, each taking 35,000 steps) for each parameter in all the cases, where a case refers to a specific combination of a given model and a data set. These raw MCMC sample chains underwent tests for convergence and independence. To ensure convergence, an initial portion of the chain was discarded to obtain a ``burned chain." We removed the first 125,000 steps (i.e., 5,000 initial steps of each walkers) to obtain the burned chain. In order to have independent and uncorrelated samples, the burned chains needed to be properly thinned. We thinned the burned chains by a factor of 0.75 times the integrated auto-correlation time. As a result, depending on the case, we obtained convergent and independent samples of varying sizes, approximately ranging from 15,000 to 20,000. All the statistical inferences about the model parameters were then obtained from these burned and thinned subsamples.
\par 
The median values of model parameters and 1-sigma confidence intervals on 
them are presented in the Tables \ref{resultstable} and \ref{resultsH0table}. 
The Figs \multiref{LCDMdist}{aTanhH0dist} display 2D contour plots 
depicting the posterior probability distribution of the parameters, 
which provide an indication of potential correlations among the parameters, as well as 1D marginalised distributions of each parameter are also shown there.%
\par
Before getting into the detailed results for individual models, we would like to highlight some overall observations regarding the model parameters in the light of Figs. \ref{Om0tension}, \ref{btension}, and \ref{H0tension}.
(i) $\Omega_{m0}$: For all models and data-set combinations, either the median values of $\Omega_{m0}$ fall within the 1-2 sigma interval of the Planck value ($\Omega_{m0,\rm Planck}=0.315\pm0.007$ \cite{Planck:2018vyg}) or, the Planck value is within the 1-2 sigma intervals of the model's median values of $\Omega_{m0}$ (for cases with or without the SH0ES prior for $H_{0}$).
(ii) $b$: There is a general trend of a shift towards lower values of $b$ when considering the SH0ES prior for $H_{0}$ compared to the corresponding cases without the SH0ES prior for the SBC$(H_{0})$, BCH$\textsc{ii}(H_{0})$, and SBCH$\textsc{ii}(H_{0})$ data-sets,
whereas opposite trends are observed for the SC$(H_{0})$ and SCH$\textsc{ii}(H_{0})$ data-sets.
(iii) $H_{0}$: Without the SH0ES prior for $H_{0}$, for all models and data-set combinations, either the median values of $H_{0}$ are within a 1-2 sigma interval of the Planck value ($H_{0,\rm Planck}=67.4\pm0.5$ \cite{Planck:2018vyg}) or, the Planck value is within a 1-2 sigma interval of the model's median values of $H_{0}$. However, with the SH0ES prior for $H_{0}$, there are slight departures towards higher values of $H_{0}$ (as expected), but they are not close to $H_{0,\rm SH0ES} = 73.04\pm 1.04$ \cite{Riess:2021jrx}.
It is worth noting that the cases with the SH0ES prior for $H_{0}$ generally have tighter bounds on the model's $H_{0}$ values (this observation does not
 hold true for $\Omega_{m0}$ or $b$). 
Additionally, there is less overall tension  among the model-fitted values of $H_{0}$ compared to the tension between $H_{0,\rm Planck}$ and $H_{0,\rm SH0ES}$.\\

\begin{table*}
\caption{Results from the MCMC fitting process(for the cases without SH0ES prior for $H_{0}$): The median values of the model parameters along with 1-sigma(68.26\%) confidence intervals on them, the transition redshift($z_{\rm t}$), the current values of EoS parameter for geometric/dark energy component($w_{\rm DE0}$) and total EoS parameter($w_{\rm tot0}$), the minimum values of $\chi^{2}$ and other quantities for making statistical inferences such as the reduced chi-square value $\chi^{2}_{\rm red} = \chi^{2}_{\rm min}/\nu$(where the number of degrees of freedom $\nu = N-k$, $N$ is the total number of data points and $k$ is the number of model parameters), the Akaike Information Criterion(AIC), the Bayesian Information Criterion(BIC), $\Delta$AIC = AIC$_{f(R)\rm\,\,model}$ - AIC$_{\Lambda\rm{CDM}\,\,model}$ and $\Delta$BIC = BIC$_{f(R)\rm model}$ - BIC$_{\Lambda\rm{CDM}\,\,model}$ --- for all the models and data set combinations considered in this work.}
\begin{tabular}{|c|c|c|c|c|c|c|c|c|c|c|c|c|c|}
\hline
Data & $\Omega_{m0}$ & $b$ & $H_{0}$ & $z_{\rm t}$ & $w_{\rm DE0}$ & $w_{\rm tot0}$ &  $\chi^{2}$  & $\chi^{2}_{\rm red}$& AIC & BIC  & $\Delta$AIC & $\Delta$BIC\\\hline\hline

$\Lambda$CDM  &     &     &     &     &     &     &     &     &     &     &     &    \\
SC  &  $0.36_{-0.018}^{+0.018}$  &  ---  &  $66.42_{-1.674}^{+1.703}$  &  $0.53_{-0.039}^{+0.040}$  &  $-1$  &  $-0.64_{-0.018}^{+0.018}$  &  1777.31  &  1.03  &  1781.31  &  1792.22  &  0.0  &  0.0 \\
SBC  &  $0.32_{-0.010}^{+0.010}$  &  ---  &  $69.20_{-0.669}^{+0.684}$  &  $0.63_{-0.025}^{+0.025}$  &  $-1$  &  $-0.68_{-0.010}^{+0.010}$  &  1815.39  &  1.03  &  1819.39  &  1830.34  &  0.0  &  0.0 \\
SCH$\textsc{ii}$  &  $0.34_{-0.016}^{+0.017}$  &  ---  &  $68.21_{-1.179}^{+1.182}$  &  $0.56_{-0.037}^{+0.039}$  &  $-1$  &  $-0.66_{-0.016}^{+0.017}$  &  2145.62  &  1.12  &  2149.62  &  2160.73  &  0.0  &  0.0 \\
BCH$\textsc{ii}$  &  $0.30_{-0.011}^{+0.011}$  &  ---  &  $68.85_{-0.608}^{+0.619}$  &  $0.68_{-0.029}^{+0.029}$  & $-1$ &  $-0.70_{-0.011}^{+0.011}$  &  408.22  &  1.69  &  412.22  &  419.19  &  0.0  &  0.0 \\
SBCH$\textsc{ii}$  &  $0.31_{-0.009}^{+0.010}$  &  ---  &  $69.24_{-0.599}^{+0.613}$  &  $0.63_{-0.024}^{+0.024}$  &  $-1$  &  $-0.69_{-0.009}^{+0.010}$  &  2180.25  &  1.12  &  2184.25  &  2195.39  &  0.0  &  0.0 \\
HS1  &     &     &     &     &     &     &     &     &     &     &     &    \\
SC  &  $0.27_{-0.041}^{+0.039}$  &  $1.46_{-0.711}^{+0.662}$  &  $67.10_{-1.704}^{+1.721}$  &  $0.78_{-0.132}^{+0.188}$  &  $-0.78_{-0.078}^{+0.070}$  &  $-0.57_{-0.033}^{+0.031}$  &  1773.75  &  1.03  &  1779.75  &  1796.11  &  -1.56  &  3.9 \\
SBC  &  $0.30_{-0.011}^{+0.011}$  &  $0.93_{-0.303}^{+0.368}$  &  $64.99_{-1.145}^{+1.175}$  &  $0.68_{-0.035}^{+0.037}$  &  $-0.84_{-0.041}^{+0.040}$  &  $-0.59_{-0.027}^{+0.027}$  &  1800.65  &  1.02  &  1806.65  &  1823.07  &  -12.74  &  -7.26 \\
SCH$\textsc{ii}$  &  $0.25_{-0.035}^{+0.037}$  &  $1.64_{-0.676}^{+0.553}$  &  $68.67_{-1.213}^{+1.181}$  &  $0.87_{-0.145}^{+0.188}$  &  $-0.77_{-0.071}^{+0.061}$  &  $-0.57_{-0.031}^{+0.029}$  &  2139.44  &  1.12  &  2145.44  &  2162.1  &  -4.18  &  1.37 \\
BCH$\textsc{ii}$  &  $0.30_{-0.012}^{+0.012}$  &  $0.23_{-0.161}^{+0.305}$  &  $67.73_{-1.165}^{+0.977}$  &  $0.67_{-0.033}^{+0.032}$  &  $-0.96_{-0.032}^{+0.065}$  &  $-0.67_{-0.028}^{+0.050}$  &  408.19  &  1.7  &  414.19  &  424.64  &  1.97  &  5.44 \\
SBCH$\textsc{ii}$  &  $0.31_{-0.010}^{+0.010}$  &  $0.57_{-0.206}^{+0.255}$  &  $66.66_{-0.927}^{+0.973}$  &  $0.65_{-0.029}^{+0.030}$  &  $-0.89_{-0.039}^{+0.037}$  &  $-0.61_{-0.027}^{+0.026}$  &  2171.26  &  1.12  &  2177.26  &  2193.97  &  -6.99  &  -1.42 \\
ST1/HS2  &     &     &     &     &     &     &     &     &     &     &     &    \\
SC  &  $0.30_{-0.034}^{+0.031}$  &  $2.75_{-1.165}^{+0.786}$  &  $67.01_{-1.716}^{+1.740}$  &  $0.75_{-0.118}^{+0.155}$  &  $-0.81_{-0.071}^{+0.085}$  &  $-0.56_{-0.035}^{+0.042}$  &  1773.24  &  1.02  &  1779.24  &  1795.61  &  -2.07  &  3.39 \\
SBC  &  $0.31_{-0.011}^{+0.011}$  &  $2.64_{-0.754}^{+0.668}$  &  $65.72_{-1.052}^{+1.038}$  &  $0.72_{-0.040}^{+0.042}$  &  $-0.82_{-0.043}^{+0.047}$  &  $-0.57_{-0.030}^{+0.032}$  &  1799.22  &  1.02  &  1805.22  &  1821.64  &  -14.17  &  -8.7 \\
SCH$\textsc{ii}$  &  $0.28_{-0.033}^{+0.030}$  &  $2.99_{-1.035}^{+0.568}$  &  $68.62_{-1.190}^{+1.234}$  &  $0.84_{-0.132}^{+0.169}$  &  $-0.78_{-0.074}^{+0.087}$  &  $-0.56_{-0.038}^{+0.046}$  &  2138.76  &  1.12  &  2144.76  &  2161.43  &  -4.86  &  0.7 \\
BCH$\textsc{ii}$  &  $0.31_{-0.013}^{+0.014}$  &  $1.04_{-0.729}^{+1.234}$  &  $67.62_{-1.130}^{+1.068}$  &  $0.69_{-0.032}^{+0.033}$  &  $-0.91_{-0.077}^{+0.074}$  &  $-0.64_{-0.059}^{+0.060}$  &  408.18  &  1.7  &  414.18  &  424.62  &  1.95  &  5.43 \\
SBCH$\textsc{ii}$  &  $0.31_{-0.010}^{+0.010}$  &  $2.03_{-0.661}^{+0.691}$  &  $66.88_{-0.847}^{+0.840}$  &  $0.69_{-0.034}^{+0.035}$  &  $-0.85_{-0.038}^{+0.039}$  &  $-0.59_{-0.027}^{+0.028}$  &  2168.85  &  1.12  &  2174.85  &  2191.57  &  -9.4  &  -3.83 \\
HS3  &     &     &     &     &     &     &     &     &     &     &     &    \\
SC  &  $0.33_{-0.028}^{+0.026}$  &  $5.02_{-2.092}^{+1.001}$  &  $66.84_{-1.708}^{+1.684}$  &  $0.70_{-0.094}^{+0.115}$  &  $-0.84_{-0.071}^{+0.105}$  &  $-0.56_{-0.038}^{+0.057}$  &  1772.23  &  1.02  &  1778.23  &  1794.6  &  -3.07  &  2.38 \\
SBC  &  $0.31_{-0.011}^{+0.011}$  &  $5.62_{-1.130}^{+0.643}$  &  $66.54_{-0.993}^{+1.030}$  &  $0.76_{-0.040}^{+0.040}$  &  $-0.79_{-0.064}^{+0.079}$  &  $-0.54_{-0.045}^{+0.055}$  &  1798.35  &  1.02  &  1804.35  &  1820.77  &  -15.04  &  -9.57 \\
SCH$\textsc{ii}$  &  $0.31_{-0.026}^{+0.026}$  &  $5.24_{-1.657}^{+0.673}$  &  $68.53_{-1.162}^{+1.216}$  &  $0.78_{-0.106}^{+0.113}$  &  $-0.80_{-0.083}^{+0.105}$  &  $-0.56_{-0.043}^{+0.062}$  &  2137.24  &  1.12  &  2143.24  &  2159.9  &  -6.38  &  -0.83 \\
BCH$\textsc{ii}$  &  $0.31_{-0.013}^{+0.013}$  &  $2.96_{-2.019}^{+2.134}$  &  $67.93_{-0.874}^{+0.824}$  &  $0.72_{-0.038}^{+0.039}$  &  $-0.90_{-0.072}^{+0.073}$  &  $-0.62_{-0.057}^{+0.056}$  &  408.21  &  1.7  &  414.21  &  424.65  &  1.99  &  5.46 \\
SBCH$\textsc{ii}$  &  $0.32_{-0.010}^{+0.010}$  &  $5.01_{-1.389}^{+0.924}$  &  $67.42_{-0.883}^{+0.827}$  &  $0.73_{-0.037}^{+0.036}$  &  $-0.83_{-0.052}^{+0.065}$  &  $-0.57_{-0.037}^{+0.046}$  &  2166.71  &  1.12  &  2172.71  &  2189.42  &  -11.54  &  -5.97 \\
EXP  &     &     &     &     &     &     &     &     &     &     &     &    \\
SC  &  $0.31_{-0.042}^{+0.036}$  &  $2.34_{-0.792}^{+0.597}$  &  $66.97_{-1.738}^{+1.729}$  &  $0.73_{-0.126}^{+0.160}$  &  $-0.83_{-0.090}^{+0.093}$  &  $-0.56_{-0.041}^{+0.044}$  &  1773.24  &  1.02  &  1779.24  &  1795.61  &  -2.07  &  3.39 \\
SBC  &  $0.31_{-0.011}^{+0.011}$  &  $2.42_{-0.390}^{+0.365}$  &  $66.05_{-1.102}^{+1.097}$  &  $0.78_{-0.042}^{+0.043}$  &  $-0.79_{-0.048}^{+0.050}$  &  $-0.54_{-0.033}^{+0.034}$  &  1799.21  &  1.02  &  1805.21  &  1821.63  &  -14.18  &  -8.71 \\
SCH$\textsc{ii}$  &  $0.28_{-0.041}^{+0.036}$  &  $2.55_{-0.659}^{+0.480}$  &  $68.68_{-1.185}^{+1.242}$  &  $0.83_{-0.137}^{+0.166}$  &  $-0.79_{-0.086}^{+0.086}$  &  $-0.55_{-0.041}^{+0.043}$  &  2138.92  &  1.12  &  2144.92  &  2161.58  &  -4.7  &  0.85 \\
BCH$\textsc{ii}$  &  $0.30_{-0.013}^{+0.014}$  &  $1.16_{-0.812}^{+0.940}$  &  $68.21_{-1.208}^{+0.869}$  &  $0.72_{-0.046}^{+0.048}$  &  $-0.93_{-0.068}^{+0.102}$  &  $-0.65_{-0.053}^{+0.080}$  &  408.22  &  1.7  &  414.22  &  424.67  &  2.0  &  5.47 \\
SBCH$\textsc{ii}$  &  $0.32_{-0.010}^{+0.010}$  &  $2.10_{-0.383}^{+0.372}$  &  $67.21_{-0.883}^{+0.899}$  &  $0.76_{-0.036}^{+0.037}$  &  $-0.83_{-0.044}^{+0.044}$  &  $-0.56_{-0.031}^{+0.031}$  &  2167.67  &  1.12  &  2173.67  &  2190.38  &  -10.58  &  -5.01 \\
TSUJI  &     &     &     &     &     &     &     &     &     &     &     &    \\
SC  &  $0.33_{-0.036}^{+0.032}$  &  $4.32_{-1.398}^{+0.828}$  &  $66.91_{-1.704}^{+1.743}$  &  $0.73_{-0.126}^{+0.160}$  &  $-0.83_{-0.090}^{+0.093}$  &  $-0.56_{-0.041}^{+0.044}$  &  1772.91  &  1.02  &  1778.91  &  1795.28  &  -2.39  &  3.06 \\
SBC  &  $0.31_{-0.011}^{+0.011}$  &  $4.74_{-0.649}^{+0.552}$  &  $66.36_{-1.067}^{+1.107}$  &  $0.78_{-0.042}^{+0.043}$  &  $-0.79_{-0.048}^{+0.050}$  &  $-0.54_{-0.033}^{+0.034}$  &  1799.12  &  1.02  &  1805.12  &  1821.54  &  -14.27  &  -8.8 \\
SCH$\textsc{ii}$  &  $0.30_{-0.035}^{+0.032}$  &  $4.60_{-1.019}^{+0.559}$  &  $68.58_{-1.196}^{+1.247}$  &  $0.83_{-0.137}^{+0.166}$  &  $-0.79_{-0.086}^{+0.086}$  &  $-0.55_{-0.041}^{+0.043}$  &  2138.49  &  1.12  &  2144.49  &  2161.15  &  -5.13  &  0.42 \\
BCH$\textsc{ii}$  &  $0.31_{-0.013}^{+0.015}$  &  $2.35_{-1.663}^{+1.877}$  &  $68.33_{-1.069}^{+0.829}$  &  $0.72_{-0.046}^{+0.048}$  &  $-0.93_{-0.068}^{+0.102}$  &  $-0.65_{-0.053}^{+0.080}$  &  408.22  &  1.7  &  414.22  &  424.66  &  2.0  &  5.47 \\
SBCH$\textsc{ii}$  &  $0.32_{-0.010}^{+0.010}$  &  $4.29_{-0.701}^{+0.590}$  &  $67.39_{-0.879}^{+0.886}$  &  $0.76_{-0.036}^{+0.037}$  &  $-0.83_{-0.044}^{+0.044}$  &  $-0.56_{-0.031}^{+0.031}$  &  2167.17  &  1.12  &  2173.17  &  2189.88  &  -11.08  &  -5.51 \\
aTanh  &     &     &     &     &     &     &     &     &     &     &     &    \\
SC  &  $0.27_{-0.042}^{+0.041}$  &  $1.69_{-0.856}^{+0.865}$  &  $67.08_{-1.707}^{+1.659}$  &  $0.79_{-0.137}^{+0.186}$  &  $-0.78_{-0.077}^{+0.065}$  &  $-0.57_{-0.031}^{+0.027}$  &  1773.73  &  1.03  &  1779.73  &  1796.1  &  -1.57  &  3.88 \\
SBC  &  $0.30_{-0.011}^{+0.012}$  &  $0.99_{-0.339}^{+0.412}$  &  $65.02_{-1.169}^{+1.158}$  &  $0.68_{-0.035}^{+0.036}$  &  $-0.84_{-0.039}^{+0.038}$  &  $-0.59_{-0.026}^{+0.026}$  &  1800.58  &  1.02  &  1806.58  &  1823.0  &  -12.81  &  -7.34 \\
SCH$\textsc{ii}$  &  $0.25_{-0.035}^{+0.036}$  &  $1.87_{-0.768}^{+0.721}$  &  $68.65_{-1.151}^{+1.188}$  &  $0.88_{-0.142}^{+0.177}$  &  $-0.76_{-0.065}^{+0.052}$  &  $-0.58_{-0.028}^{+0.025}$  &  2139.45  &  1.12  &  2145.45  &  2162.12  &  -4.17  &  1.39 \\
BCH$\textsc{ii}$  &  $0.30_{-0.011}^{+0.012}$  &  $0.24_{-0.169}^{+0.334}$  &  $67.68_{-1.190}^{+0.986}$  &  $0.67_{-0.033}^{+0.031}$  &  $-0.95_{-0.037}^{+0.065}$  &  $-0.66_{-0.031}^{+0.051}$  &  408.2  &  1.7  &  414.2  &  424.64  &  1.97  &  5.45 \\
SBCH$\textsc{ii}$  &  $0.31_{-0.010}^{+0.010}$  &  $0.59_{-0.225}^{+0.290}$  &  $66.67_{-0.938}^{+0.958}$  &  $0.65_{-0.029}^{+0.030}$  &  $-0.88_{-0.038}^{+0.036}$  &  $-0.61_{-0.027}^{+0.026}$  &  2171.09  &  1.12  &  2177.09  &  2193.81  &  -7.16  &  -1.59 \\
\hline
\end{tabular}

\label{resultstable}
\end{table*}

\begin{table*}
\caption{Results from the MCMC fitting process(for the cases with SH0ES prior for $H_{0}$): The median values of the model parameters along with 1-sigma(68.26\%) confidence intervals on them, the transition redshift($z_{\rm t}$), the current values of EoS parameter for geometric/dark energy component($w_{\rm DE0}$) and total EoS parameter($w_{\rm tot0}$), the minimum values of $\chi^{2}$ and other quantities for making statistical inferences such as the reduced chi-square value $\chi^{2}_{\rm red} = \chi^{2}_{\rm min}/\nu$(where the number of degrees of freedom $\nu = N-k$, $N$ is the total number of data points and $k$ is the number of model parameters), the Akaike Information Criterion(AIC), the Bayesian Information Criterion(BIC), $\Delta$AIC = AIC$_{f(R)\rm\,\,model}$ - AIC$_{\Lambda\rm{CDM}\,\,model}$ and $\Delta$BIC = BIC$_{f(R)\rm model}$ - BIC$_{\Lambda\rm{CDM}\,\,model}$ --- for all the models and data set combinations considered in this work.}
\begin{tabular}{|c|c|c|c|c|c|c|c|c|c|c|c|c|c|}
\hline
Data & $\Omega_{m0}$ & $b$ & $H_{0}$ & $z_{\rm t}$ & $w_{\rm DE0}$ & $w_{\rm tot0}$ &  $\chi^{2}$  & $\chi^{2}_{\rm red}$& AIC & BIC  & $\Delta$AIC & $\Delta$BIC\\\hline\hline

$\Lambda$CDM  &     &     &     &     &     &     &     &     &     &     &     &    \\
SC$H_{0}$  &  $0.34_{-0.016}^{+0.016}$  &  ---  &  $71.28_{-0.887}^{+0.898}$  &  $0.58_{-0.037}^{+0.038}$  &  $-1$  &  $-0.66_{-0.016}^{+0.016}$  &  1788.16  &  1.03  &  1792.16  &  1803.07  &  0.0  &  0.0 \\
SBC$H_{0}$  &  $0.32_{-0.010}^{+0.010}$  &  ---  &  $70.40_{-0.589}^{+0.591}$  &  $0.61_{-0.024}^{+0.025}$  &  $-1$  &  $-0.68_{-0.010}^{+0.010}$  &  1824.91  &  1.04  &  1828.91  &  1839.86  &  0.0  &  0.0 \\
SCH$\textsc{ii}H_{0}$  &  $0.33_{-0.015}^{+0.015}$  &  ---  &  $70.98_{-0.794}^{+0.799}$  &  $0.60_{-0.036}^{+0.038}$  &  $-1$  &  $-0.67_{-0.015}^{+0.015}$  &  2154.83  &  1.13  &  2158.83  &  2169.94  &  0.0  &  0.0 \\
BCH$\textsc{ii}H_{0}$  &  $0.30_{-0.011}^{+0.011}$  &  ---  &  $69.97_{-0.546}^{+0.547}$  &  $0.66_{-0.028}^{+0.028}$  &  $-1$  &  $-0.70_{-0.011}^{+0.011}$  &  420.25  &  1.74  &  424.25  &  431.23  &  0.0  &  0.0 \\
SBCH$\textsc{ii}H_{0}$  &  $0.32_{-0.009}^{+0.010}$  &  ---  &  $70.25_{-0.541}^{+0.552}$  &  $0.62_{-0.023}^{+0.023}$  &  $-1$  &  $-0.68_{-0.009}^{+0.010}$  &  2190.08  &  1.13  &  2194.08  &  2205.23  &  0.0  &  0.0 \\
HS1  &     &     &     &     &     &     &     &     &     &     &     &    \\
SC$H_{0}$  &  $0.24_{-0.031}^{+0.036}$  &  $1.55_{-0.630}^{+0.499}$  &  $71.43_{-0.889}^{+0.897}$  &  $0.89_{-0.145}^{+0.168}$  &  $-0.77_{-0.067}^{+0.054}$  &  $-0.59_{-0.029}^{+0.027}$  &  1782.1  &  1.03  &  1788.1  &  1804.47  &  -4.06  &  1.4 \\
SBC$H_{0}$  &  $0.32_{-0.010}^{+0.010}$  &  $0.15_{-0.101}^{+0.134}$  &  $69.70_{-0.799}^{+0.735}$  &  $0.61_{-0.025}^{+0.026}$  &  $-0.97_{-0.019}^{+0.032}$  &  $-0.66_{-0.016}^{+0.022}$  &  1824.39  &  1.04  &  1830.39  &  1846.81  &  1.48  &  6.95 \\
SCH$\textsc{ii}H_{0}$  &  $0.23_{-0.027}^{+0.033}$  &  $1.67_{-0.581}^{+0.432}$  &  $71.18_{-0.823}^{+0.780}$  &  $0.96_{-0.149}^{+0.158}$  &  $-0.76_{-0.061}^{+0.048}$  &  $-0.59_{-0.029}^{+0.026}$  &  2146.44  &  1.12  &  2152.44  &  2169.1  &  -6.39  &  -0.84 \\
BCH$\textsc{ii}H_{0}$  &  $0.31_{-0.011}^{+0.011}$  &  $0.05_{-0.035}^{+0.068}$  &  $69.76_{-0.575}^{+0.578}$  &  $0.65_{-0.028}^{+0.029}$  &  $-0.99_{-0.005}^{+0.011}$  &  $-0.69_{-0.013}^{+0.014}$  &  420.28  &  1.74  &  426.28  &  436.73  &  2.03  &  5.51 \\
SBCH$\textsc{ii}H_{0}$  &  $0.32_{-0.010}^{+0.010}$  &  $0.14_{-0.091}^{+0.121}$  &  $69.69_{-0.688}^{+0.642}$  &  $0.62_{-0.024}^{+0.025}$  &  $-0.98_{-0.016}^{+0.028}$  &  $-0.66_{-0.015}^{+0.020}$  &  2189.75  &  1.13  &  2195.75  &  2212.46  &  1.67  &  7.24 \\
ST1/HS2  &     &     &     &     &     &     &     &     &     &     &     &    \\
SC$H_{0}$  &  $0.27_{-0.032}^{+0.029}$  &  $2.85_{-0.971}^{+0.601}$  &  $71.45_{-0.873}^{+0.901}$  &  $0.86_{-0.132}^{+0.168}$  &  $-0.79_{-0.069}^{+0.084}$  &  $-0.57_{-0.035}^{+0.044}$  &  1781.71  &  1.03  &  1787.71  &  1804.08  &  -4.45  &  1.01 \\
SBC$H_{0}$  &  $0.32_{-0.010}^{+0.010}$  &  $0.83_{-0.445}^{+0.578}$  &  $69.51_{-0.799}^{+0.800}$  &  $0.63_{-0.029}^{+0.030}$  &  $-0.94_{-0.050}^{+0.045}$  &  $-0.63_{-0.034}^{+0.033}$  &  1823.2  &  1.04  &  1829.2  &  1845.62  &  0.28  &  5.75 \\
SCH$\textsc{ii}H_{0}$  &  $0.26_{-0.031}^{+0.030}$  &  $3.03_{-0.889}^{+0.440}$  &  $71.19_{-0.820}^{+0.802}$  &  $0.94_{-0.144}^{+0.179}$  &  $-0.77_{-0.074}^{+0.084}$  &  $-0.57_{-0.038}^{+0.047}$  &  2145.93  &  1.12  &  2151.93  &  2168.6  &  -6.89  &  -1.34 \\
BCH$\textsc{ii}H_{0}$  &  $0.31_{-0.011}^{+0.011}$  &  $0.21_{-0.147}^{+0.238}$  &  $69.85_{-0.579}^{+0.577}$  &  $0.66_{-0.028}^{+0.028}$  &  $-1.00_{-0.003}^{+0.015}$  &  $-0.69_{-0.012}^{+0.016}$  &  420.25  &  1.74  &  426.25  &  436.71  &  2.0  &  5.48 \\
SBCH$\textsc{ii}H_{0}$  &  $0.32_{-0.009}^{+0.010}$  &  $0.75_{-0.427}^{+0.521}$  &  $69.53_{-0.723}^{+0.727}$  &  $0.64_{-0.027}^{+0.027}$  &  $-0.94_{-0.046}^{+0.046}$  &  $-0.64_{-0.031}^{+0.033}$  &  2188.66  &  1.13  &  2194.66  &  2211.38  &  0.58  &  6.15 \\
HS3  &     &     &     &     &     &     &     &     &     &     &     &    \\
SC$H_{0}$  &  $0.30_{-0.026}^{+0.025}$  &  $4.98_{-1.683}^{+0.715}$  &  $71.42_{-0.885}^{+0.901}$  &  $0.80_{-0.107}^{+0.114}$  &  $-0.82_{-0.077}^{+0.100}$  &  $-0.57_{-0.041}^{+0.057}$  &  1780.99  &  1.03  &  1786.99  &  1803.36  &  -5.17  &  0.29 \\
SBC$H_{0}$  &  $0.33_{-0.010}^{+0.010}$  &  $3.38_{-1.363}^{+1.213}$  &  $69.55_{-0.671}^{+0.678}$  &  $0.67_{-0.033}^{+0.033}$  &  $-0.90_{-0.030}^{+0.035}$  &  $-0.60_{-0.024}^{+0.026}$  &  1820.05  &  1.03  &  1826.05  &  1842.47  &  -2.87  &  2.61 \\
SCH$\textsc{ii}H_{0}$  &  $0.28_{-0.024}^{+0.024}$  &  $5.12_{-1.303}^{+0.539}$  &  $71.14_{-0.779}^{+0.817}$  &  $0.85_{-0.108}^{+0.108}$  &  $-0.79_{-0.082}^{+0.097}$  &  $-0.56_{-0.045}^{+0.059}$  &  2144.92  &  1.12  &  2150.92  &  2167.59  &  -7.91  &  -2.35 \\
BCH$\textsc{ii}H_{0}$  &  $0.31_{-0.012}^{+0.013}$  &  $0.57_{-0.402}^{+1.553}$  &  $69.78_{-0.579}^{+0.605}$  &  $0.67_{-0.029}^{+0.030}$  &  $-0.99_{-0.006}^{+0.073}$  &  $-0.68_{-0.016}^{+0.055}$  &  420.23  &  1.74  &  426.23  &  436.68  &  1.98  &  5.45 \\
SBCH$\textsc{ii}H_{0}$  &  $0.32_{-0.010}^{+0.010}$  &  $3.23_{-1.289}^{+1.198}$  &  $69.49_{-0.622}^{+0.594}$  &  $0.68_{-0.032}^{+0.032}$  &  $-0.90_{-0.029}^{+0.034}$  &  $-0.61_{-0.023}^{+0.025}$  &  2185.47  &  1.13  &  2191.47  &  2208.19  &  -2.61  &  2.96 \\
EXP  &     &     &     &     &     &     &     &     &     &     &     &    \\
SC$H_{0}$  &  $0.27_{-0.040}^{+0.036}$  &  $2.48_{-0.662}^{+0.473}$  &  $71.46_{-0.891}^{+0.888}$  &  $0.85_{-0.140}^{+0.172}$  &  $-0.80_{-0.084}^{+0.085}$  &  $-0.56_{-0.040}^{+0.042}$  &  1781.83  &  1.03  &  1787.83  &  1804.2  &  -4.32  &  1.13 \\
SBC$H_{0}$  &  $0.33_{-0.010}^{+0.010}$  &  $1.41_{-0.466}^{+0.381}$  &  $69.75_{-0.758}^{+0.732}$  &  $0.68_{-0.038}^{+0.034}$  &  $-0.90_{-0.042}^{+0.039}$  &  $-0.61_{-0.031}^{+0.030}$  &  1820.84  &  1.03  &  1826.84  &  1843.26  &  -2.08  &  3.4 \\
SCH$\textsc{ii}H_{0}$  &  $0.26_{-0.038}^{+0.036}$  &  $2.62_{-0.572}^{+0.392}$  &  $71.22_{-0.801}^{+0.807}$  &  $0.93_{-0.147}^{+0.189}$  &  $-0.77_{-0.084}^{+0.084}$  &  $-0.56_{-0.041}^{+0.043}$  &  2146.11  &  1.12  &  2152.11  &  2168.78  &  -6.71  &  -1.16 \\
BCH$\textsc{ii}H_{0}$  &  $0.31_{-0.011}^{+0.011}$  &  $0.47_{-0.318}^{+0.412}$  &  $69.92_{-0.571}^{+0.562}$  &  $0.67_{-0.032}^{+0.033}$  &  $-1.00_{-0.005}^{+0.050}$  &  $-0.68_{-0.017}^{+0.036}$  &  420.24  &  1.74  &  426.24  &  436.7  &  2.0  &  5.47 \\
SBCH$\textsc{ii}H_{0}$  &  $0.32_{-0.010}^{+0.010}$  &  $1.36_{-0.449}^{+0.368}$  &  $69.68_{-0.662}^{+0.658}$  &  $0.69_{-0.036}^{+0.033}$  &  $-0.90_{-0.041}^{+0.038}$  &  $-0.61_{-0.030}^{+0.029}$  &  2185.98  &  1.13  &  2191.98  &  2208.69  &  -2.1  &  3.47 \\
TSUJI  &     &     &     &     &     &     &     &     &     &     &     &    \\
SC$H_{0}$  &  $0.29_{-0.035}^{+0.032}$  &  $4.44_{-1.031}^{+0.569}$  &  $71.45_{-0.896}^{+0.892}$  &  $0.85_{-0.140}^{+0.172}$  &  $-0.80_{-0.084}^{+0.085}$  &  $-0.56_{-0.040}^{+0.042}$  &  1781.6  &  1.03  &  1787.6  &  1803.97  &  -4.56  &  0.89 \\
SBC$H_{0}$  &  $0.33_{-0.010}^{+0.010}$  &  $3.13_{-0.873}^{+0.740}$  &  $69.82_{-0.723}^{+0.717}$  &  $0.68_{-0.038}^{+0.034}$  &  $-0.90_{-0.042}^{+0.039}$  &  $-0.61_{-0.031}^{+0.030}$  &  1819.49  &  1.03  &  1825.49  &  1841.91  &  -3.43  &  2.04 \\
SCH$\textsc{ii}H_{0}$  &  $0.27_{-0.034}^{+0.032}$  &  $4.58_{-0.856}^{+0.447}$  &  $71.19_{-0.811}^{+0.787}$  &  $0.93_{-0.147}^{+0.189}$  &  $-0.77_{-0.084}^{+0.084}$  &  $-0.56_{-0.041}^{+0.043}$  &  2145.82  &  1.12  &  2151.82  &  2168.49  &  -7.0  &  -1.45 \\
BCH$\textsc{ii}H_{0}$  &  $0.31_{-0.011}^{+0.012}$  &  $1.06_{-0.731}^{+1.069}$  &  $69.91_{-0.570}^{+0.559}$  &  $0.67_{-0.032}^{+0.033}$  &  $-1.00_{-0.005}^{+0.050}$  &  $-0.68_{-0.017}^{+0.036}$  &  420.22  &  1.74  &  426.22  &  436.67  &  1.97  &  5.44 \\
SBCH$\textsc{ii}H_{0}$  &  $0.32_{-0.010}^{+0.010}$  &  $3.05_{-0.847}^{+0.713}$  &  $69.73_{-0.649}^{+0.658}$  &  $0.69_{-0.036}^{+0.033}$  &  $-0.90_{-0.041}^{+0.038}$  &  $-0.61_{-0.030}^{+0.029}$  &  2184.7  &  1.12  &  2190.7  &  2207.41  &  -3.38  &  2.19 \\
aTanh  &     &     &     &     &     &     &     &     &     &     &     &    \\
SC$H_{0}$  &  $0.24_{-0.030}^{+0.035}$  &  $1.75_{-0.752}^{+0.627}$  &  $71.47_{-0.879}^{+0.868}$  &  $0.90_{-0.140}^{+0.155}$  &  $-0.77_{-0.066}^{+0.047}$  &  $-0.59_{-0.027}^{+0.024}$  &  1782.11  &  1.03  &  1788.11  &  1804.48  &  -4.05  &  1.41 \\
SBC$H_{0}$  &  $0.32_{-0.010}^{+0.010}$  &  $0.16_{-0.104}^{+0.139}$  &  $69.65_{-0.781}^{+0.764}$  &  $0.61_{-0.026}^{+0.026}$  &  $-0.97_{-0.022}^{+0.034}$  &  $-0.65_{-0.018}^{+0.023}$  &  1824.32  &  1.04  &  1830.32  &  1846.74  &  1.4  &  6.88 \\
SCH$\textsc{ii}H_{0}$  &  $0.23_{-0.027}^{+0.033}$  &  $1.86_{-0.705}^{+0.555}$  &  $71.17_{-0.770}^{+0.770}$  &  $0.96_{-0.144}^{+0.151}$  &  $-0.76_{-0.059}^{+0.041}$  &  $-0.59_{-0.026}^{+0.023}$  &  2146.48  &  1.12  &  2152.48  &  2169.15  &  -6.34  &  -0.79 \\
BCH$\textsc{ii}H_{0}$  &  $0.31_{-0.011}^{+0.011}$  &  $0.05_{-0.034}^{+0.066}$  &  $69.76_{-0.587}^{+0.579}$  &  $0.66_{-0.029}^{+0.029}$  &  $-0.99_{-0.006}^{+0.012}$  &  $-0.69_{-0.013}^{+0.015}$  &  420.27  &  1.74  &  426.27  &  436.73  &  2.02  &  5.5 \\
SBCH$\textsc{ii}H_{0}$  &  $0.32_{-0.009}^{+0.010}$  &  $0.14_{-0.092}^{+0.128}$  &  $69.64_{-0.690}^{+0.677}$  &  $0.62_{-0.024}^{+0.024}$  &  $-0.97_{-0.019}^{+0.031}$  &  $-0.66_{-0.016}^{+0.022}$  &  2189.68  &  1.13  &  2195.68  &  2212.4  &  1.6  &  7.17 \\
\hline
\end{tabular}

\label{resultsH0table}
\end{table*}

\subsection{Constraints on $\Lambda$CDM Model}
\begin{figure}
\centering
\includegraphics[scale=0.30]{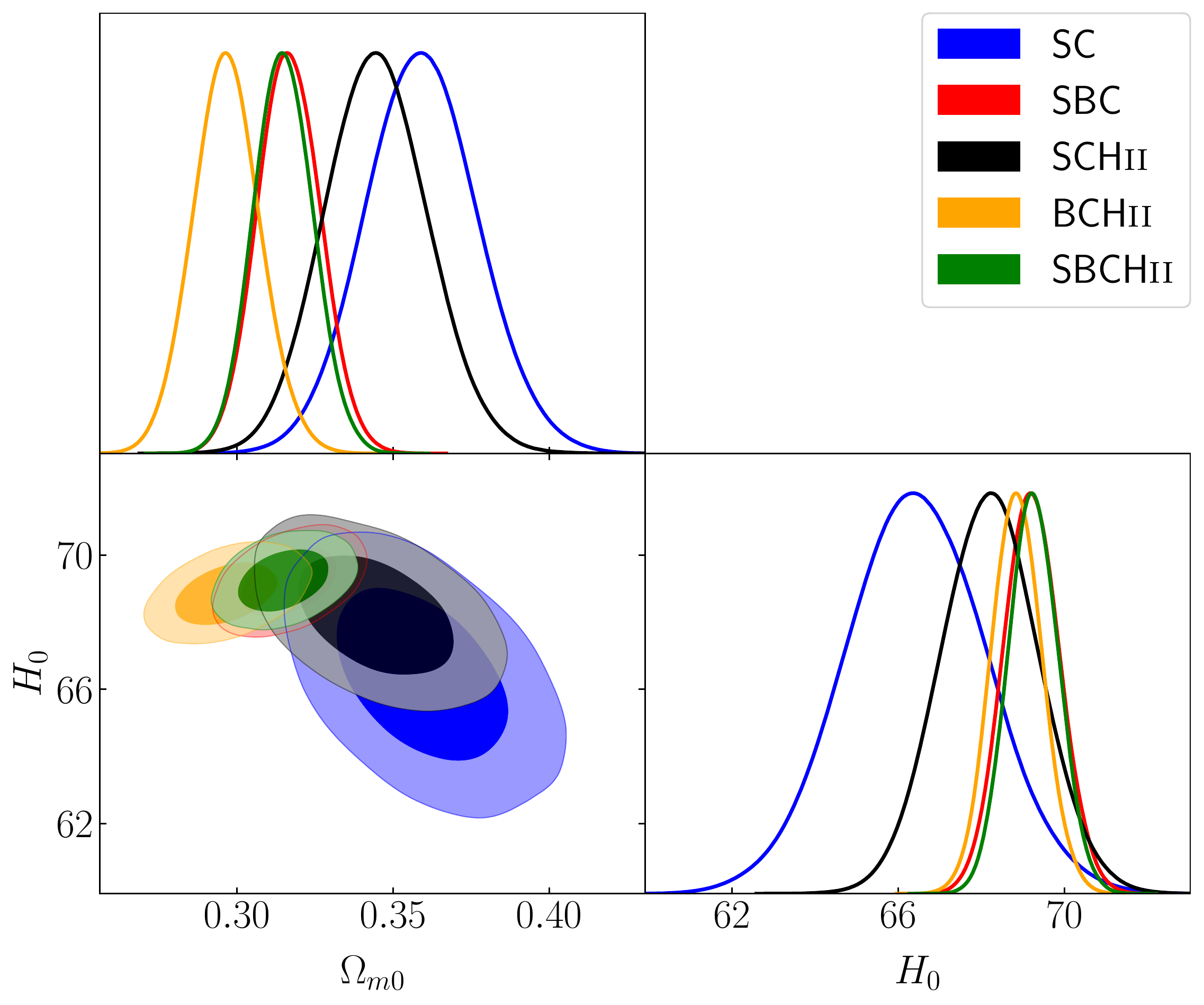}
\caption{The $\Lambda$CDM Model(without SH0ES prior for $H_{0}$): The posterior probability distribution plots of fitted parameters. The color correspondence for different data-set combinations can be seen in the figure legends. The darker and lighter shades of colors represent 1-sigma(68.26\%) and 2-sigma(95.44\%) confidence intervals, respectively.}
\label{LCDMdist}
\end{figure}

\begin{figure}
\centering
\includegraphics[scale=0.30]{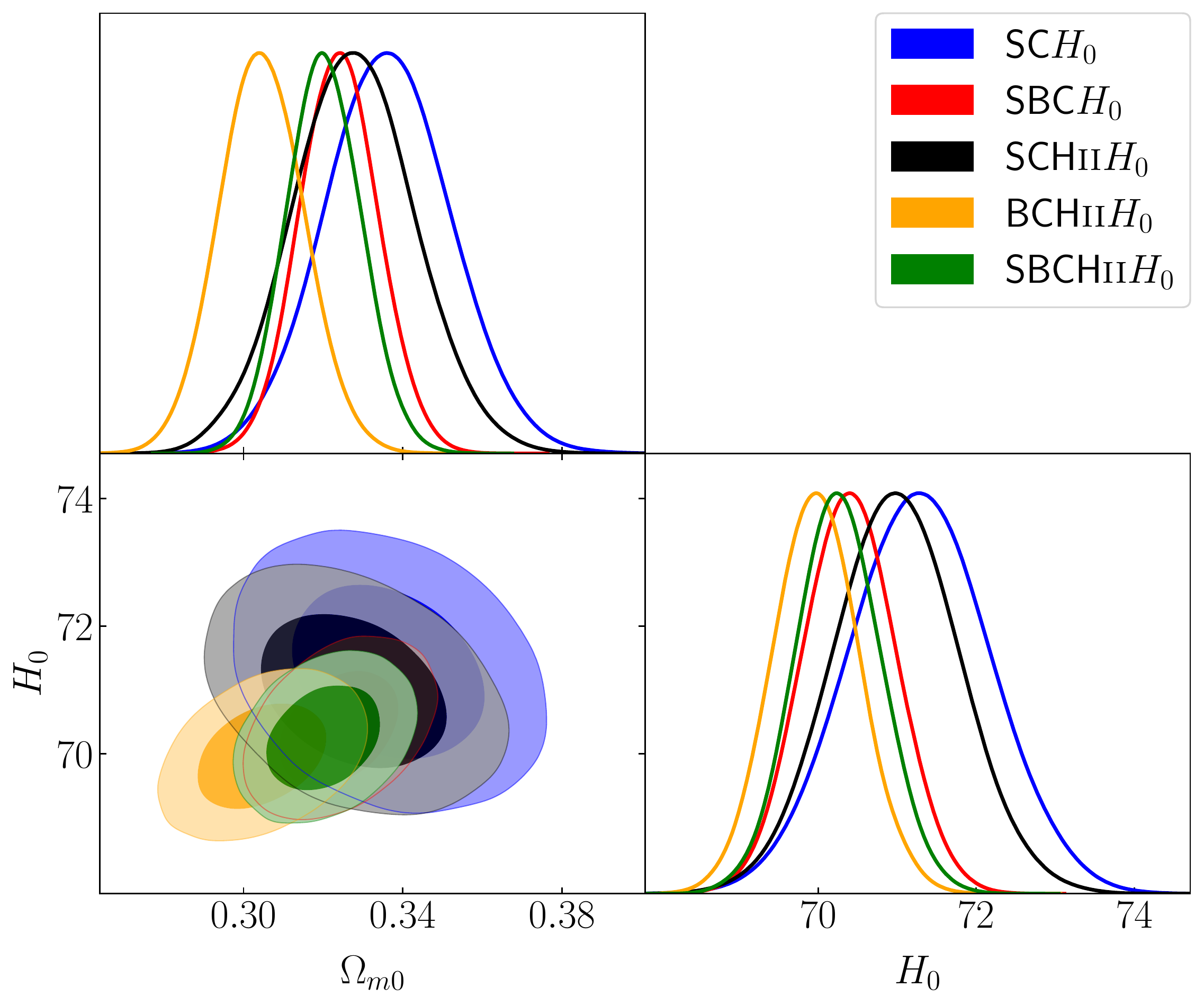}
\caption{The $\Lambda$CDM Model(with SH0ES prior for $H_{0}$): The posterior probability distribution plots of fitted parameters. The color correspondence for different data-set combinations can be seen in the figure legends. The darker and lighter shades of colors represent 1-sigma(68.26\%) and 2-sigma(95.44\%) confidence intervals, respectively.}
\label{LCDMH0dist}
\end{figure}
We have also included the results for the $\Lambda$CDM model, which is commonly used as a benchmark for assessing $f(R)$ models. We consider a two-parameter $\Lambda$CDM model,  
with the Hubble parameter given by $H = H_{0}\sqrt{\Omega_{m0}(1+z)^{3} + (1-\Omega_{m0})}$, where the two parameters $\Omega_{m0}$ and $H_{0}$  
stand for the matter density and the Hubble parameter at the present epoch,
respectively.
The median values along with 1-sigma confidence intervals for these parameters 
are presented in Tables \ref{resultstable} and \ref{resultsH0table}. 
The posterior probability distribution of $\Omega_{m0}$ and $H_{0}$,
both for the respective cases - without and with the SH0ES prior for $H_{0}$ -
are depicted in Figs \ref{LCDMdist} and \ref{LCDMH0dist}.
For all data sets (except SC), the values of $\Omega_{m0}$ are compatible with $\Omega_{m0,{\rm Planck}}$ within 1-2(3) sigma. There is a tension of approximately 1.5-3 sigma between the values of $H_{0}$ when comparing the cases with and without the SH0ES prior for $H_{0}$. There exists more or less same order of tensions between the fitted values of $H_{0}$ and   $H_{0,{\rm Planck}}$ or $H_{0,{\rm SH0ES}}$.\\

\subsection{Constraints on Hu-Sawicki Model and Starobinsky Model}
\begin{figure}
\centering
\includegraphics[scale=0.30]{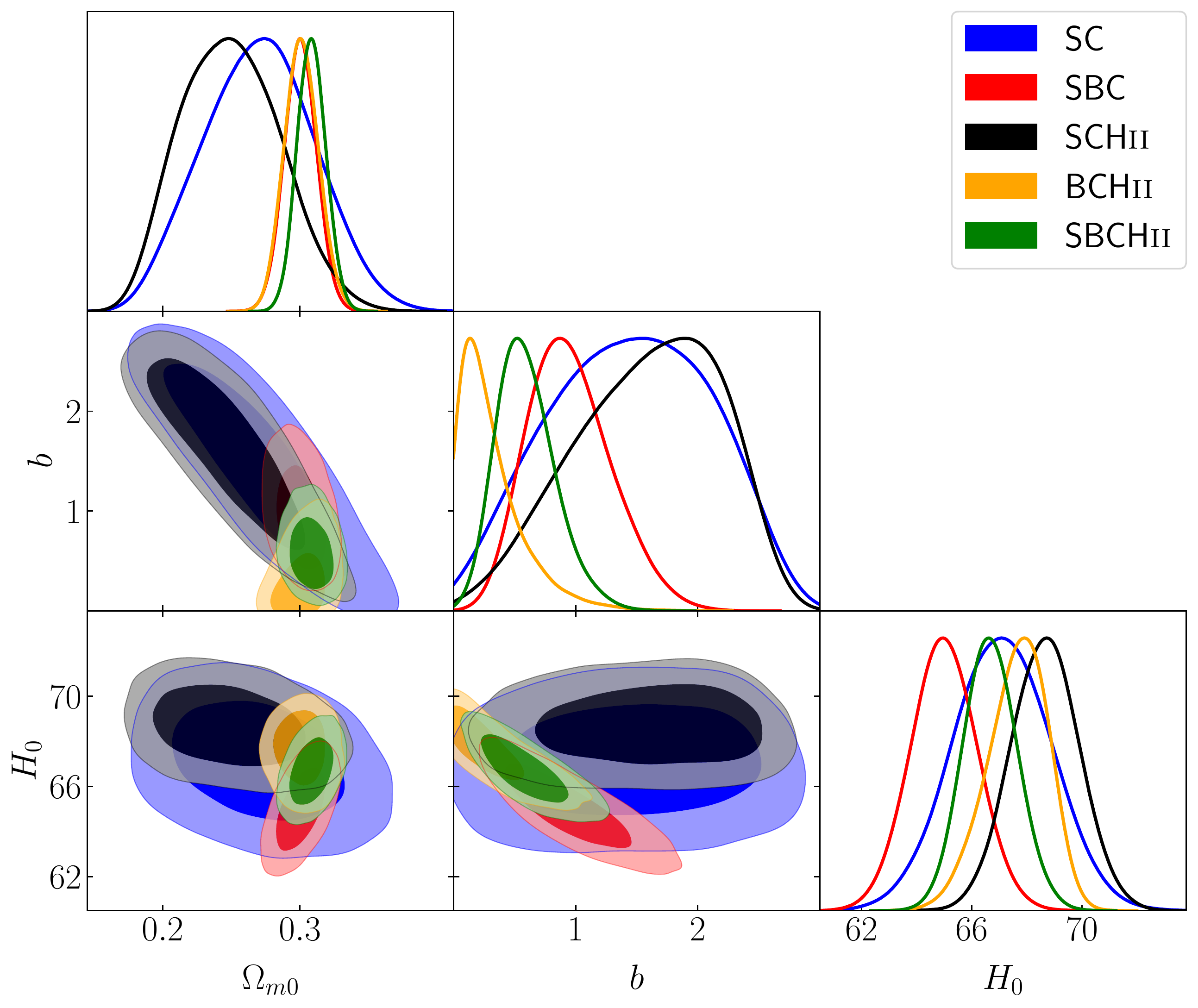}
\caption{The Hu-Sawicki Model($n_{_{\rm HS}}=1$, without SH0ES prior for $H_{0}$): The posterior probability distribution plots of fitted parameters. The color correspondence for different data-set combinations can be seen in the figure legends. The darker and lighter shades of colors represent 1-sigma(68.26\%) and 2-sigma(95.44\%) confidence intervals, respectively.}
\label{HS1dist}
\end{figure}

\begin{figure}
\centering
\includegraphics[scale=0.30]{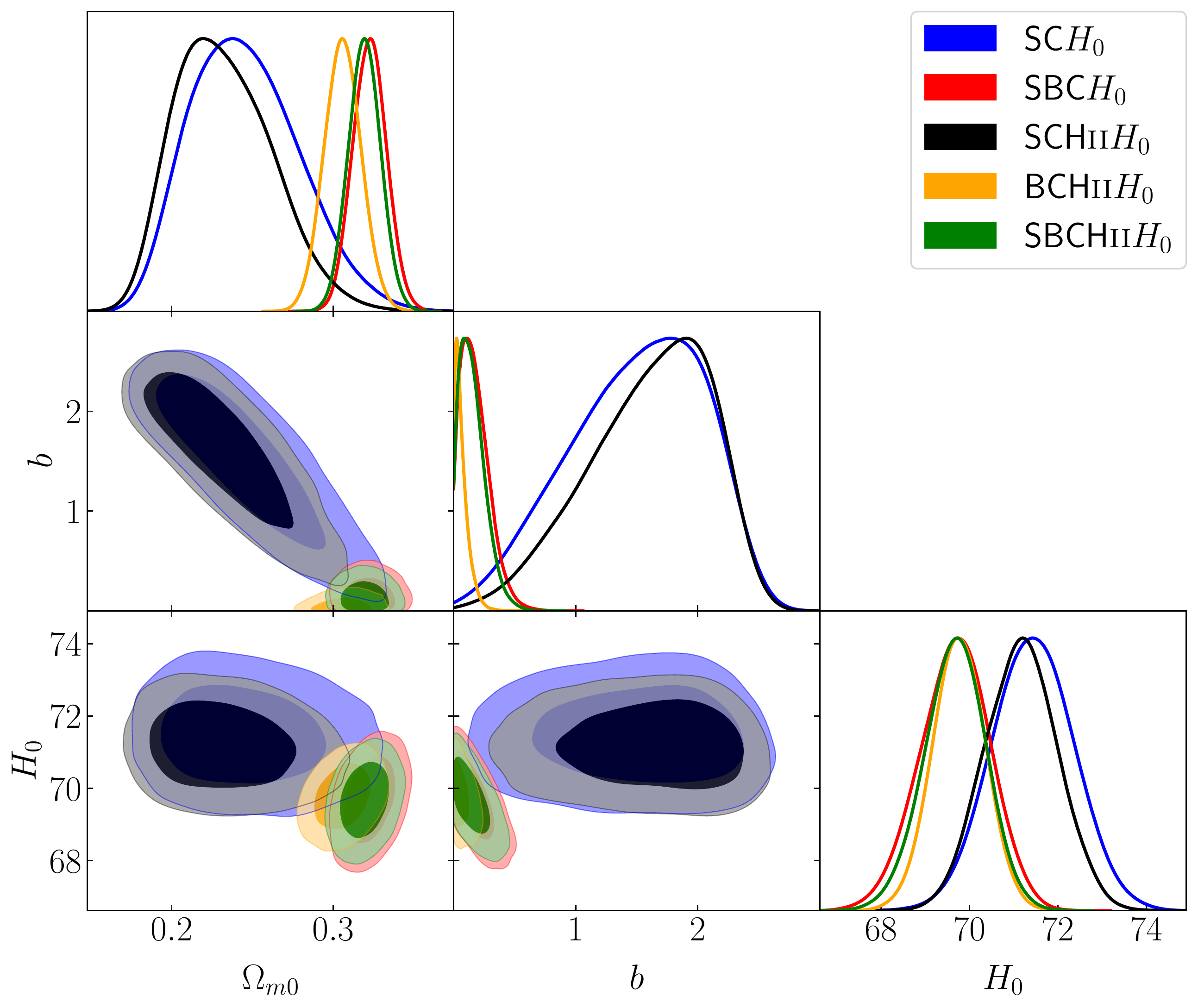}
\caption{The Hu-Sawicki Model($n_{_{\rm HS}}=1$, with SH0ES prior for $H_{0}$): The posterior probability distribution plots of fitted parameters. The color correspondence for different data-set combinations can be seen in the figure legends. The darker and lighter shades of colors represent 1-sigma(68.26\%) and 2-sigma(95.44\%) confidence intervals, respectively.}
\label{HS1H0dist}
\end{figure}

\begin{figure}
\centering
\includegraphics[scale=0.30]{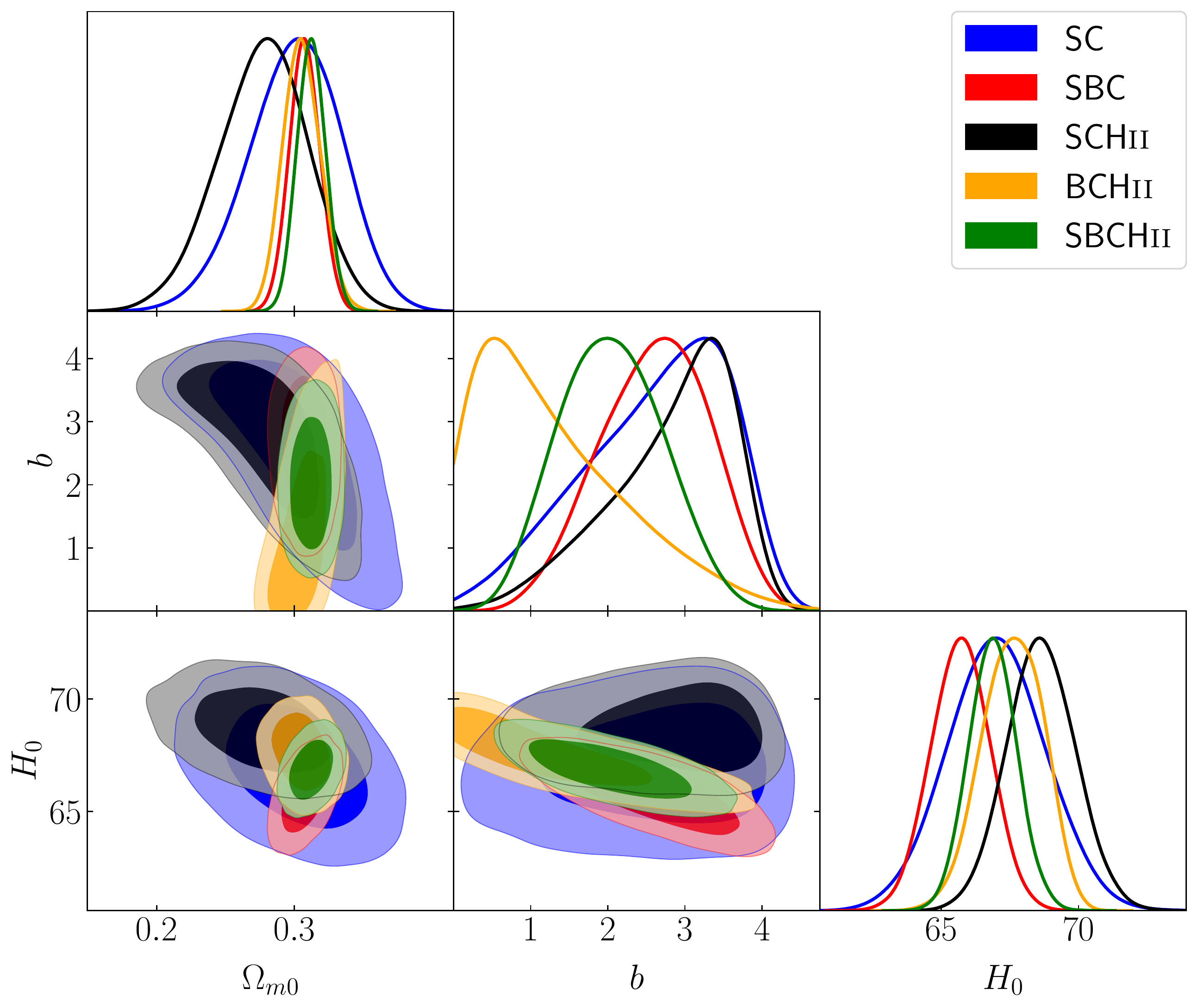}
\caption{The Starobinsky Model($n_{_{\rm S}}=1$, without SH0ES prior for $H_{0}$): The posterior probability distribution plots of fitted parameters. The color correspondence for different data-set combinations can be seen in the figure legends. The darker and lighter shades of colors represent 1-sigma(68.26\%) and 2-sigma(95.44\%) confidence intervals, respectively.}
\label{Starodist}
\end{figure}

\begin{figure}
\centering
\includegraphics[scale=0.30]{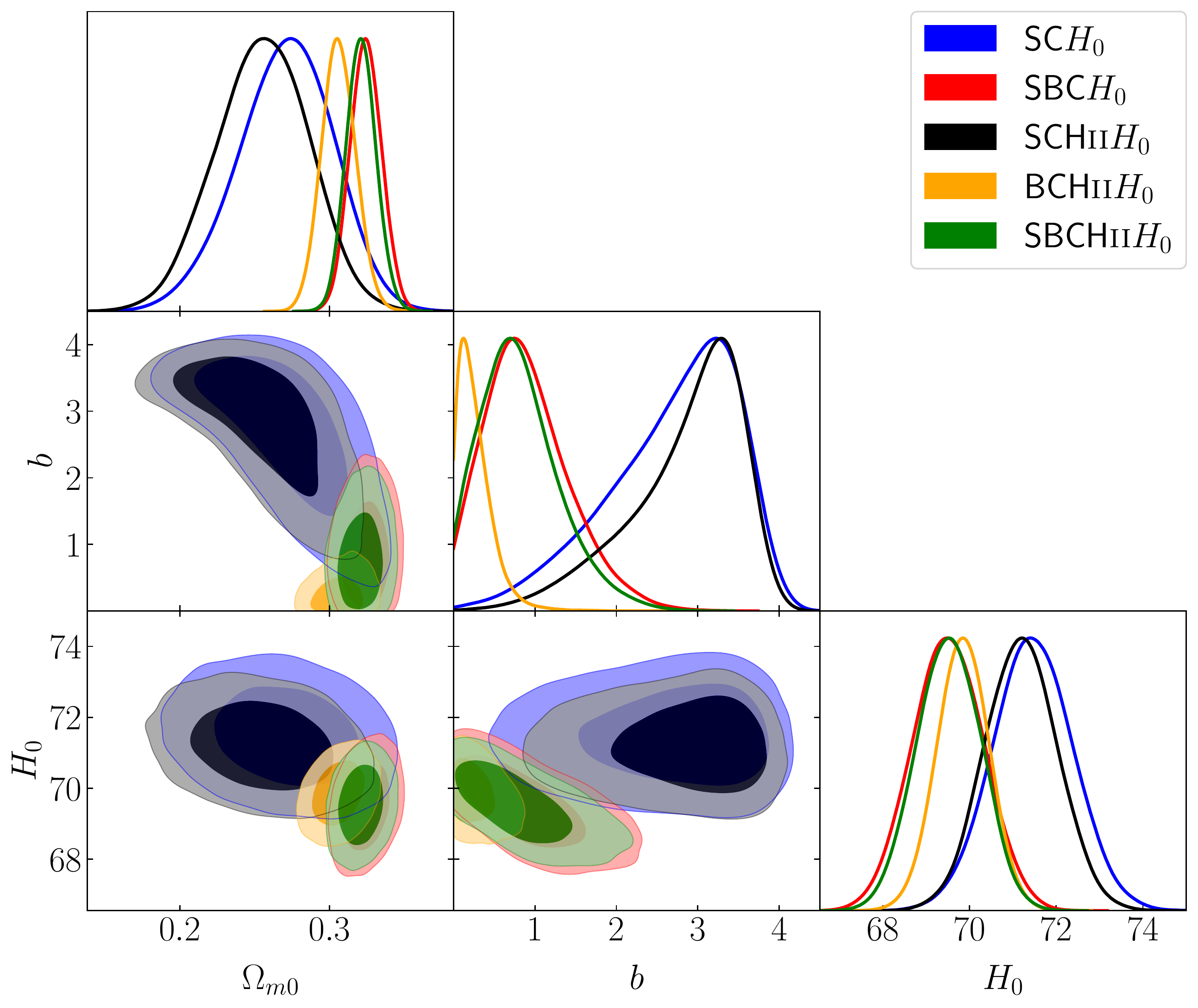}
\caption{The Starobinsky Model($n_{_{\rm S}}=1$, with SH0ES prior for $H_{0}$): The posterior probability distribution plots of fitted parameters. The color correspondence for different data-set combinations can be seen in the figure legends. The darker and lighter shades of colors represent 1-sigma(68.26\%) and 2-sigma(95.44\%) confidence intervals, respectively.}
\label{StaroH0dist}
\end{figure}

\begin{figure}
\centering
\includegraphics[scale=0.30]{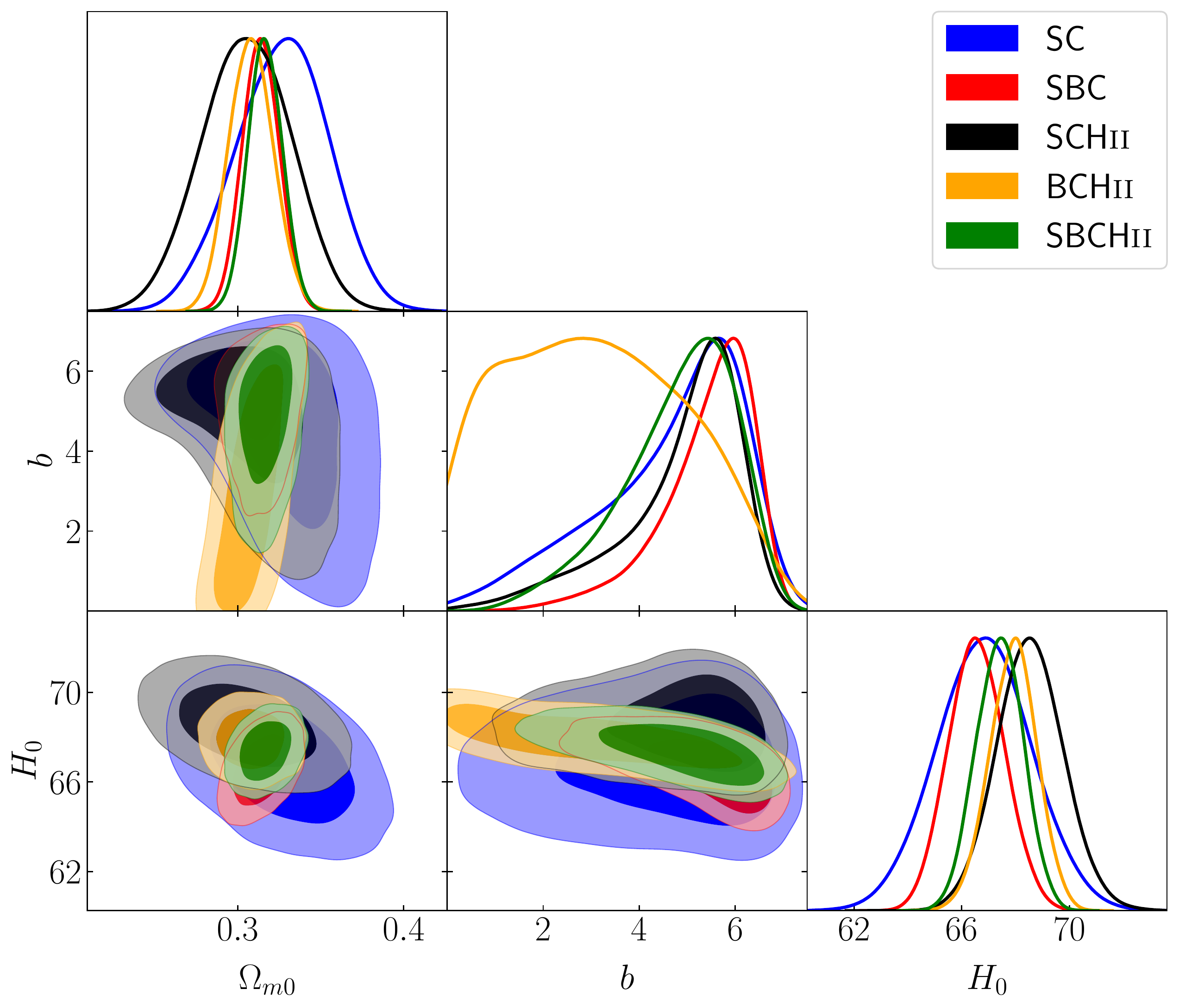}
\caption{The Hu-Sawicki Model($n_{_{\rm HS}}=3$, without SH0ES prior for $H_{0}$): The posterior probability distribution plots of fitted parameters. The color correspondence for different data-set combinations can be seen in the figure legends. The darker and lighter shades of colors represent 1-sigma(68.26\%) and 2-sigma(95.44\%) confidence intervals, respectively.}
\label{HS3dist}
\end{figure}

\begin{figure}
\centering
\includegraphics[scale=0.30]{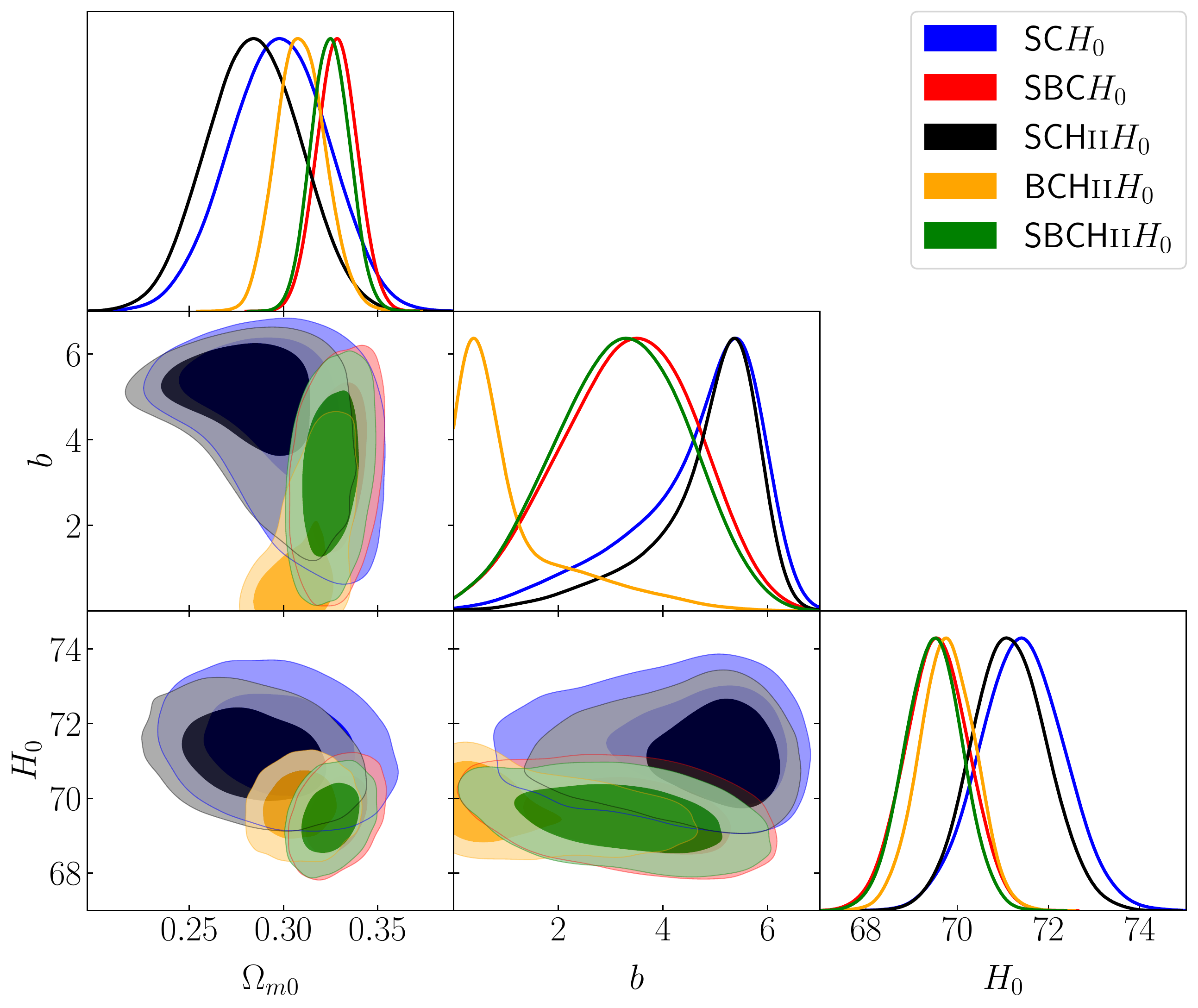}
\caption{The Hu-Sawicki Model($n_{_{\rm HS}}=3$, with SH0ES prior for $H_{0}$): The posterior probability distribution plots of fitted parameters. The color correspondence for different data-set combinations can be seen in the figure legends. The darker and lighter shades of colors represent 1-sigma(68.26\%) and 2-sigma(95.44\%) confidence intervals, respectively.}
\label{HS3H0dist}
\end{figure}
Previous studies (see \cite{Nunes:2016drj} and references therein) have established that there exists a degeneracy between $n_{_{\rm HS}}/n_{_{\rm S}}$ and $\Omega_{m0}$. One,
therefore , often works with fixed values of $n_{_{\rm HS}}$ and $n_{_{\rm S}}$  to address this degeneracy. In \cite{Capozziello:2007eu}, it has been
worked out that $n_{_{\rm HS}},\,n_{_{\rm S}}>0.9$, while \cite{Fu:2010zza} proposed that $n_{_{\rm HS}}$ and $n_{_{\rm S}}$ should be integers. 
The difficulty in constraining  higher $n_{_{\rm HS}}/n_{_{\rm S}}$ values due to 
numerical instability issues while solving the modified Hubble equations is also 
well known. Consequently, it has become common practice to work with   
$n_{_{\rm HS}}=1$ and $n_{_{\rm S}}=1$.
In our study we have also explored the case of $n_{_{\rm HS}}=3$. Note that 
in the re-parameterized version using the deviation parameter $b$, the Hu-Sawicki model with  $n_{_{\rm HS}}=2$ is equivalent to the Starobinsky model with 
$n_{_{\rm S}}=1$.\\

The quantitative results of fitting the Hu-Sawicki model 
(Eq. \ref{HSmodel1}) with   $n_{_{\rm HS}}=1,3$ and
the Starobinsky model (Eq. \ref{STmodel1}) with $n_{_{\rm S}}=1$, 
following the procedure 
discussed in Sec. \ref{ocd}, 
are presented in the Tables \ref{resultstable} and 
\ref{resultsH0table}. 
The 2D contour plots of the posterior probability distribution 
of the model parameters and 1D marginalised distribution of each 
of the parameters are shown in the Figs. \ref{HS1dist} and \ref{HS1H0dist} 
(for $n_{_{\rm HS}}=1$), Figs. \ref{Starodist} and \ref{StaroH0dist} 
(for $n_{_{\rm HS}}=2$ or, $n_{_{\rm S}}=1$), Figs. \ref{HS3dist} and \ref{HS3H0dist} 
(for $n_{_{\rm HS}}=3$),  for the respective cases - without and with the SH0ES 
prior for $H_{0}$.\\

The Fig. \ref{Om0tension} illustrates that as
$n_{_{\rm HS}}$ increases (from 1 to 3)
the  values of $\Omega_{m0}$  approach closer to those from the $\Lambda$CDM model
(and/or to $\Omega_{m0,\,\rm Planck}$) and the constraints on them become tighter. 
This applies to both cases with and without the SH0ES prior for $H_{0}$. 
We may observe from Fig. \ref{H0tension} that when the SH0ES prior for $H_{0}$ is not considered, increasing $n_{_{\rm HS}}$ causes the model values of $H_{0}$ 
to approach $H_{0,\,{\rm Planck}}$, except for the SC dataset where there is 
almost no change with varying $n_{_{\rm HS}}$. However, the constraints from the BCH$\textsc{ii}$ dataset are not considered reliable due to high $\chi^{2}/{\nu}$. 
When the SH0ES prior is included, there is hardly any change in the model values of $H_{0}$ with respect to  $n_{_{\rm HS}}$, and they lie around $69.5-71.5$ km\,s$^{-1}$Mpc$^{-1}$. As expected, the parameter $b$ increases as $n_{_{\rm HS}}$  increases (see Fig. \ref{btension}). 
Also, the constraints on $b$ become weaker. We may infer that for higher values of $n_{_{\rm HS}}$ (say 4, 5, ...) and $n_{_{\rm S}}$ (say 2, 3, ..), 
the constraints on $b$ would become even more weaker. This implies that with further increase in  $n_{_{\rm HS}}$ or  $n_{_{\rm S}}$ , one can obtain a model that closely resembles the $\Lambda$CDM model in terms of its predictions for the physical parameters $\Omega_{m0}$ and $H_{0}$, despite the constraints on $b$ moving towards larger values. 
In a sense, this undermines the purpose of exploring $f(R)$ models further, and there is a strategic motivation to limit the exploration to $n_{_{\rm HS}}=1,\,2$ only, putting aside the consideration of computational difficulties in constraining for the higher values of 
$n_{_{\rm HS}}$ or $n_{_{\rm S}}$.
For all three models, there are instances where $b=0$ is only very marginally allowed(i.e., within a 2-3 sigma limit), indicating distinguishable models from the $\Lambda$CDM model. This finding is also an important result of the present work.

\subsection{Constraints on Exponential Model}
\begin{figure}
\centering
\includegraphics[scale=0.30]{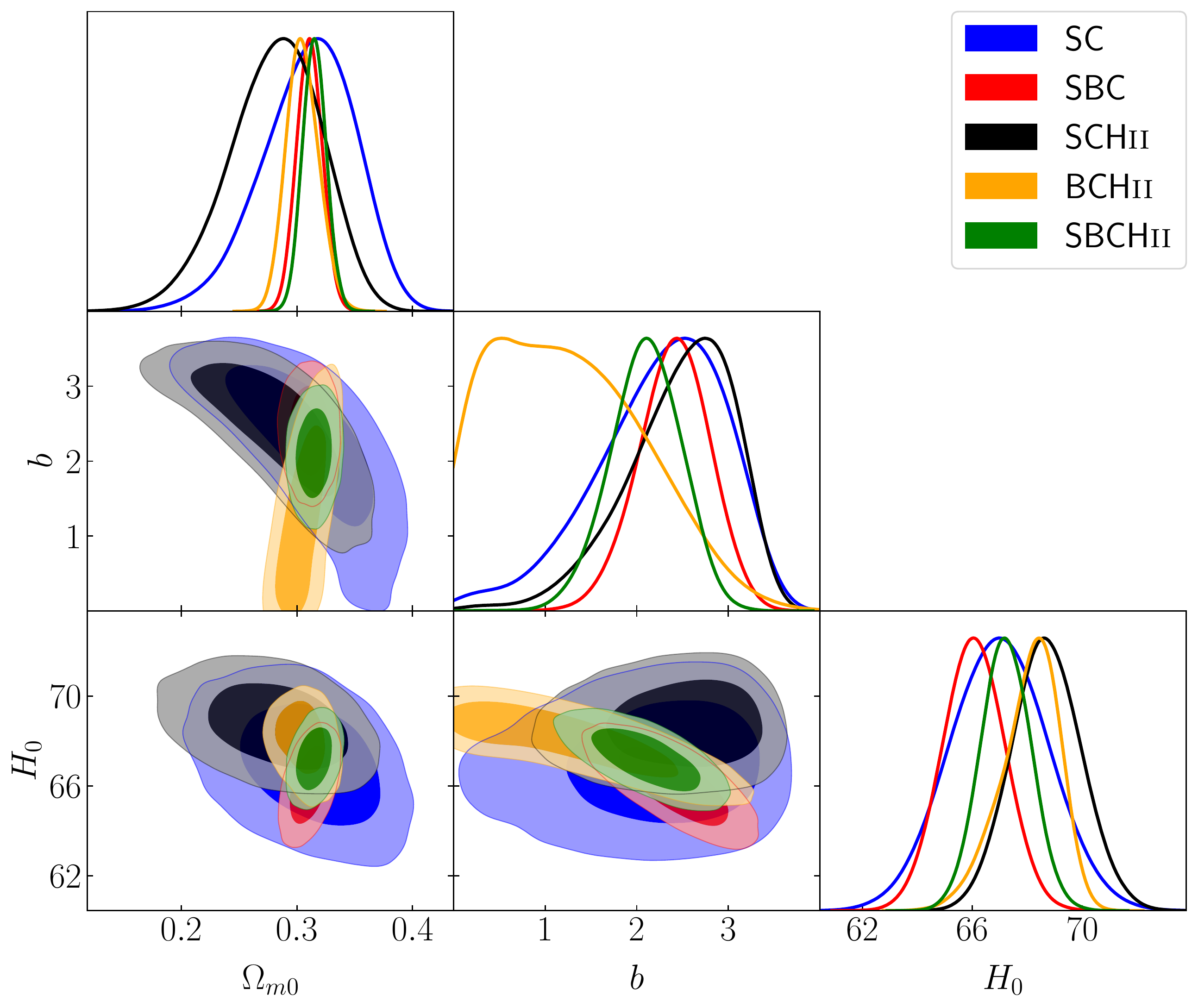}
\caption{The Exponential Model(without SH0ES prior for $H_{0}$): The posterior probability distribution plots of fitted parameters. The color correspondence for different data-set combinations can be seen in the figure legends. The darker and lighter shades of colors represent 1-sigma(68.26\%) and 2-sigma(95.44\%) confidence intervals, respectively.}
\label{Expdist}
\end{figure}

\begin{figure}
\centering
\includegraphics[scale=0.30]{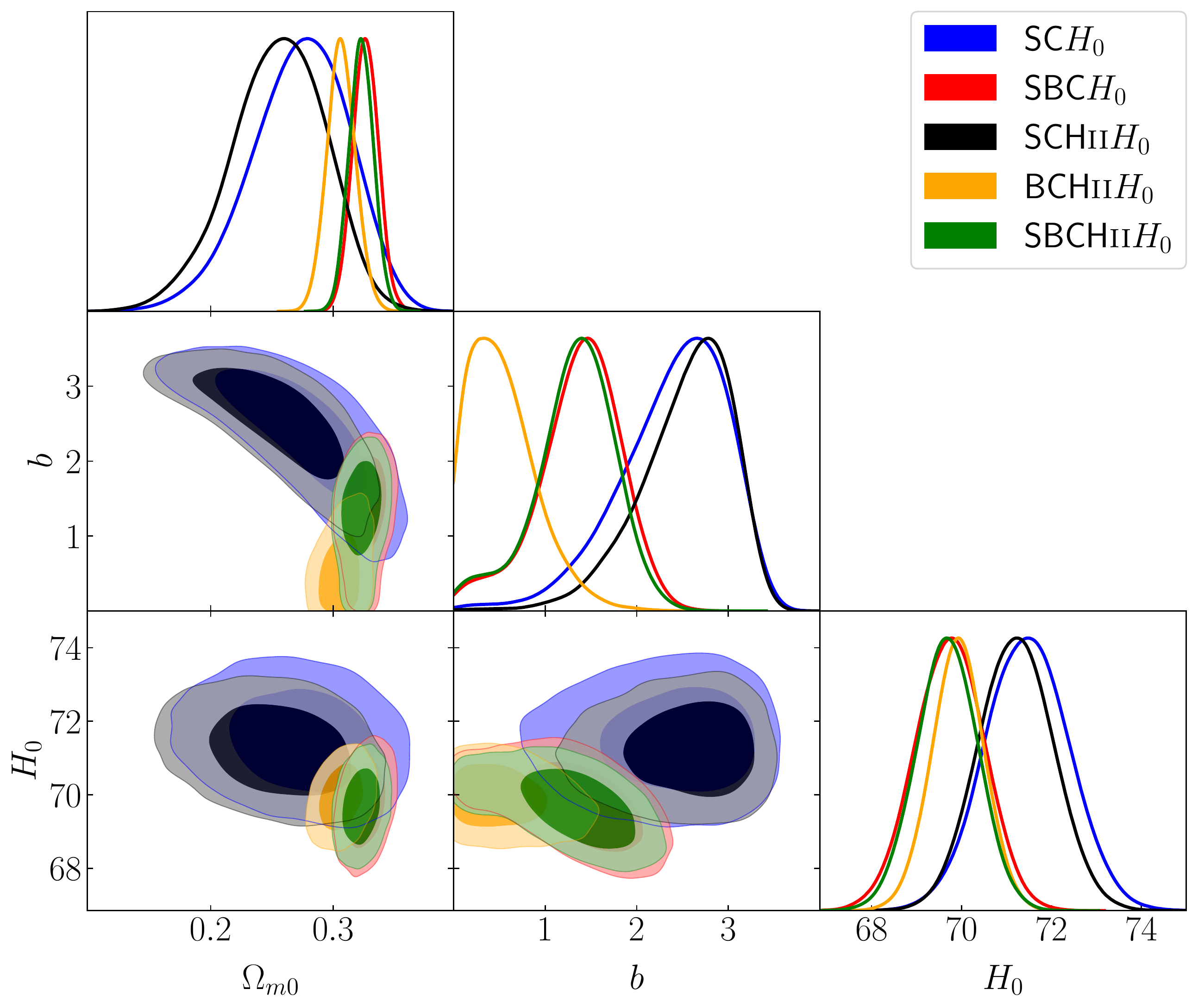}
\caption{The Exponential Model(with SH0ES prior for $H_{0}$): The posterior probability distribution plots of fitted parameters. The color correspondence for different data-set combinations can be seen in the figure legends. The darker and lighter shades of colors represent 1-sigma(68.26\%) and 2-sigma(95.44\%) confidence intervals, respectively.}
\label{ExpH0dist}
\end{figure}

In the Tables  \ref{resultstable} and \ref{resultsH0table}, we
present the median values and 1-sigma (68.26\%) confidence intervals 
for the parameters of the exponential model (Eq. \ref{expmodel1}). 
The corresponding 2D contour plots illustrating the posterior probability distribution of the parameters, along with the 1D marginal distributions for each parameter, are depicted in Figs \ref{Expdist} and \ref{ExpH0dist} for the cases - without and with the SH0ES prior for $H_{0}$, respectively.
We observe  from   Figs \ref{Expdist}, \ref{ExpH0dist}, and \ref{btension}   
that except for the BCH$\textsc{ii}(H_{0})$ dataset, 
the  parameter $b$ clearly deviates from zero, 
with $b=0$   barely allowed. 
Despite this, the values of the parameters $\Omega_{m0}$ and $H_{0}$ 
(as illustrated in Figs. \ref{Om0tension} and \ref{H0tension}) 
are reasonably close to the standard values derived from Planck constraints 
\cite{Planck:2018vyg}. The significance of this result becomes further clear
in the subsequent section on model comparisons. 
When considering the SH0ES prior for $H_{0}$, the range of model values of $H_{0}$  is $\sim 69.7$ to $71.5$ km\,s$^{-1}$Mpc$^{-1}$.

\subsection{Constraints on Tsujikawa Model}

\begin{figure}
\centering
\includegraphics[scale=0.30]{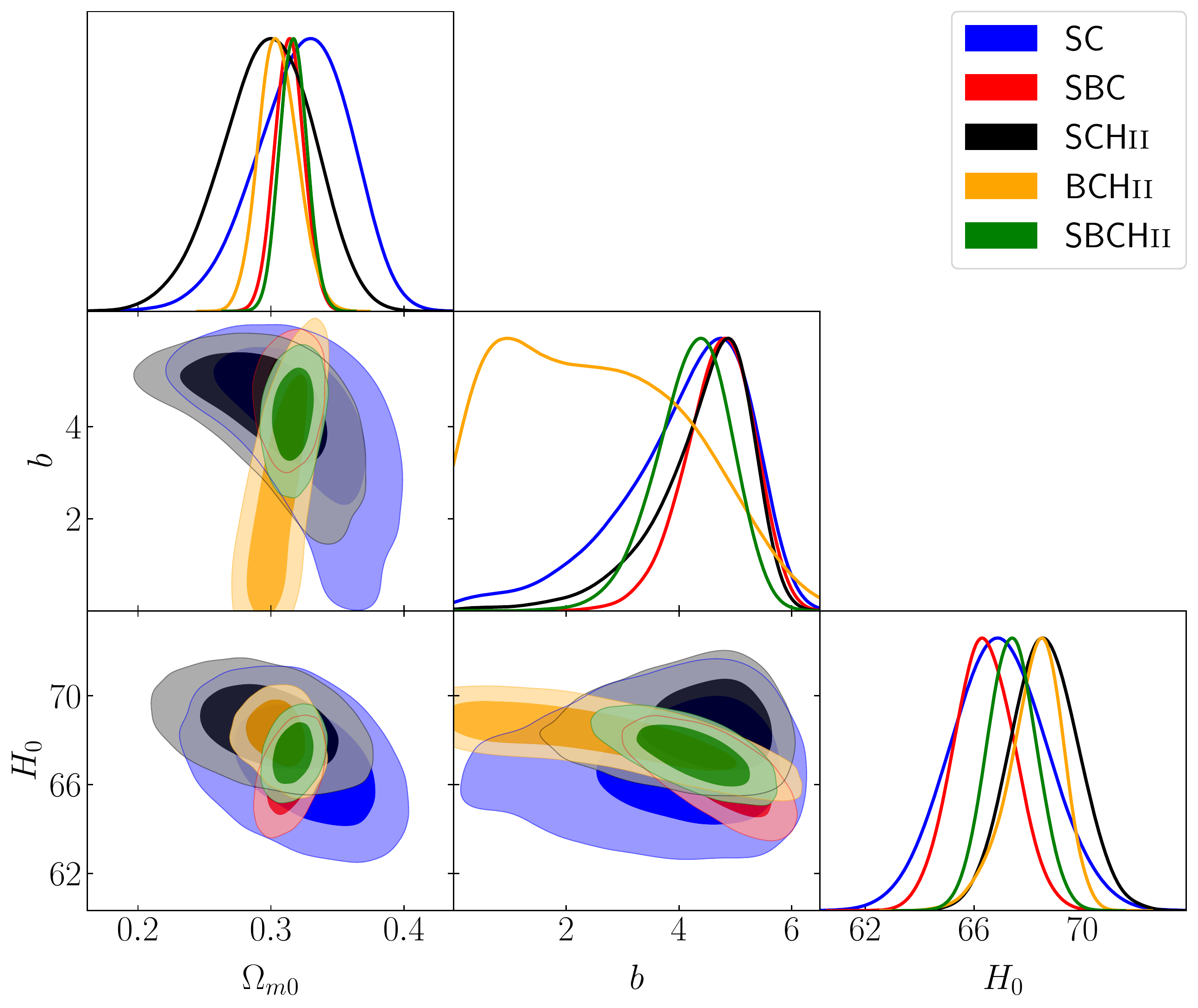}
\caption{The Tsujikawa Model(without SH0ES prior for $H_{0}$): The posterior probability distribution plots of fitted parameters. The color correspondence for different data-set combinations can be seen in the figure legends. The darker and lighter shades of colors represent 1-sigma(68.26\%) and 2-sigma(95.44\%) confidence intervals, respectively.}
\label{Tsujidist}
\end{figure}

\begin{figure}
\centering
\includegraphics[scale=0.30]{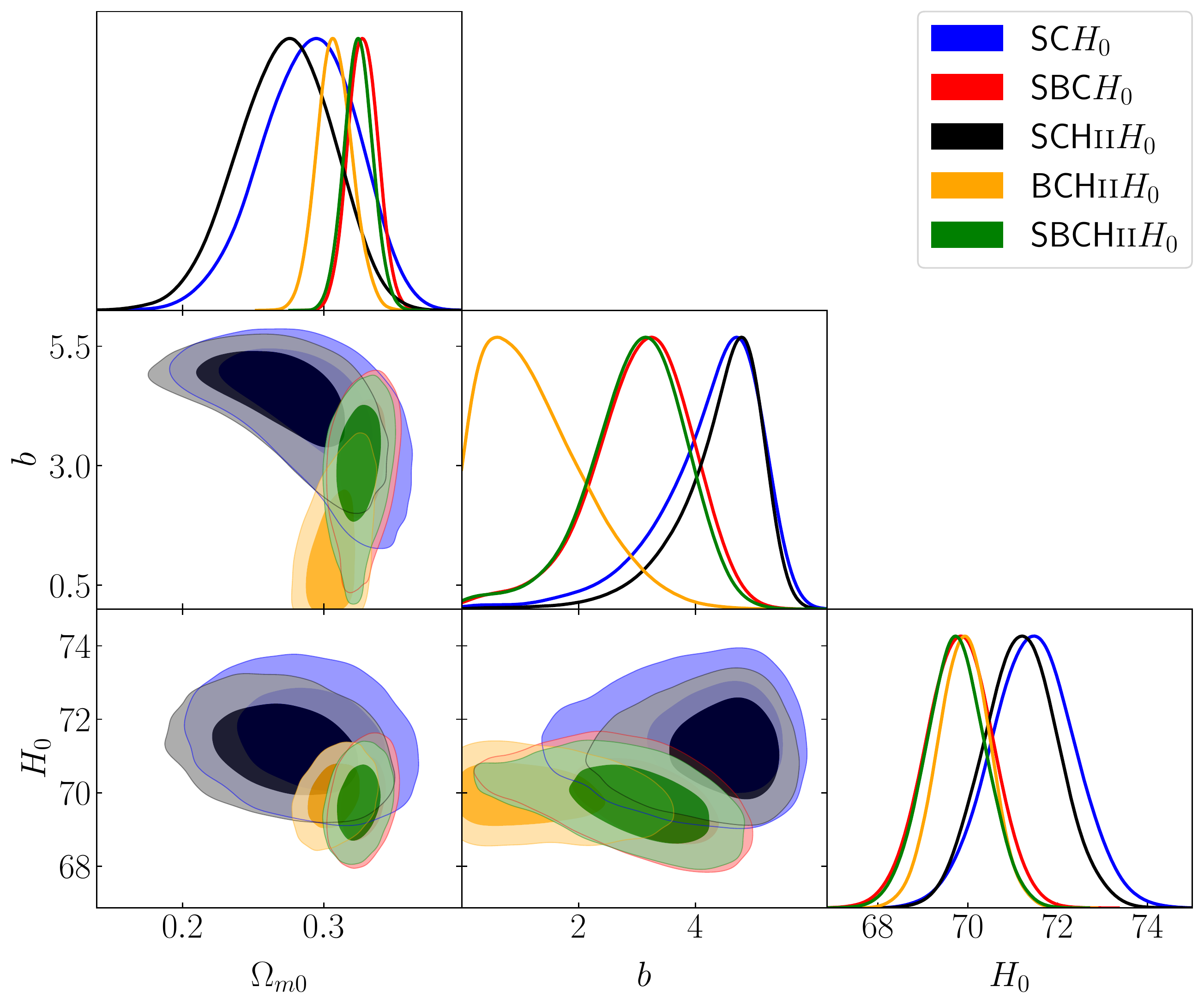}
\caption{The Tsujikawa Model(with SH0ES prior for $H_{0}$): The posterior probability distribution plots of fitted parameters. The color correspondence for different data-set combinations can be seen in the figure legends. The darker and lighter shades of colors represent 1-sigma(68.26\%) and 2-sigma(95.44\%) confidence intervals, respectively.}
\label{TsujiH0dist}
\end{figure}

The constraints on the parameters of the Tsujikawa model described by 
Eq. \ref{tsujimodel1} are presented in the Tables \ref{resultstable} 
and \ref{resultsH0table}, as well as in Figs. \ref{Tsujidist} and \ref{TsujiH0dist}. 
These constraints are shown separately for cases without and with the SH0ES prior 
on $H_{0}$. Similar to the findings of the exponential model, 
here also we observe that the deviation parameter $b$ is mostly nonzero, 
with only a very marginal allowance for it to be zero, across all the datasets, 
except BCH$\textsc{ii}(H_{0})$. We disregard the results from BCH$\textsc{ii}(H_{0})$ 
on account of very high value of the reduced $\chi^{2}_{\rm min}/\nu$. 
Hence, we obtain an $f(R)$ that is clearly distinguishable from the $\Lambda$CDM model. 
The fitted values of $\Omega_{m0}$ and $H_{0}$ also fall within reasonable 
limits compared to $\Omega_{m0,{\rm Planck}}$ and $H_{0,{\rm Planck}}$ 
(or $H_{0,{\rm SH0ES}}$). With SH0ES prior for $H_{0}$, the model's median values 
for $H_{0}$ range from approximately 69.7 to 71.5 km s$^{-1}$Mpc$^{-1}$, 
which introduces a tension of 2-3 sigma with $H_{0,{\rm Planck}}$ 
or $H_{0,{\rm SH0ES}}$.

\subsection{Constraints on arcTanh Model}

\begin{figure}
\centering
\includegraphics[scale=0.30]{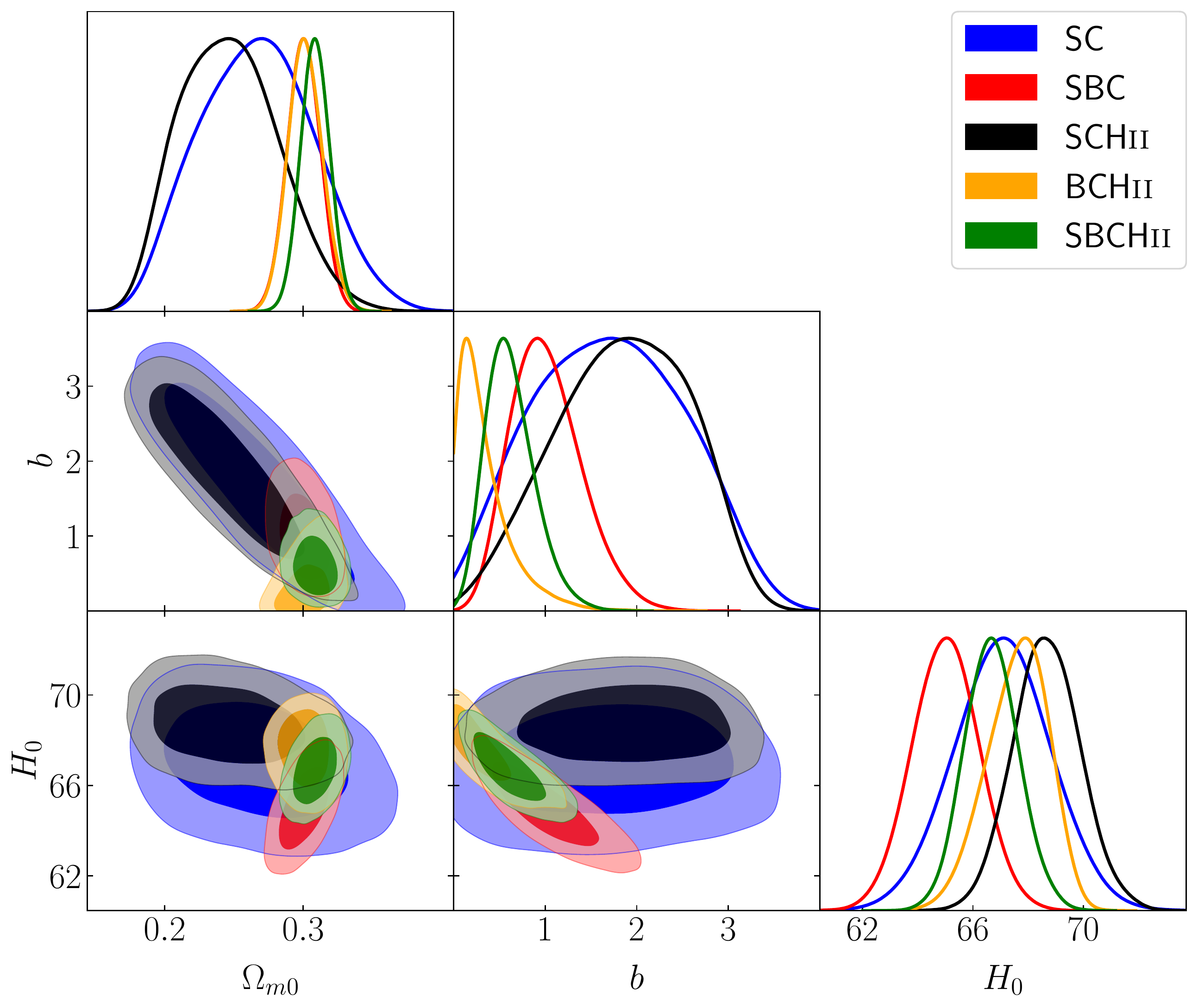}
\caption{The arcTanh Model(without SH0ES prior for $H_{0}$): The posterior probability distribution plots of fitted parameters. The color correspondence for different data-set combinations can be seen in the figure legends. The darker and lighter shades of colors represent 1-sigma(68.26\%) and 2-sigma(95.44\%) confidence intervals, respectively.}
\label{aTanhdist}
\end{figure}

\begin{figure}
\centering
\includegraphics[scale=0.30]{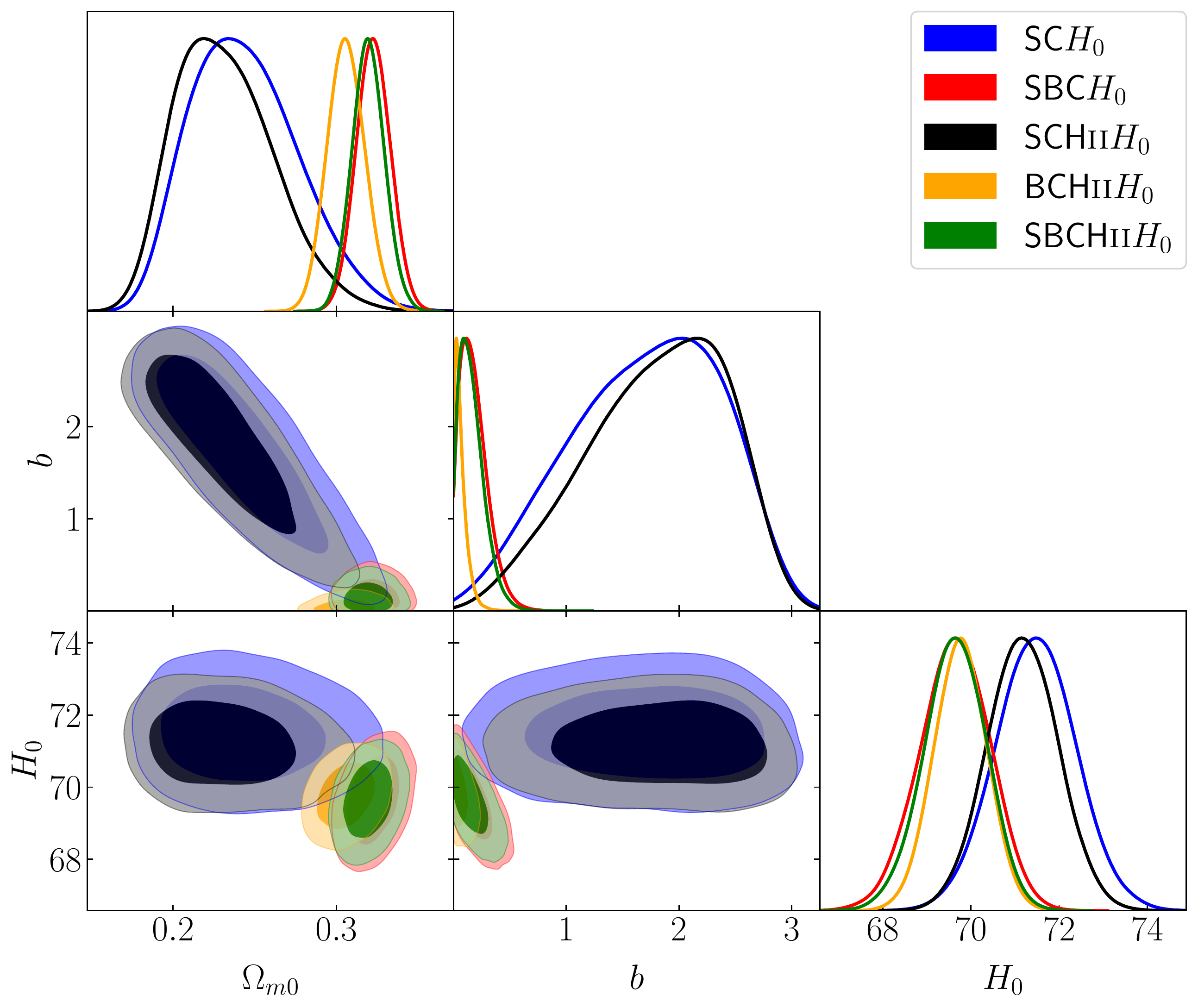}
\caption{The arcTanh Model(with SH0ES prior for $H_{0}$): The posterior probability distribution plots of fitted parameters. The color correspondence for different data-set combinations can be seen in the figure legends. The darker and lighter shades of colors represent 1-sigma(68.26\%) and 2-sigma(95.44\%) confidence intervals, respectively.}
\label{aTanhH0dist}
\end{figure}
The results obtained by constraining the aTanh model are presented in the  Tables \ref{resultstable} and \ref{resultsH0table}, as well as in the Figs. \ref{aTanhdist} and \ref{aTanhH0dist}, separately for cases without and with the SH0ES prior for $H_{0}$. In several cases, excluding BCH$\textsc{ii}(H_{0})$, the parameter $b=0$ falls well within the 1-2 sigma range. However, there are instances where $b=0$ lies outside the 2-sigma range. In the subsequent section on model comparison, we will observe that these latter cases are also more favored, making this outcome significant. The fitted values of $\Omega_{m0}$ for the model fall within the 1-2 sigma range of $\Omega_{m0,{\rm Planck}}$. Furthermore, $H_{0,{\rm Planck}}$ is approximately within the 1-2 sigma range of that predicted by the model when the SH0ES prior is not considered. Upon inclusion of the SH0ES prior for $H_{0}$, the model's predicted median values for $H_{0}$ range from approximately 69.65 to 71.47 km\,s$^{-1}$Mpc$^{-1}$, which are in 2-3 sigma tension with   $H_{0,{\rm Planck}}$ or $H_{0,{\rm SH0ES}}$.\\

It is important to highlight that  the  outcomes of this model 
are not significantly different   
from those of the HS1 model. This similarity would not sound surprising,
if we examine Eqs \ref{HSmodel1} and \ref{aTanhmodel} along with the fact that 
$\mathrm{arctanh}(\Lambda/R)\approx{\Lambda/R}$ for $\Lambda/R\ll1$
(which holds true in this case since $\Lambda\sim0.7H_{0}^{2}$ and $R\gtrsim8H_{0}^{2}$).   



\section{Model Comparison}
\label{comparison}

The standard statistical tools commonly used to assess model fitting (and model comparison) in cosmology 
include the reduced chi-square statistics ($\chi^{2}_{\nu}$), the Akaike Information 
Criterion (AIC) \cite{AIC}, and the Bayesian Information Criterion (BIC) \cite{BIC} 
(also see \cite{Liddle:2004nh, Liddle:2007fy, Burnham2004}).
The last six columns of Tables \ref{resultstable} and \ref{resultsH0table} 
contain essential quantities related to the current analysis, 
specifically pertaining to the utilization of these statistical tools.
The reduced chi-square statistics is defined as $\chi^{2}_{\nu} = \chi^{2}_{\rm min}/{\nu}$ whereas the AIC and the BIC are respectively defined by the following equations
\begin{eqnarray}
{\rm AIC} &=& -2\ln\mathcal{L}_{\rm max} + 2k\,,
\label{AICeq}\\
{\rm BIC} &=& -2\ln\mathcal{L}_{\rm max} + k\ln{N}.
\label{BICeq}
\end{eqnarray}
The number of degrees of freedom, represented by the symbol $\nu$, is determined by 
subtracting the number of model parameters ($k$) from the total number of data points ($N$). 
The minimum value of $\chi^{2}$ is denoted as $\chi^{2}_{\rm min}$ and is connected to the 
maximum likelihood ($\mathcal{L}_{\rm max}$) through the relation $\chi^{2}_{\rm min} = 
-2\ln\mathcal{L}_{\rm max}$. While comparing multiple competing models using a given data set, 
the model with the lowest values of $\chi^{2}$, AIC, and BIC is considered to be more 
favoured  by the data. However, it is insufficient to rely solely on 
$\chi^{2}$ due to the 
principle of Occam's razor, which emphasizes the importance of considering the number of model parameters.
 Typically, for a nested model, as the number of parameters increases, the fit improves, leading to a decrease in $\chi^{2}_{\rm min}$ (or an increase in likelihood), regardless of the relevance of the newly included parameter(s). Both AIC and BIC incorporate a penalty term ($2k$ and $k\ln{N}$, respectively) that penalizes models with more parameters, in addition to the $\chi^{2}_{\rm min}$ term, taking into account any improvement in fit. Thus
 a balance between the quality of fit and the complexity of the model
 is achieved through AIC and BIC. Comparing the Eqs. \ref{AICeq} and \ref{BICeq}  for AIC and BIC,  respectively, it can be observed that the penalty for models with a greater number of parameters is more severe in BIC than in AIC. Unfortunately, the conclusions derived from applying these two criteria may sometimes disagree. In such situations of disagreement, it is necessary to investigate whether any violations of the assumptions on which AIC and BIC are based have occurred (for details see \cite{Liddle:2004nh,Liddle:2007fy,Burnham2004}).

To perform a comparative analysis of two models, a useful  measure  
$\Delta{X} = X_{2} - X_{1}$, computed as the difference between the values of $X$ (= AIC or BIC) for the two models: 1 and 2,  facilitates a quantitative assessment of the evidence 
supporting the preference of model 1 over model 2.  A rule of thumb which is
commonly followed as a guideline to indicate the degree of strength of the evidence 
with which the model 2 (model 1) is favoured over the other,  is as follows:
 (i) $0<\Delta{X}\leqslant 2$ ($-2\leqslant\Delta{X}<0$):   weak evidence,
 (ii) $2<\Delta{X}\leqslant 6$ ($-6\leqslant\Delta{X}<-2$):    positive   evidence,
 (iii) $6<\Delta{X}\leqslant 10$ ($-10\leqslant\Delta{X}<-6$):  strong   evidence,  
  (iv) $\Delta{X}> 10$ ($\Delta{X}<-10$): very strong (highly pronounced) evidence.\\

As can be seen from Tables  \ref{resultstable} and \ref{resultsH0table},  
the reduced $\chi^{2}$ values for the BCH$\textsc{ii}(H_{0})$ data-set do not come
close to the value 1, for all the models. Consequently, these cases are not considered
 in our subsequent discussions related to model comparisons. The purpose of
 inclusion of these cases here is to  
 illustrate that not all possible combinations of data-sets yield significantly
 meaningful results.\\

When we analyze different data-sets without considering the SH0ES prior for $H_{0}$, 
we observe interesting patterns in Table \ref{resultstable}. The  $\Delta$AICs values
indicate varying degrees of evidence ranging from weak to very strong in favor of $f(R)$ 
models compared to the $\Lambda$CDM model.
On the other hand, the values of $\Delta$BIC 
suggest evidence that is either weak or positive against $f(R)$ models in certain cases, while in other cases, the evidence is weak, positive, or even strong in favour of $f(R)$ models.
The strongest support for all $f(R)$ models is from the SBC data-set, followed by the SBCH$\textsc{ii}$ data-set. In both cases, the trends of $\Delta$AICs and $\Delta$BICs align, 
guiding us in the same direction.\\

When the SH0ES prior for $H_{0}$ is included, the results, as shown in Table \ref{resultsH0table}, exhibit some differences. The overall support for $f(R)$ models weakens compared to the previous case. Based on consideration of the $\Delta$AICs, the maximum support for any $f(R)$ model is now only strong (no longer very strong), and weak evidence against the $f(R)$ models is observed in a few cases. On the other hand, analyzing the $\Delta$BICs for data-sets with the SH0ES prior for $H_{0}$ reveals weak, positive, or strong evidence against the $f(R)$ models when compared to the $\Lambda$CDM model.
Among the data-sets with the SH0ES prior for $H_{0}$, the strongest evidence against the $f(R)$ models is observed in the SBCH$\textsc{ii}H_{0}$ data-set, followed by the SBC$H_{0}$ data-set. Furthermore, Table \ref{resultsH0table} highlights that when a SH0ES prior for $H_{0}$ is employed, the evidence against models such as HS3, EXP, and TSUJI (which are more similar to the $\Lambda$CDM model) is weak or mildly positive according to $\Delta$BIC. Conversely, for models such as HS1, aTanh, and ST (which are less or least similar to the $\Lambda$CDM model), the evidence against them is strong when evaluated using $\Delta$BIC. Consequently, we can conclude that with a SH0ES prior for $H_{0}$, the $\Delta$BIC criterion disfavours models that significantly deviate from the $\Lambda$CDM model.\\

A crucial observation to note is that the cases of $f(R)$ models with strong or very strong evidence for being preferred over the $\Lambda$CDM model (as indicated by $\Delta$AIC and/or $\Delta$BIC) also correspond to the cases where a non-zero value of the deviation parameter $b$ is favored, with the possibility of a very marginal acceptance of $b=0$. This finding is an important result of this study, which can be observed from Figs. \multiref{HS1dist}{aTanhH0dist} and Tables \ref{resultstable} and \ref{resultsH0table}.

\section{Accelerated Expansion of the Universe}
\label{accel}
In this section we   discuss the findings regarding the relevant quantities that describe the current accelerated expansion of the universe.  
 The behavior of the deceleration parameter ($q(z)$), which is defined as $q(z)\equiv -a\ddot{a}/\dot{a}^{2}=-\ddot{a}/(H^{2}a)$, provides valuable insight into whether the cosmic expansion is accelerating ($q<0$) or decelerating ($q>0$), serving as an indicator of the respective phases. Through several model-independent measurements of $q(z)$,
  it has been observed that $q(z=0)<0$ and $q(z>z_{\rm t})>0$ \cite{Rani:2015lia,Jesus:2019nnk}. The quantity $z_{\rm t}$,
called  transition redshift, indicates the redshift at which the universe has 
undergone  transition  from a decelerated  phase to an accelerated  phase of expansion.
\par
The total equation-of-state (EoS) parameter ($w_{\rm eff}$) 
as a function of redshift provides valuable information about the evolution of the universe
through the eras of  radiation domination, matter domination, and the subsequent 
late time accelerated phase of expansion dubbed as dark energy era.
 During these three phases, $w_{\rm eff}$ takes values approximately 
 equal to $1/3$, $0$, and $-1$ (for $\Lambda$CDM). On the other hand, the EoS parameter 
 ($w_{\rm DE}$) associated with the dark energy component remains constant at $-1$ for all values of redshift (both in the past and future) in  $\Lambda$CDM model. 
However, in the case of any viable $f(R)$ model, 
value of the parameter $w_{\rm DE}$   can cross   the 
so called ``phantom-divide" line ($w_{\rm DE}(z)=-1$) multiple times before finally settling at $-1$ in the far future \cite{jaime2012fr,Bamba:2010iy}.\\

\begin{figure}
\centering
\includegraphics[scale=0.4]{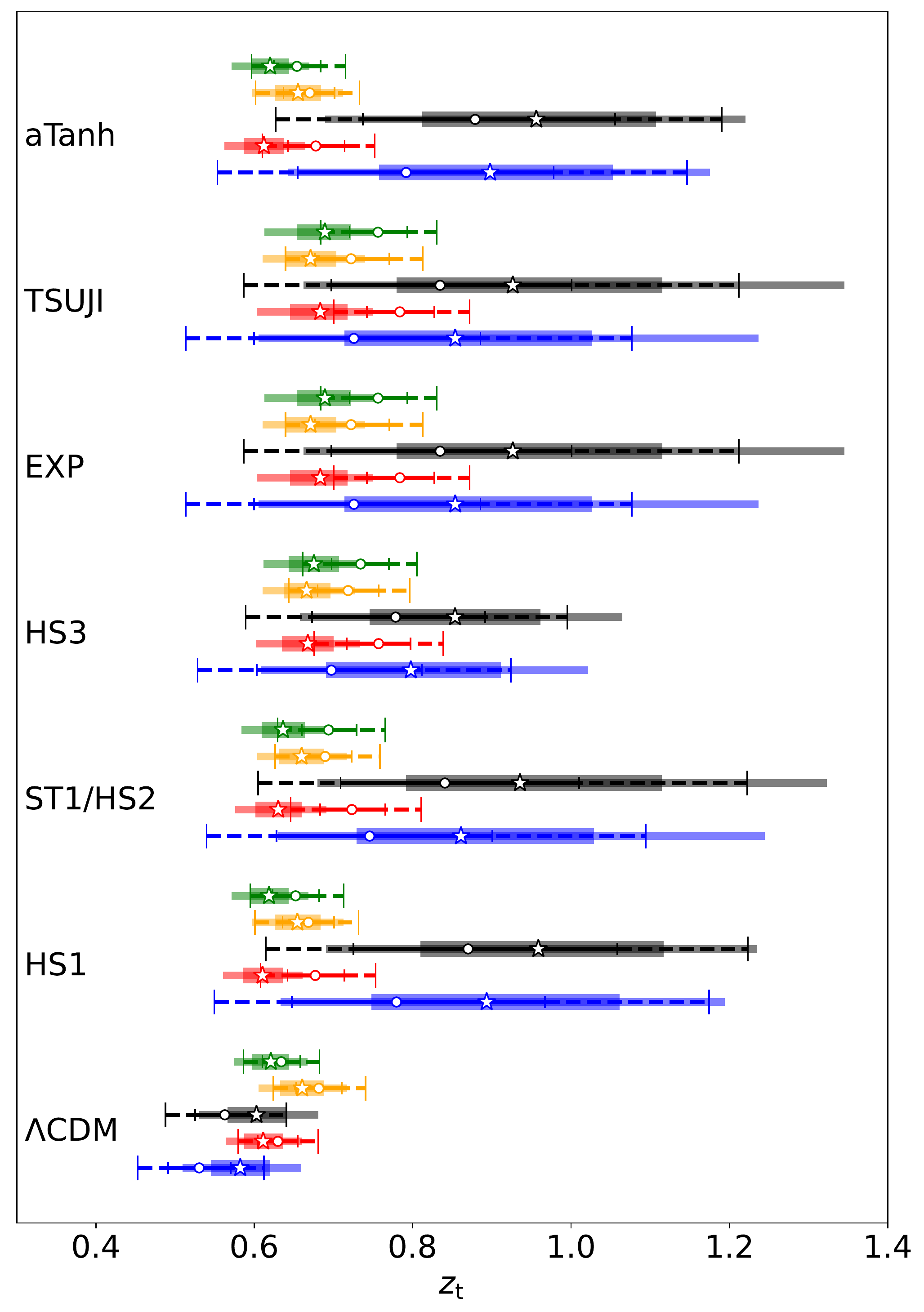}
\caption{Transition Redshift: The blank star and blank circle markers represent median values for the cases with and without SH0ES prior for $H_{0}$, respectively. Different colors represent different data-set combinations and these are same as in any of the parameter distribution plots(e.g. Fig. \ref{HS1dist}) or see 2nd paragraph of Sec.\ref{acc}. The thick and thin horizontal colored bars represent 1-sigma(68.26\%) and 2-sigma(95.44\%) confidence intervals for the cases with SH0ES prior. The colored continuous/dashed lines represent 1-sigma(68.26\%, with smaller cap size) and 2-sigma(95.44\%, with bigger cap size) confidence intervals for the cases without SH0ES prior.}
\label{z_trans}
\end{figure}

\begin{figure}
\centering
\includegraphics[scale=0.32]{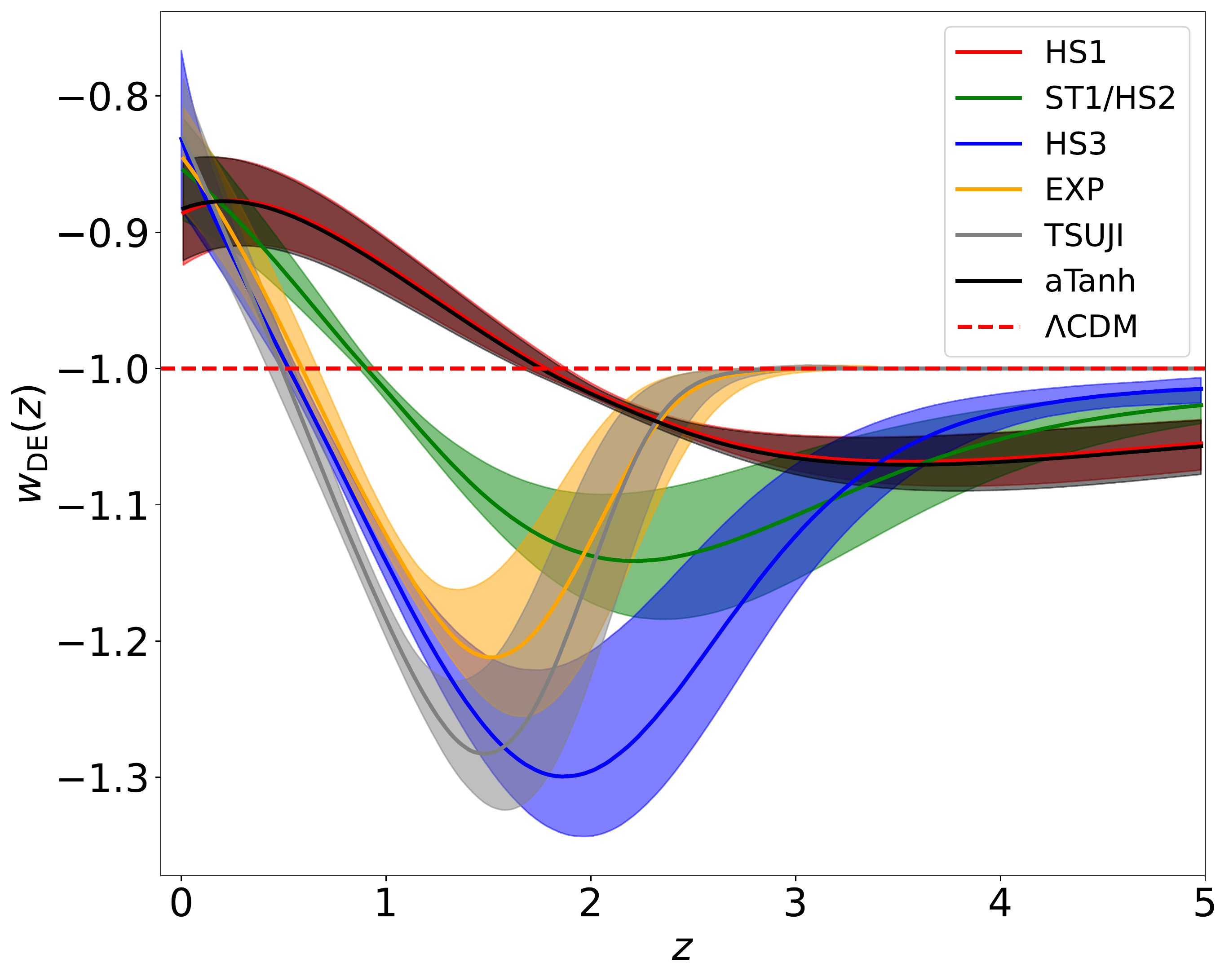}
\caption{Evolution of EoS parameter of the ``effective geometric dark energy'' for all $f(R)$ models obtained from the  data-set SBCH$\textsc{ii}$. Different colored solid lines show median values for different models as indicated by legends whereas with same color shaded areas show 1-sigma confidence interval.}
\label{wDE_evol}
\end{figure}

\begin{figure}
\centering
\includegraphics[scale=0.32]{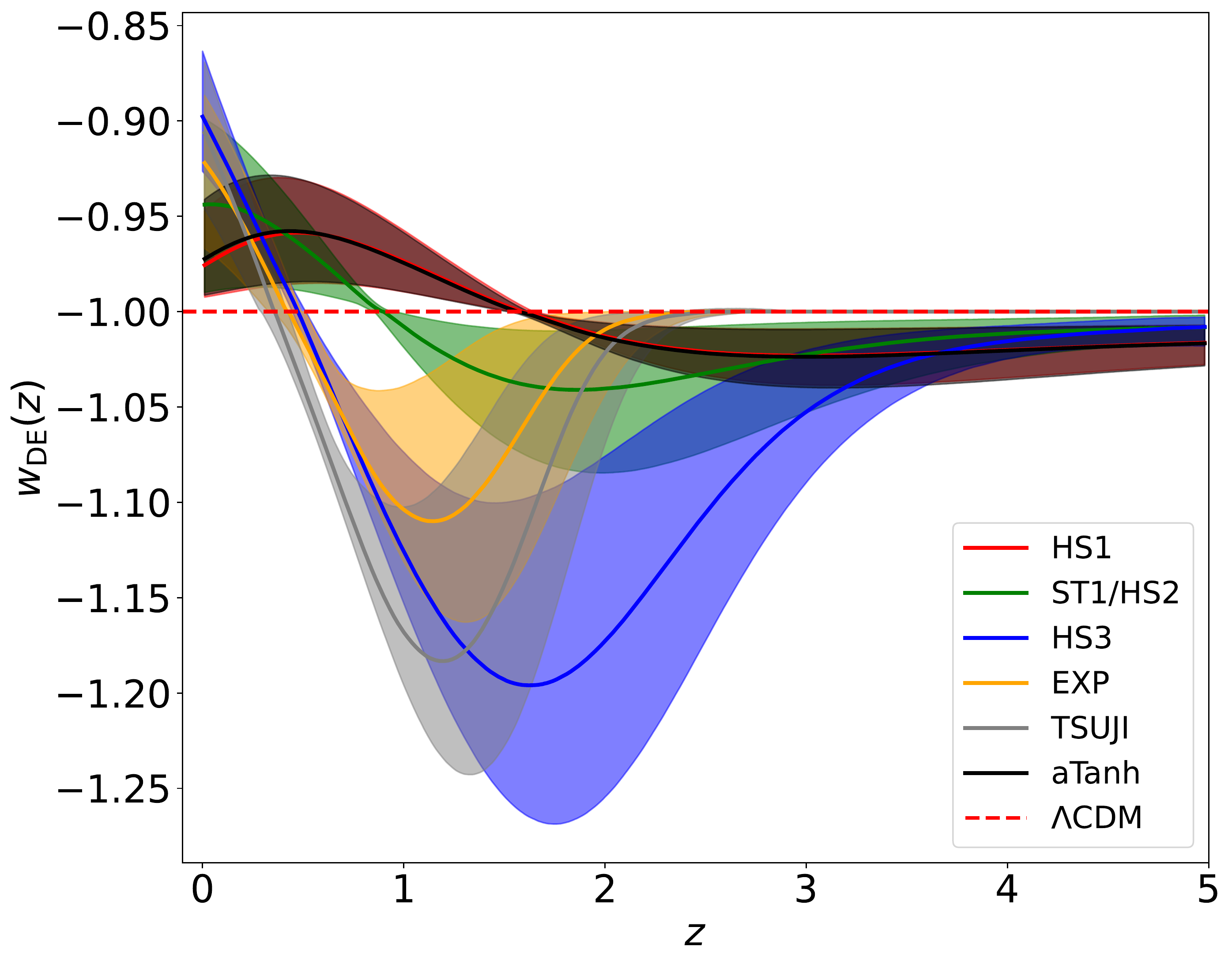}
\caption{Evolution of EoS parameter of the ``effective geometric dark energy'' for all $f(R)$ models obtained from the  data-set SBCH$\textsc{ii}H_{0}$. Different colored solid lines show median values for different models as indicated by legends whereas shaded areas with colors show 1-sigma confidence interval.}
\label{wDE_evolH0}
\end{figure}

In the Tables \ref{resultstable} and \ref{resultsH0table} we present the median values 
and 1-sigma confidence intervals for the quantities $z_{\rm t}$, $w_{\rm DE,0}$, and $w_{\rm eff,0}$ for two cases - without and with the SH0ES prior on $H_0$. In Fig \ref{z_trans}, we display the values of $z_{\rm t}$ for all models and data-sets. This figure shows   the median values of $z_{\rm t}\sim 0.5-1$ for all the models, which is compatible with the model-independent predictions for $z_{\rm t}$  \cite{Rani:2015lia,Jesus:2019nnk}. In the Figs. \ref{wDE_evol} and \ref{wDE_evolH0} we have plotted the evolution of $w_{\rm DE}$ with redshift based on the best constraints obtained for the models, (which happens to be the constraints from the data-set SCH$\textsc{ii}(H_{0})$).
In all $f(R)$ models, 
the values  of $w_{\rm DE}(z)$ mark a crossover from the so-called ``phantom regime" 
in the distant past to the so-called ``quintessence regime" in the more recent past, with the present-epoch values ($w_{\rm DE,0}$) falling within the quintessence region.  
 When the SH0ES prior on $H_0$ is included, the deviations of $w_{\rm DE}(z)$ for $f(R)$ models from the $\Lambda$CDM model are reduced and the transition from the phantom regime to the quintessence regime happens more recently compared to the situation without the SH0ES prior.
\par
For any viable $f(R)$ model,  the $w_{\rm DE}(z)$ curve
is expected to  eventually reach a stable, constant value of $-1$ in the distant past 
- a behavior which is necessary to account for the matter-dominated era 
in the context of the model.
From  Figs. \ref{wDE_evol} and \ref{wDE_evolH0}, we can readily observe that the Tsujikawa model and the Exponential model exhibit a clear convergence to $-1$, while for the other
models  this convergence takes place in a more remote past, characterized by higher values of $z$.
The trend for possible future crossing into phantom regime is clearly visible for the HS1 model, the Starobinsky model and the aTanh model, whereas this trend is not as apparent in the other models. However, theoretical studies  \cite{jaime2012fr,Bamba:2010iy}  have indicated that these models are also expected to cross the phantom divide in the future, albeit in an oscillatory manner.
Although it would be an intriguing future project to extend the analysis of the $w_{\rm DE}(z)$ curve up to $z\rightarrow-1(a\rightarrow\infty)$, it is beyond the scope of the present work. This is due to the potential limitations of the employed ODE system, which may not accurately capture the oscillatory features in such a far future regime.
\par
\begin{figure}
\centering
\includegraphics[scale=0.32]{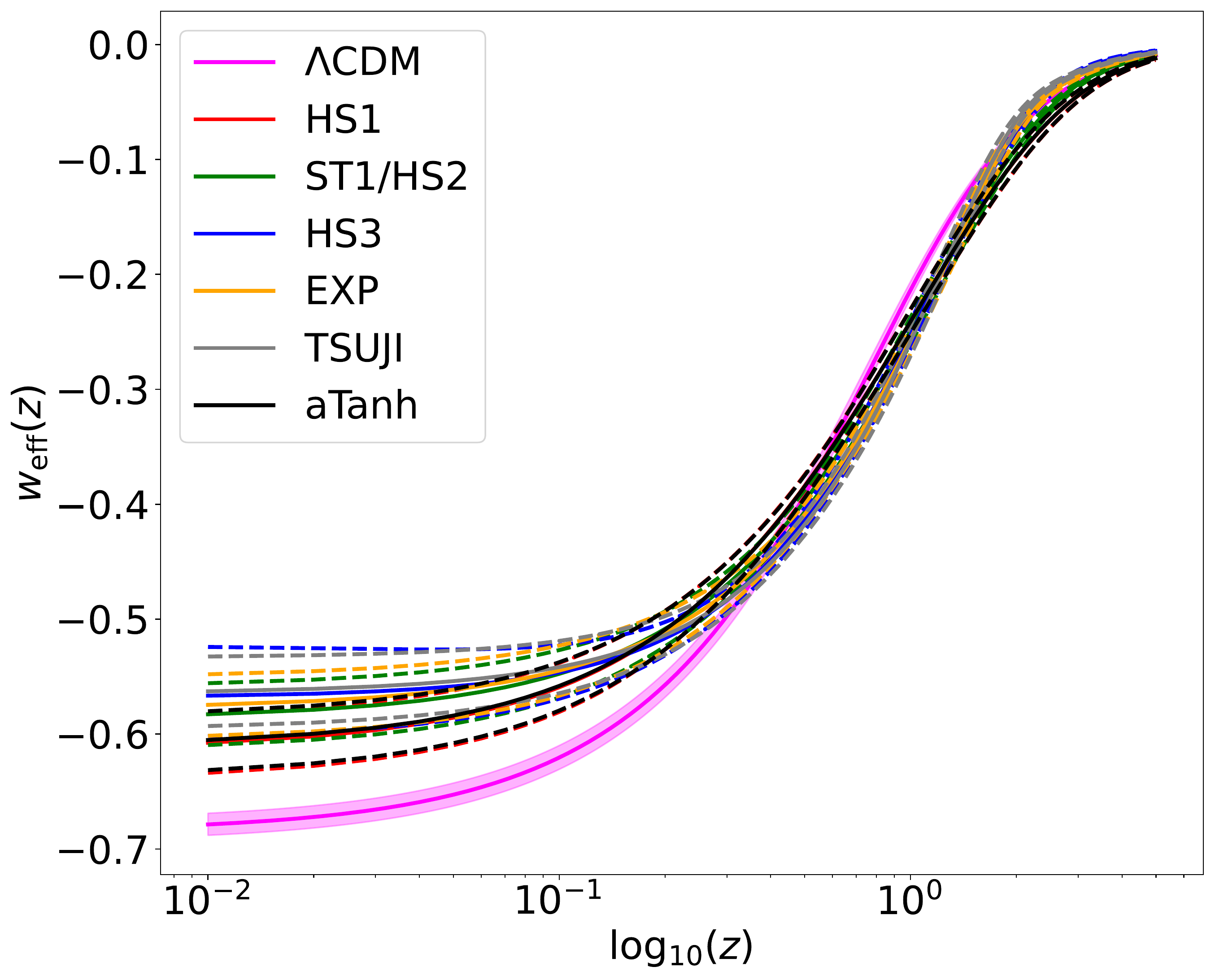}
\caption{Evolution of the total EoS parameter for all $f(R)$ models and the $\Lambda$CDM model obtained from the  data-set SBCH$\textsc{ii}$. Different colored solid lines show median values for different models as indicated by legends whereas dashed lines(or shaded area) with colors show 1-sigma confidence interval. For better contrast we choose $\log_{10}(z)$ as independent variable.}
\label{weff_evol}
\end{figure}
\begin{figure}
\centering
\includegraphics[scale=0.32]{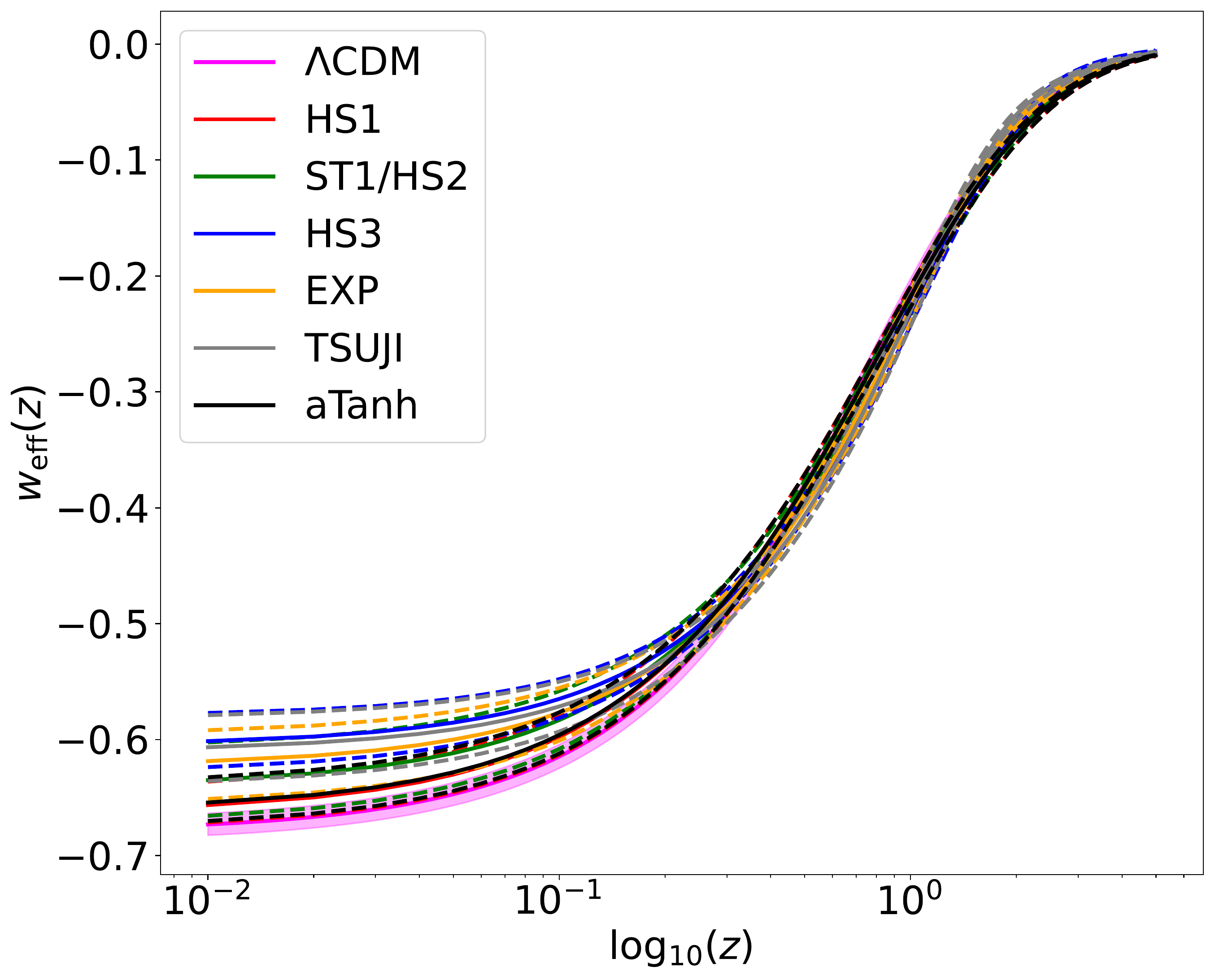}
\caption{Evolution of the total EoS parameter for all $f(R)$ models and the $\Lambda$CDM model obtained from the  data-set SBCH$\textsc{ii}H_{0}$. Different colored solid lines show median values for different models as indicated by legends whereas dashed lines(or shaded area) with colors show 1-sigma confidence interval. For better contrast we choose $\log_{10}(z)$ as independent variable.}
\label{weff_evolH0}
\end{figure}

The behavior of $w_{\rm eff}(z)$ is depicted in Figs. \ref{weff_evol} and \ref{weff_evolH0}, revealing a transition from a matter-dominated era (represented by $w_{\rm eff}(z)\rightarrow 0$) to a dark energy dominated era around $z\sim 5$ for all the $f(R)$ models. Considering the SH0ES prior for $H_{0}$ leads to relatively lower values of $w_{\rm eff,0}$ for all the $f(R)$ models compared to the cases without the prior. Furthermore, the discrepancy between the predictions of $w_{\rm eff}(z)$ obtained from the $\Lambda$CDM model and those from the $f(R)$ models reduces when the SH0ES prior is included.\\

\section{Conclusions}
\label{concl}

In this study, we used the recently released Pantheonplus SNIa data along with cosmic chronometer measurements, baryonic acoustic oscillation measurements, H\textsc{ii} starburst galaxies data, and the local measurement of the Hubble parameter, to constrain popularly studied $f(R)$ models in the metric formalism. The redshift coverage of these data-sets is  $0.0012 \lesssim z \lesssim 2.55$, and the sparse region beyond $z>1.4$ is complemented by the data from H\textsc{ii} galaxies. The inclusion of H\textsc{ii} galaxies data has shown improvements in parameter constraints, even though its use in constraining $f(R)$ models is not yet widespread.
In light of the so-called ``Hubble tension",
we have also examined cases with vis-a-vis without a SH0ES prior for $H_{0}$ --- a feature which is very often lacking in many earlier studies (see references in Sec \ref{intro}). It is noteworthy to find that incorporating this prior for $H_{0}$, based on local measurements, generally leads to reduction in the differences between predictions of $f(R)$ models and the $\Lambda$CDM model

The six viable $f(R)$ models examined in this study are the Hu-Sawicki model with $n_{{\rm HS}} = 1, 3$, the Starobinsky model ($n_{{\rm S}} = 1)$, the exponential, the Tsujikawa and the arcTanh model. Reparametrisation of these $f(R)$ models using the deviation parameter $b$, shows that in the limit 
$b \to 0$, these models converge to the standard 
$\Lambda$CDM model.
Each of these $f(R)$ models is characterized by three parameters: the deviation parameter $b$, the present-epoch values of the matter density parameter $\Omega_{m0}$, and the Hubble parameter $H_{0}$.\\

Regardless of whether the SH0ES prior for $H_0$ is considered, the $\Omega_{m0}$ values for all the examined models fall within the 1-2 sigma range of the Planck constrained value ($\Omega_{m0,\rm Planck} = 0.315\pm 0.007$) or, conversely, the Planck value lies within the 1-2 sigma range of the model values. 
When no SH0ES prior for $H_{0}$ is considered, the fitted values of $H_0$ for the models are also  within the 1-2 sigma range of that from the Planck constraint ($H_{0,\rm Planck} = 67.4\pm 0.5$), 
or alternatively, the Planck value lies within the 1-2 sigma range 
of the fitted model values. 
When incorporating the SH0ES prior on $H_0$, the fitted model values of $H_0$ tend to lie within the 2-3 sigma range on the lower side of the SH0ES measurement of $H_{0,\rm SH0ES} = 73.04\pm 1.04$. An additional impact resulting from the SH0ES prior on $H_0$ is the tightening of constraints on all three model parameters for all the models and data-sets.\\

In our study, we have found instances of all the investigated $f(R)$ models where the deviation parameter $b$ significantly deviates from zero, indicating that $b=0$ is only marginally allowed.  It is worth highlighting that 
these cases also correspond to the situations where $\Delta$AIC and/or $\Delta$BIC based analysis suggest that the $f(R)$ models are (very) strongly favoured over the $\Lambda$CDM model.
Based on our knowledge of previous studies (referred in Sec \ref{intro}), it can be stated that up until now, the support for $f(R)$ models has mainly been weak or positive (based on $\Delta$AIC and/or $\Delta$BIC). However, our current research yields (very) strong support for $f(R)$ models, along with support for non-zero values of the deviation parameter $b$. This indicates that the viability of $f(R)$ models have not yet been ruled out  by the available cosmological data sets.\\

We find that the  relevant quantities characterizing the (accelerated) expansion of the universe --- $z_{\rm t}$, $w_{\rm eff,0}$, and $w_{\rm DE,0}$ --- estimated
in this study are compatible  with their model-independent estimates from earlier works. All the models examined in this study,   predict that the universe underwent a transition from a decelerated phase of expansion to an accelerated phase of expansion in the recent past, occurring at $z_{\rm t}\sim 0.5-1$. Furthermore, the current values of $w_{\rm DE}$ fall within the quintessential region, having recently crossed over from the phantom region (around $z\sim 0.5-2$, depending on the specific models).\\

The instances of strong evidence in favor of $f(R)$ models accompanied by clear non-zero $b$(or only very marginal allowance for $b=0$), agreements between the derived quantities $z_{\rm t}$, $w_{\rm eff,0}$, and $w_{\rm DE,0}$ here and their expected values from model-independent predictions, suggest that the analyzed cosmological data sets
allow room for the consideration of viability of $f(R)$ models as an explanation for the observed late time cosmic acceleration. In fact, it would be worthwhile to conduct even further investigations of these models using future data sets.

  
\section*{Data Availability}
The observed cosmological data such as SNIa, CC, BAO, H\textsc{ii} starburst galaxies and local measurement of $H_{0}$ are publicly available --- the references to which are cited in the text. The simulated data generated in this work are available from the corresponding author, KR, upon reasonable request.

\section*{Acknowledgements}
K.R. would like to thank HoD, Dept. of Comp.
Sc., RKMVERI, for providing computational facilities.
A.C. would like to thank Indian Institute
of Technology, Kanpur, for supporting this work by
means of Institute Post-Doctoral Fellowship (Ref. No.
DF/PDF197/2020-IITK/970).





\appendix

\section{The Modified Friedmann Equations as a System of First Order ODEs}
\label{odesol}
\begin{figure}
\centering
\includegraphics[scale=0.35]{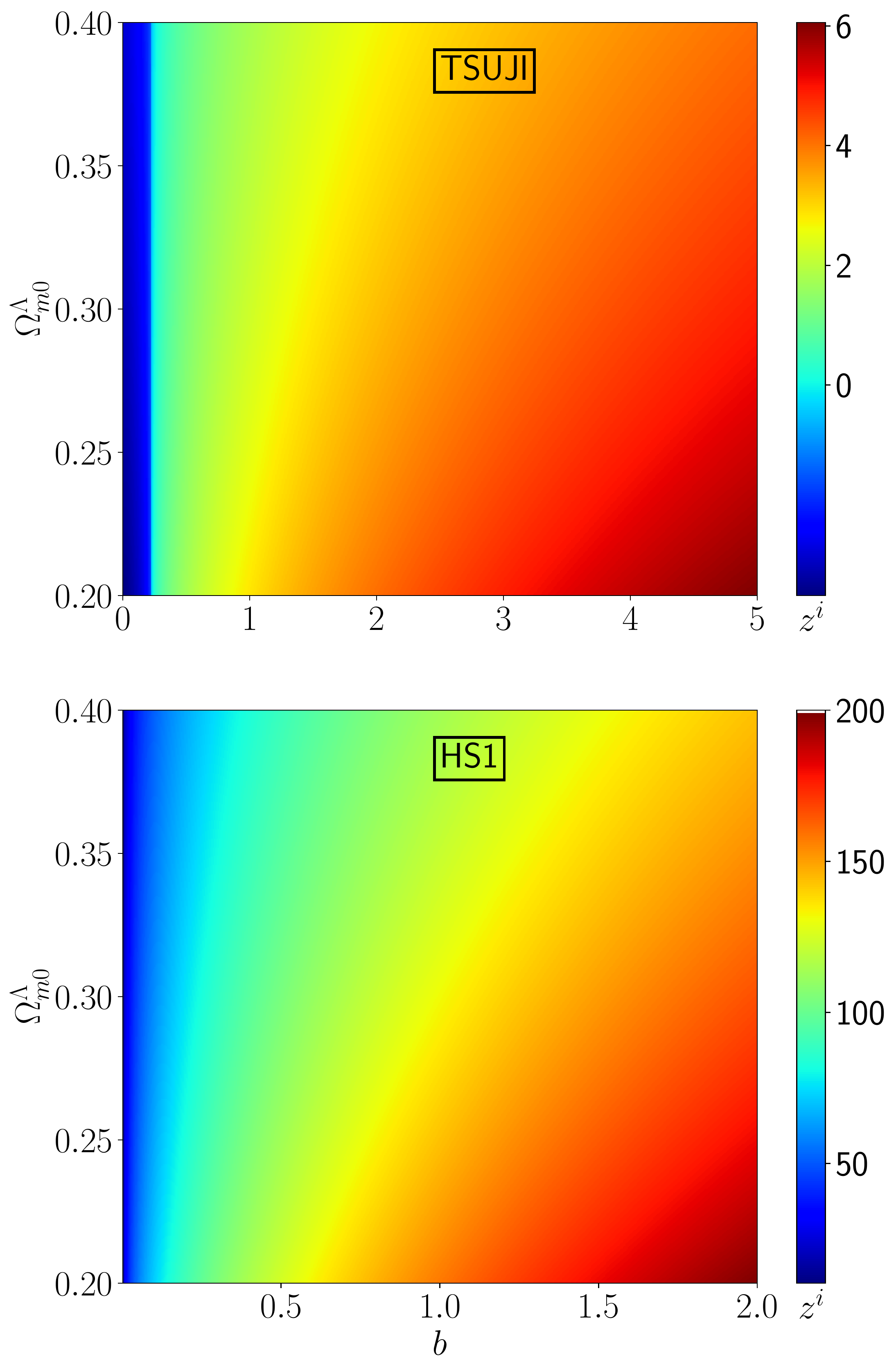}
\caption{The color projection plot of relation Eq. \ref{zswitch} which show $z^{i}(\Om0L,\,b)$ below which the solutions need to be switched from the $\Lambda$CDM to $f(R)$(at $z^{i}$ we obtain the initial conditions for the ODE system Eqs. \ref{O-1} to \ref{O-6} from $\Lambda$CDM approximations). Top panel is plot is for the Tsujikawa model and the bottom panel plot is for the Hu-Sawicki model($n_{_{\rm HS}}=1$).}
\label{zswitch2D}
\end{figure}

The strategies employed to solve the system of Eqs. \ref{Ricci}, \ref{FE111} and 
\ref{FE222} have been proposed and utilized in various earlier studies\cite{Amendola:2006we, Carloni:2007br,Abdelwahab:2011dk,delaCruz-Dombriz:2015tye,Odintsov:2017tbc,Odintsov:2017qif,Farrugia:2021zwx}.
Although there has been proposals for using
model-dependent schemes \cite{Odintsov:2017qif,Leizerovich:2021ksf} too,
in our current work, we were able to obtain stable solutions for all the considered $f(R)$ models using a single scheme.  
This scheme involves transforming the ODEs into a system of first-order ODEs using dynamical variables, which aviods numerical instability and reduces overall computational cost. Based on the literature reviewed, we define the following dimensionless variables
\be
\begin{split}
&X_{1} = \frac{R}{6(\eta\H0L)^{2}},\,\,X_{2} = \frac{\dot{R}F^{\prime}}{(\eta\H0L)F},\,\,X_{3} = \frac{f}{6(\eta\H0L)^{2}F}\,,\\
&O_{\rm m} = \frac{\Om0L(1+z)^{3}}{\eta^{2}F},\,\,O_{\rm r} = \frac{\Or0L(1+z)^{4}}{\eta^{2}F}\,,\\
& H = \eta\H0L,\,\,\Gamma = \frac{F}{RF^{\prime}},\,\,r = \frac{R}{R_{\star}}\,,
\end{split}
\label{dvar}
\ee
where $z$ denotes redshift, $\Om0L$ and $\Or0L$ respectively represent
 the matter and radiation density at present epoch, and $\eta=H/H_{0}^{\Lambda}$. The superscript $\Lambda$ denotes quantities inferred  in a $\Lambda$CDM paradigm. 
The parameter $R_{\star}$ has the dimension of Ricci scalar, 
and it's value gets fixed by the choice of a specific $f(R)$ model
(e.g. for the Hu-Sawicki model $R_{\star}\equiv\mu^{2}$, for the Starobinsky model 
$R_{\star}\equiv{R}_{\rm S}$, for the exponential model $R_{\star}\equiv 1/\beta$ and for the Tsujikawa model $R_{\star}\equiv R_{\rm T}$). 
The fact that any $f(R)$ model is expected to approach $\Lambda$CDM cosmology at sufficiently high redshifts (say $z^{i}$, initial redshift)  
allows us to set the necessary initial conditions 
for the  ODE  system. The determination of $z^i$ will be discussed later in this section.\\

In terms of the above dynamical/dimensionless variables defined in Eq. \ref{dvar}, the 
Eqs. \ref{Ricci} and \ref{FE111} can be written as
\be
\frac{d\eta}{dz} = \frac{\eta}{(1+z)}(2 - X_{1})\,,
\ee 
and,
\be
1 = O_{\rm m} + O_{\rm r} + X_{1} - X_{2} - X_{3}\,,
\label{chauchy_init}
\ee
respectively. Thus it becomes evident from the expression in 
Eq. \ref{chauchy_init} that it serves the purpose of setting the 
initial conditions and monitoring the solutions at each steps
of integration. We can also express the equation for effective geometric dark energy(Eq. \ref{w_DE1}) in terms of above variables as 
\be
w_{\rm DE} = \frac{w_{\rm tot} - O_{\rm r}F/3}{1 - (O_{\rm m}+O_{\rm r})F}\,.
\label{w_DE2}
\ee
We  further define $N = -\log(1+z)$, which allows us to rewrite the system of first-order  ODEs  that need to be solved as follows:
\begin{eqnarray}
\frac{dX_{1}}{dN} &=& X_{1}\left(X_{2}\Gamma - 2X_{1}+4\right)\,,
\label{O-1}\\
\frac{dX_{2}}{dN} &=& (O_{\rm m} + 2X_{1} - X_{2} - 4X_{3} - X_{1}X_{2}  - X_{2}^{2}),
\label{O-2}\\
\frac{dX_{3}}{dN}&=& ( X_{1}X_{2}\Gamma -X_{2}X_{3} + 4X_{3} -2X_{1}X_{3})\,,
\label{O-3}\\
\frac{dO_{\rm m}}{dN} &=& -O_{\rm m}(-1 + 2X_{1} + X_{2})\,, 
\label{O-4}\\
\frac{dr}{dN} &=& X_{2}\Gamma{r}\,,
\label{O-5}\\
{\rm and},\nonumber\\
\frac{dO_{\rm r}}{dN} &=& -O_{\rm r}(2X_{1} + X_{2})\,,
\label{O-6}
\end{eqnarray}
where $\Gamma = F/(RF')$. We choose $\Or0L=2.9656\times10^{-4}\Om0L$, and with this choice,  Eq. \ref{O-6}  becomes redundant as it reduces to Eq. \ref{O-4}. Therefore, we only need to solve the system of ODEs consisting of  Eqs. \ref{O-1} to \ref{O-5}   and not Eq. \ref{O-6}. The reason for considering $\Or0L$  (and hence $O_r$) will be explained later in this section. Ultimately, by solving the above system, we obtain $H = \sqrt{rR_{\star}/(6X_1)}$. Using this expression for $H(z)$, we can determine other cosmological observables as defined in Sec. \ref{ocd}. While we perform model-fitting with parameters $\Om0L$, $b$ and $H^{\Lambda}_{0}$, we eventually obtain the relevant parameters for the $f(R)$ models, \textbf{viz.} $\Omega_{m0}^{f(R)}$, $b$, and $H_{0}^{f(R)}$ (using Eq. \ref{LfR2}), where $H_{0}^{f(R)}$ is the numerical solution of the aforementioned $H$ at $z=0$.\\

The initial conditions needed to solve the system of Eqs. \ref{O-1} to \ref{O-5} are given by
\be
\begin{split}
X_{1}^{i}&=\frac{\Om0L(1+z^{i})^{3} + 4\Ol0L}{2\eta^{i\,2}}\,,\\
X_{2}^{i} &= 0\,,\\
X_{3}^{i}&=\frac{\Om0L(1+z^{i})^{3} + 2\Ol0L}{2\eta^{i\,2}}\,,\\
O_{\rm m}^{i} &= 1 - O_{\rm r}^{i} - X_{1}^{i} + X_{2}^{i} + X_{3}^{i}\,,\\
r^{i}&=\frac{3(\H0L)^{2}}{R_{\star}}\left[(\Om0L(1+z^{i})^{3} + 4\Ol0L\right]\,,
\end{split}
\ee
where $\eta^{i\,2} = \Om0L(1+z^{i})^{3} + \Or0L(1+z^{i})^{4} + \Ol0L$, $O_{\rm r}^{i} = \frac{\Or0L(1+z^{i})^{4}}{\eta^{i\,2}}$ and $\Ol0L = (1 - \Om0L - \Or0L)$.
\par
In order to estimate $z_{i}$ we set $f(R(z_{i})) = R - 2\Lambda(1-\epsilon)$ where $\epsilon\ll1$. Here the expressions for $R$ and $\Lambda$ correspond to those in the $\Lambda$CDM cosmology. With this one obtains the following expression for $z_i$ (see \cite{Odintsov:2017qif,Farrugia:2021zwx} for  derivation):
\be
z_{i} = \left[\frac{4\Ol0L}{\Om0L}\left(\frac{b}{4\nu_{\rm f}} - 1 \right) \right]^{1/3} - 1\,,
\label{zswitch}
\ee
where $\nu_{\rm f}$ for different models (f) are given by following expressions:
\be
\begin{split}
&\nu_{_{\rm HS}} = \frac{1}{(1/\epsilon - 1)^{1/n_{_{\rm HS}}}},\,\,\,\nu_{_{\rm ST}} = \sqrt{(1/\epsilon)^{1/n_{_{\rm S}}} - 1}\,,\\
&\nu_{_{\rm E}} = -1/\ln\epsilon\,,\,\,\,\nu_{_{\rm T}}= 1/\mathrm{arctanh}(1-\epsilon)\,,\,\,\,\nu_{_{\rm aTanh}}\approx \epsilon\,.
\end{split}
\ee
If, during MCMC sampling, the value of $z^i$ for a particular set of parameters ($\Om0L,\,b$) is found to be smaller than the maximum redshift ($z_{\rm max}$) of the data, we switch from the $\Lambda$CDM solution within the range $[z^i,z_{\rm max}]$ to the $f(R)$ solution for the range $[0,z^i]$ at $z^i$. Depending on $f(R)$ models,
for some samples of ($\Om0L,\,b$) Eq. \ref{zswitch} may yield $z^i \leq 0$. This indicates that for such situations the $\Lambda$CDM cosmology is the solution for all $z>0$(for example, see the top panel plot in Fig. \ref{zswitch2D}).\\

Prior to running the MCMC code for a large number of samples, 
we conducted a preliminary run with a smaller sample size 
(approximately 50,000) to determine suitable values for $\epsilon$. 
This avoids long computation times without comporimsing with 
accuracy of the results. Based on the findings of this pilot project, 
we choose $\epsilon = 10^{-10}$ for the exponential and the Tsujikawa models, 
$\epsilon=10^{-8}$ for the Starobinsky model and the Hu-Sawicki model with 
$n_{_{\rm HS}}=3$, and $\epsilon=10^{-6}$ for the Hu-Sawicki model with 
$n_{_{\rm HS}}=1$ and the aTanh model.\\

We have included a color projection plot in Fig. \ref{zswitch2D} illustrating the
relationship described in Eq . \ref{zswitch} for the Tsujikawa model 
and the Hu-Sawicki model ($n_{_{\rm HS}}=1$). From this plot, 
it can be observed
that for some samples of ($\Om0L,\,b$), 
the value of $z^{i}$ can reach as high as 
$ ~200$. In such cases, when we have high $z^{i}$ 
one must include $\Or0L$ in the definition of $\eta^{i}$ 
even though $\Or0L$ is not a free parameter and 
one has data-set comprising   lower  redshifts (as in our case).
Through our preliminary investigations, we observed that excluding $\Or0L$ 
led to numerical instability and inaccuracies when solving 
the system of ODEs described by Eqs. \ref{O-1} to \ref{O-6}. 
Therefore, based on both theoretical principles and our practical experience, 
we made the decision to include $\Or0L$ in our analysis.

\section{BAO and CC Data}
The data-sets for BAO and CC used in this work is given in the Tables \ref{BAOdata} and \ref{CCdata}, respectively. The covariance matrices, taken from respective cited references, for BAO data points 21-25, 27-28 and 29-30, respectively, are given in the following equations:
\be
C_{1} = \begin{bmatrix}
624.707 &  23.729 &  325.332 &  8.34963 &  157.386 &  3.57778\\
23.729 &  5.60873 &  11.6429 &  2.33996 &  6.39263 &  0.968056\\
325.332 &  11.6429 &  905.777 &  29.3392 &  515.271 &  14.1013\\
8.34963 &  2.33996 &  29.3392 &  5.42327 &  16.1422 &  2.85334\\
157.386 &  6.39263 &  515.271 &  16.1422 &  1375.12 &  40.4327\\
3.57778 &  0.968056 &  14.1013 &  2.85334 &  40.4327 &  6.25936\\
\end{bmatrix},\label{Cov_Alam}
\ee
\be
C_{2} = \begin{bmatrix}
0.0911 & -0.0338\\
-0.0338 & 0.22\\
\end{bmatrix},\label{Cov_Bautista}
\ee
and,
\be
C_{3} = \begin{bmatrix}
0.3047 & 0.1707\\
0.1707 & 0.6373\\
\end{bmatrix}.\label{Cov_Neveux}
\ee
\begin{table}
\caption{The BAO data collected from many cited sources. The three needed covariance matrices for data-points (i)21-25 (ii)27-28 and (iii)29-30 can be found in the respective references and are given here in Eqs. \ref{Cov_Alam}, \ref{Cov_Bautista}, and \ref{Cov_Neveux}, respectively.}
\begin{tabular}{|l| c | c | c | c |}
\hline 
No. & $z_{\rm eff}$ &  Value & Observable  & Reference  \\ 
\hline
1 & $0.106$ & $0.336  \pm 0.015$  & $r_d/D_V$  & \cite{Beutler:2011hx}\\
2 & $0.275$ & $0.1390 \pm 0.0037$ & $r_d/D_V$ & \cite{SDSS:2009ocz}\\
\hline
3   &   0.122   &   $ 3.65   \pm   0.115 $ & $D_V/r_d$ & \cite{Carter:2018vce}\\
4   &   0.15   &   $ 4.47   \pm   0.168 $   &  $D_V/r_d$  & \cite{Ross:2014qpa}\\
5   &   0.32   &   $ 8.47   \pm   0.167 $    & $D_V/r_d$ & \cite{Tojeiro:2014eea}\\
6 & 0.35  & $8.88   \pm 0.17$     & $D_V/r_d$ & \cite{Padmanabhan:2012hf}\\
7   &   0.44   &   $ 11.55   \pm   0.559 $ &  $D_V/r_d$  &	\cite{Kazin:2014qga}\\
8 & $0.57$  & $13.67  \pm 0.22$   &  $D_V/r_d$ & \cite{Anderson:2012sa}\\
9   &   0.6   &   $ 14.95   \pm   0.68 $  &  $D_V/r_d$  & \cite{Kazin:2014qga}\\
10   &   0.72   &   $ 16.08   \pm   0.406 $    & $D_V/r_d$ & \cite{Bautista:2017wwp}\\
11   &   0.73   &   $ 16.93   \pm   0.579 $ &  $D_V/r_d$  & \cite{Kazin:2014qga}\\
12   &   1.52   &   $ 26.0   \pm   0.995 $ &  $D_V/r_d$  & \cite{Ata:2017dya}\\
 \hline
13 & $0.54$  & $9.212  \pm 0.41$   & $D_A/r_d$    & \cite{Seo:2012xy}\\
14   &   0.697   &   $ 10.39   \pm   0.496 $     & $D_A/r_d$ & \cite{Sridhar:2020czy}\\
15 & $0.81$  & $10.75\pm0.43$ & $D_A/r_d$ &\cite{DES:2017rfo}\\
16   &   0.874   &   $ 11.37   \pm   0.693 $    & $D_A/r_d$ & \cite{Sridhar:2020czy}\\
\hline
17 & 2.3	& 	$9.07 \pm 0.31 $  &  $D_H/r_d$  & 	\cite{Bautista:2017zgn}\\
18 & $2.34$  & $8.86   \pm 0.29$   & $D_H/r_d $ & \cite{deSainteAgathe:2019voe}\\
\hline
19 & 2.36 & 9.0$\pm 0.3$& $c/(r_{d}H)$ & \cite{BOSS:2013igd}\\
\hline 
20 & $2.4$	& 	$36.6 \pm 1.2$  &  $D_M/r_d$  & 	\cite{duMasdesBourboux:2017mrl}\\
\hline 
21 & $0.38$ & 1512.39 & $D_M\left(r_{d,{\rm fid}}/r_d\right)$ & \cite{BOSS:2016wmc}\\
22 & $0.38$ & 81.2087 & $H(z)\left(r_d/r_{d,{\rm fid}}\right)$  & \cite{BOSS:2016wmc}\\
23 & $0.51$ & 1975.22 & $D_M\left(r_{d,{\rm fid}}/r_d\right)$  & \cite{BOSS:2016wmc}\\
24 & $0.51$ & 90.9029 & $H(z)\left(r_d/r_{d,{\rm fid}}\right)$ & \cite{BOSS:2016wmc}\\
25 & $0.61$  & 2306.68 & $D_M\left(r_{d,{\rm fid}}/r_d\right)$ & \cite{BOSS:2016wmc}\\
26 & $0.61$  & 98.9647  & $H(z)\left(r_d/r_{d,{\rm fid}}\right)$ & \cite{BOSS:2016wmc}\\
\hline
27  & $0.698$	& 	$ 17.65 \pm 0.3$    &  $D_M/r_d$   & \cite{Bautista:2020ahg}\\
28 & $0.698$	& 	$ 19.77 \pm 0.47$    &  $D_H/r_d$   & \cite{Bautista:2020ahg}\\
\hline
29 & $1.48$	& 	$  13.23 \pm 0.47$    & $D_H/r_d$   & \cite{Neveux:2020voa} \\
30 & $1.48$	& 	$  30.21 \pm 0.79$    &  $D_M/r_d$   & \cite{Neveux:2020voa} \\
\hline
\end{tabular}
\label{BAOdata}
\end{table}

\begin{table}
\caption{The CC data collected from many cited sources.}
\begin{tabular}{|l| c | c | c |}
\hline
$z$ & $H(z)$ & $\sigma_{H(z)}$ & Reference\\
\hline
0.07 & 69.0 & 19.6 &  \cite{Zhang:2012mp}\\
0.09 & 69 & 12 &  \cite{Simon:2004tf}\\
0.12 & 68.6 & 26.2 &  \cite{Zhang:2012mp}\\
0.17 & 83 & 8 &  \cite{Simon:2004tf}\\
0.179 & 75 & 4 &  \cite{Moresco:2012jh}\\
0.199 & 75 & 5 &  \cite{Moresco:2012jh}\\
0.20 & 72.9 & 29.6 &  \cite{Zhang:2012mp}\\
0.27 & 77 & 14 &  \cite{Simon:2004tf}\\
0.28 & 88.8 & 36.6 &  \cite{Zhang:2012mp}\\
0.352 & 83 & 14 &  \cite{Moresco:2012jh}\\
0.38 & 83 & 13.5 &  \cite{Moresco:2016mzx}\\
0.4 & 95 & 17 &  \cite{Simon:2004tf}\\
0.4004 & 77 & 10.2 &  \cite{Moresco:2016mzx}\\
0.425 & 87.1 & 11.2 &  \cite{Moresco:2016mzx}\\
0.445 & 92.8 & 12.9 &  \cite{Moresco:2016mzx}\\
0.47 & 89.0 & 49.6 &  \cite{Ratsimbazafy:2017vga}\\
0.4783 & 80.9 & 9 &  \cite{Moresco:2016mzx}\\
0.48 & 97 & 62 &  \cite{Stern:2009ep}\\
0.593 & 104 & 13 &  \cite{Moresco:2012jh}\\
0.68 & 92 & 8 &  \cite{Moresco:2012jh}\\
0.75 & 98.8 & 33.6 &  \cite{Borghi:2021rft}\\
0.781 & 105 & 12 &  \cite{Moresco:2012jh}\\
0.875 & 125 & 17 &  \cite{Moresco:2012jh}\\
0.88 & 90 & 40 &  \cite{Stern:2009ep}\\
0.9 &  117 &  23 &  \cite{Simon:2004tf}\\
1.037 & 154 & 20 &  \cite{Moresco:2012jh}\\
1.3 & 168 & 17 &  \cite{Simon:2004tf}\\
1.363 & 160 & 33.6 &  \cite{Moresco:2015cya}\\
1.43 & 177 & 18 &  \cite{Simon:2004tf}\\
1.53 & 140 & 14 &  \cite{Simon:2004tf}\\
1.75 & 202 & 40 &  \cite{Simon:2004tf}\\
1.965 & 186.5 & 50.4 &  \cite{Moresco:2015cya}\\
\hline
\end{tabular}
\label{CCdata}
\end{table}


\bibliographystyle{unsrtnat} 
\bibliography{ref_metric_2023}
\end{document}